\documentclass[twocolumn,superscriptaddress,aps,prb,longbibliography]{revtex4-2}
\usepackage[utf8]{inputenc}
\usepackage[T1]{fontenc}
\usepackage{newtxtext,newtxmath,amsmath,mathtools}
\usepackage{graphicx}
\usepackage[colorlinks]{hyperref}
\usepackage{dcolumn}
\usepackage{bm}
\usepackage{microtype}
\usepackage{physics}
\usepackage{xcolor}
\usepackage[caption=false]{subfig}
\usepackage{orcidlink}
\usepackage{relsize}

\begin{document}

\title{Quantum selection of order and dynamic properties of Kitaev-Heisenberg ferromagnet on a triangular lattice}

\author{Kaushal K. Kesharpu\orcidlink{0000-0003-4933-6819}}
\email{kesharpu@theor.jinr.ru}

\author{Pavel A. Maksimov\orcidlink{0000-0001-9264-7888}}
\email{maksimov@theor.jinr.ru}

\affiliation{Bogolyubov Laboratory of Theoretical Physics, Joint Institute for Nuclear Research, 
Dubna, Moscow region 141980, Russia}

\date{\today}

\begin{abstract}
Recent interest in monolayer materials motivated a search for two-dimensional ferromagnets with sizable spin-orbit coupling. Magnetic anisotropy of exchange Hamiltonian, induced by spin-orbit coupling, may not only stabilize long-range order, but also in turn can be a source of frustration and accidental degeneracy, which is the case for the Kitaev-Heisenberg model. Here we present an extensive study of ground state and excitations of ferromagnetic anisotropic-exchange Kitaev-Heisenberg model on a triangular lattice using order-by-disorder and augmented spin-wave theory calculations. It is shown that while bond-dependent terms of the model do not affect the ground state classically, quantum fluctuations select preferred magnetization direction of the ferromagnetic state and significantly alter classical phase diagram. Anisotropic terms of the magnetic Hamiltonian also give rise to magnon-magnon interactions that lead to spontaneous decay and spectral renormalization, which we illustrate using non-linear spin-wave theory.

\end{abstract}

\maketitle


\section{Introduction}
\label{sec:introduction}
Discovery of two-dimensional ferromagnets \cite{Gong2017,Fei2018,Deng2018,Zhang_2019,Parkin_2021} sparked a large field of research with a wealth of materials being studied \cite{Gong_2019_review,Soriano_2020_review}. Even though these materials are confined to one layer and Mermin-Wagner theorem would prohibit spontaneous symmetry breaking in the isotropic case \cite{MerminWagner}, magnetic order is nonetheless stabilized by easy-axis anisotropy of exchange Hamiltonian \cite{Huang2017}. While this anisotropy is a result of strong spin-orbit coupling (SOC) \cite{HYKee_2021}, in some cases SOC can induce other types of anisotropy, which introduce frustration to the magnetic system \cite{Witczak_Krempa14}, instead of order.

Frustration, a situation where all of the interactions in the system cannot be satisfied simultaneously, is actually ubiquitous in nature and can be found not only in magnetic crystals 
 \cite{Toulouse_frustration,Vannimenus_1977} but also in biological systems \cite{Ferreiro_2018} and glasses \cite{Shintani2006}. In the specific case of frustrated magnetic systems it is the different contributions of the exchange Hamiltonian that cannot be all minimized concurrently \cite{ramirez_strongly_1994}. Such a coexistence of competing interactions may lead to a degeneracy of the ground state \cite{wannier_antiferromagnetism._1950}, which in turn can destabilize long-range order in favor of a quantum spin liquid state \cite{ShastrySutherland,balents_review,savary_balents}.

While in some cases this degeneracy is an intrinsic property of the Hamiltonian and yields an exotic ground state \cite{White_kagome_2011,zhu_white15,iqbal16_j1j2}, there are situations where the degeneracy is an artifact of a semiclassical approach and hence is accidental~\cite{villain1980-Ordereffect}.
In the latter case additional contributions to the energy of the system beyond the mean-field procedure can lift the degeneracy through thermal and quantum fluctuations via order-by-disorder mechanism \cite{shender1982-Antiferromagneticgarnets,henley1987-Orderingdisorder,henley1989-Orderingdue,chandra_coleman_larkin,chubukov_j1j2}.

Notable cases of order-by-disorder effect include stabilization of a $1/3$ magnetization plateau by fluctuations in a triangular lattice antiferromagnet~\cite{chubukov91,alicea_plateau,Takano_2011}, as well as order selection in other geometrically frustrated lattices~\cite{chubukov_kagome,Chern_kagome_2013,AFkagome1,Zh_pyrochlore,McClarty_pyrochlore_2014,Jaubert_pyrochlore_2015,Lines_fcc,Zh_fcc1,Zh_fcc2}. Moreover, magnetic Hamiltonian can be frustrated not only by the geometry of the lattice but also by anisotropic contributions beyond Heisenberg model. For instance, compass models \cite{nussinov2015-Compassmodels} that contain bond-dependent interactions are known to exhibit accidental degeneracy, which is lifted by quantum fluctuations \cite{Giniyat_ObD_2001}. One of the most well-studied compass models is a Kitaev honeycomb model \cite{KITAEV2006}, which allows for exact spin liquid solution and contains excitations with topological properties. While adding Heisenberg interaction destabilizes spin liquid and promotes long range order, the direction of spins in the Kitaev-Heisenberg model can only be obtained when quantum \cite{chaloupka2010} or thermal \cite{Perkins_KH_ObD_2016} fluctuations are included. It was also shown that further-neighbor and other bond-dependent terms can leave accidental degeneracy intact and necessitate calculations of quantum corrections \cite{rousochatzakis2015-PhaseDiagram,Coldea_Co_2020}.

\begin{figure}
\centering
\includegraphics[width=\columnwidth]{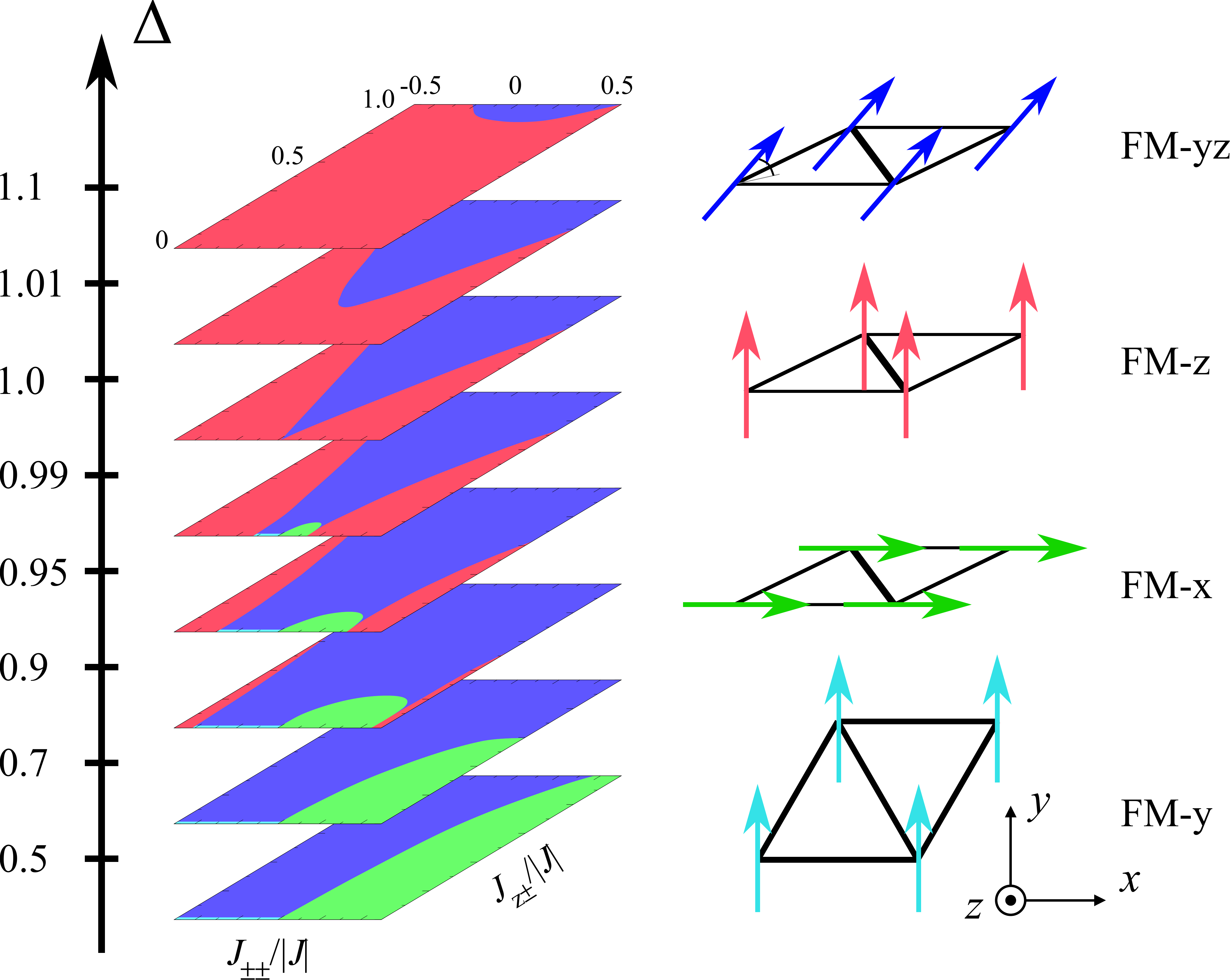}
\caption{Schematic spin-wave theory (SWT) phase diagram of the ferromagnetic regime of model \eqref{HJpm} for various values of $XXZ$ anisotropy $\Delta$. Four stable ferromagnetic configurations are shown. Note that FM-y state is only stable for $J_{z\pm}=0$.}
\label{fig_states}
\end{figure}
In general, Kitaev-Heisenberg model can also be defined on other lattices \cite{Kimchi14}, such as triangular lattice \cite{Ioannis_Z2}. It was shown to host an expansive phase diagram with various ordered and spin liquid states~\cite{Wang17,Li_CSL,Punk,prx_anisotropic,Tohyama}. Generally, these materials, where Kitaev interactions are realized, have a configuration of edge-sharing ligand octahedra \cite{Jackeli}. This symmetry allows for additional interactions that generalize the model to the extended Kitaev-Heisenberg model \cite{rau_jkg,rau2014trigonal,Chen3,Rau_tr}, also referred to as anisotropic-exchange model \cite{prx_anisotropic}.

In this paper, we study a ferromagnetic region of the phase diagram of anisotropic-exchange triangular lattice magnets \cite{prx_anisotropic} to explore its ground state configuration and excitations. While most of triangular lattice materials with bond-dependent interactions, such as YbMgGaO$_4$ \cite{SciRep,Chen1,MM,MM2,us_mimicry},YbZnGaO$_4$ \cite{Wen_freeze18,YZGO_2021,Wen_YZGO_2021,Blundell_YZGO_2022} and the delafossite family \cite{Liu_2018,Wilson_NaYbO2_2019,Wu2022,Xie2023,KErSe2_INS_2023,Scheie2024,CsCeSe2_Nikitin_2024,Xie_2024}, are antiferromagnets, NaRuO$_2$ \cite{Shikano_2004,Ortiz2023} was proposed to have strong nearest-neighbor ferromagnetic exchange along with significant Kitaev interactions \cite{Razpopov2023,Bhattacharyya2023}. There are also other triangular lattice ferromagnets \cite{Moller_2012,Rawl_2017}, and therefore a study of phase diagram and excitations of the ferromagnetic regime of extended Kitaev-Heisenberg model is of timely interest.

Order-by-disorder effect in Kitaev-Heisenberg model on the triangular lattice have been studied before \cite{Ioannis_Z2,Trebst_tr,Avella}, and it was shown that quantum fluctuations select cubic axes as a preferred magnetization direction. Here we study the full extended Kitaev-Heisenberg model with four parameters allowed by the symmetry of edge-sharing octahedra, common for the class of materials mentioned above, and show that depending on the ratio of exchange parameters, other spin directions can also be chosen. 

In particular, we show that the bond-dependent part of the Hamiltonian does not contribute to the classical energy, and in the isotropic limit of bond-independent part of the model quantum fluctuations select various states, including out-of-plane canted state, whose canting angle depends on the ratio of exchange integrals. Naively, $XXZ$ anisotropy would prefer out-of-plane or in-plane configuration but continuity argument suggests that the canted state should survive even for the easy-plane or easy-axis anisotropy. We confirm that using minimally-augmented spin-wave theory (MAGSWT) \cite{shengtao_j1j3} and density matrix renormalization group (DMRG) \cite{white_density_1992} calculations. Our results are summarized in Fig.~\ref{fig_states}. Moreover, we show that anisotropic bond-dependent terms contribute to magnon interaction self-energy, which can be significant in some parts of the Brillouin zone and yields strong spectral redistribution.

This paper is structured as follows. In Sec.~\ref{sec:Delta1} we discuss accidental degeneracy for the isotropic case of bond-independent terms and present the phase diagram with quantum fluctuations included. In Sec.~\ref{sec:MAGSWT} we present calculations for $XXZ$-anisotropic model using a modified version of MAGSWT and show the evolution of the phase diagram. In Sec.~\ref{sec:spectrum} we calculate magnetic spectrum with $1/S$ corrections included and present dynamical structure factor. We conclude and discuss our results in Sec.~\ref{sec:conclusions}.

\section{Ground state selection for $\Delta=1$}
\label{sec:Delta1}
\begin{figure}
\centering
\includegraphics[width=0.75\columnwidth]{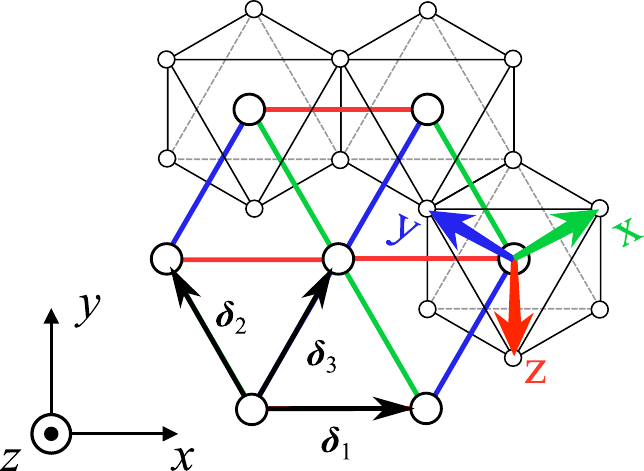}
\caption{Triangular lattice with edge-sharing ligand octahedra and nearest-neighbor vectors. Crystallographic and cubic axes are also shown.}
\label{fig_axes}
\end{figure}
The model that we study is motivated by two-dimensional triangular lattice with edge-sharing ligand octahedra \cite{Chen1,rau_jkg,rau2014trigonal,prx_anisotropic}, and its symmetry allows for four terms of the exchange Hamiltonian:
\begin{align}
{\cal H}&=\sum_{\langle ij\rangle_\alpha}
J \left(S^{x}_i S^{x}_j+S^{y}_i S^{y}_j+\Delta S^{z}_i S^{z}_j\right)\nonumber\\
+&2 J_{\pm \pm} \left[ \left( S^x_i S^x_j - S^y_i S^y_j \right) \tilde{c}_\alpha -\left( S^x_i S^y_j+S^y_i S^x_j\right) \tilde{s}_\alpha \right]\nonumber\\
 &+J_{z\pm}\left[ \left( S^y_i S^z_j +S^z_i S^y_j \right) \tilde{c}_\alpha -\left( S^x_i S^z_j+S^z_i S^x_j\right)\tilde{s}_\alpha \right],
\label{HJpm}
\end{align}
where $\tilde{c}_\alpha$ and $\tilde{s}_\alpha$ are equal to $\cos \tilde{\varphi}_\alpha$ and $\sin \tilde{\varphi}_\alpha$ with $\tilde{\varphi}_\alpha=\{0,2\pi/3,-2\pi/3\}$ for three types of nearest-neighbor bonds. This model is defined relative to $\{x,y,z\}$ axes shown in Figs.~\ref{fig_states} and \ref{fig_axes}. Through the axes rotation (see Appendix \ref{app:ham}) the anisotropic-exchange model \eqref{HJpm} is equivalent to the extended Kitaev-Heisenberg model
\begin{align}
\mathcal{H}=&\sum_{\langle ij \rangle_\gamma} \Big[{\rm J} {\bf S}_i \cdot {\bf S}_j 
+K S^\gamma_i S^\gamma_j +\Gamma \big(S^\alpha_i S^\beta_j +S^\beta_i S^\alpha_j\big)\nonumber \\
\label{eq_H_JKG}
&\ \ \ \ \ +\Gamma' \big(S^\gamma_i S^\alpha_j+
S^\gamma_i S^\beta_j+S^\alpha_i S^\gamma_j +S^\beta_i S^\gamma_j\big)\Big],
\end{align}
where $\{\alpha,\beta,\gamma\}=\{{\rm y,z,x}\}$ for the bond ${\bm \delta}_1$, and interaction on other bonds is obtained through cyclic permutation. The extended Kitaev-Heisenberg model is defined relative to the cubic axes tied to the ideal ligand octahedra $\{\text{x,y,z}\}$, as shown in Fig.~\ref{fig_axes}. Therefore, we use the terms ``anisotropic-exchange model'' and ``extended Kitaev-Heisenberg model'' interchangeably. 

Note that the model \eqref{HJpm} can be split into bond-independent $XXZ$ interaction and two bond-dependent interactions, $J_{\pm\pm}$ and $J_{z\pm}$. First, we study its phase diagram in the isotropic limit of the $XXZ$ interaction. In a ferromagnetic state for $J<0$, since the energy of interactions does not depend on bond direction, in a classical approach, $S\rightarrow \infty$, the contribution of bond-dependent terms is equal to sum of $\cos \varphi_\alpha$ and $\sin\varphi_\alpha$, which is zero. Therefore, even though the Hamiltonian does not have a continuous degeneracy, its classical energy $E_\text{cl}$ is degenerate, and  the moment direction remains unspecified on a classical level. 

However, since this degeneracy is accidental, it will be lifted by higher-order contributions. In order to calculate these corrections, one needs to perform a spin-wave expansion. We use  Holstein-Primakoff transformation \cite{hp1940} 
\begin{align}
S^x&\approx\sqrt{\frac{S}{2}} \left(a -\frac{a^\dagger a a}{4S} +\text{H.c.} \right),\nonumber\\
S^y&\approx\frac{1}{i}\sqrt{\frac{S}{2}} \left(a -\frac{a^\dagger a a}{4S} -\text{H.c.} \right),\nonumber\\
S^z&=S-a^\dagger a,
\label{eq_HP}
\end{align}
where we omitted higher order terms, and $\{x,y,z\}$ define local reference frame after the axes rotation
\begin{align}
S^x&\rightarrow S^x \sin \theta\cos \varphi-S^y \sin \varphi+S^z \cos\theta\cos\varphi,\nonumber\\
S^y&\rightarrow S^x \sin \theta\sin \varphi+S^y \cos \varphi+S^z \cos\theta\sin\varphi,\nonumber\\
S^z&\rightarrow S^z \sin \theta -S^x \cos \theta.
\end{align}
Here, $\varphi$ is the in-plane azimuthal angle, and $\theta$ is the out-of-plane canting angle.
The spin-wave expansion can be presented as a series with terms $\mathcal{H}^{(n)}$ with different number of bosons $n$:
\begin{align}
\mathcal{H}=\mathcal{H}^{(0)}+\mathcal{H}^{(1)}+\mathcal{H}^{(2)}+\mathcal{H}^{(3)}+\mathcal{H}^{(4)}+\dots
\label{eq_Hexp}
\end{align}

Quadratic Hamiltonian after Fourier transform yields
\begin{align}
\mathcal{H}^{(2)}=\sum_\mathbf{k} A^{\phantom \dagger}_\mathbf{k} a^\dagger_\mathbf{k}a^{\phantom \dagger}_\mathbf{k}-\frac{1}{2}\left(B^{\phantom \dagger}_\mathbf{k} a^\dagger_\mathbf{k} a^\dagger_\mathbf{-k}+B^*_\mathbf{k}a^{\phantom \dagger}_\mathbf{k} a^{\phantom \dagger}_\mathbf{-k}\right).
\label{eq_H2}
\end{align}
\begin{figure}
\centering
\includegraphics[width=\columnwidth]{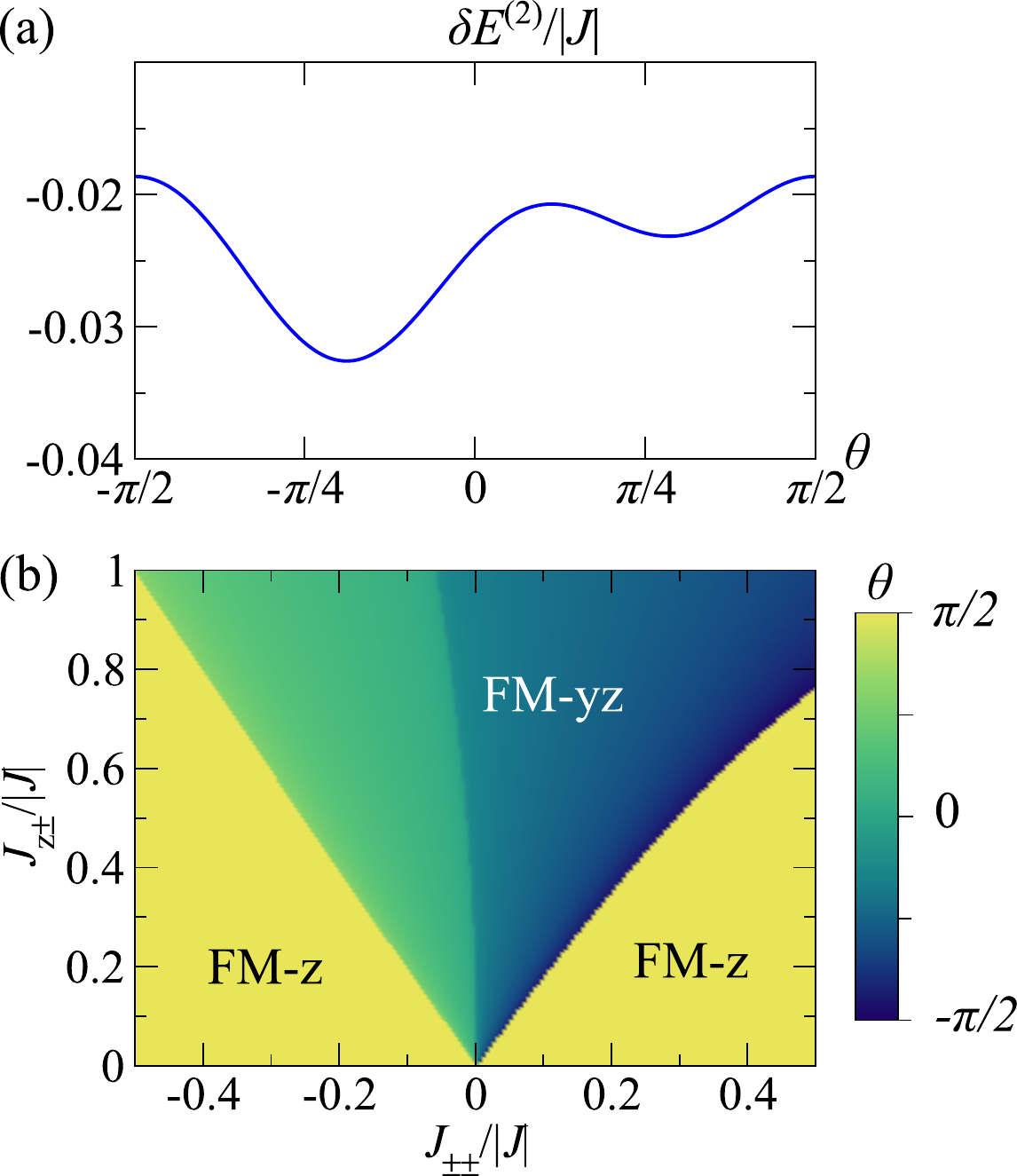}
\caption{(a) Zero-point energy for $J_{\pm\pm}=0.2|J|$, $J_{z\pm}=0.6|J|$, $\Delta=1$ and $\varphi=\pi/2$ as a function of out-of-plane angle $\theta$. (b) Phase diagram for $\Delta=1$ and  $\varphi=\pi/2$ shown as intensity plot of out-of-plane canting angle $\theta$, calculated using linear spin-wave theory with quantum fluctuations included.}
\label{fig_delta1}
\end{figure}
The expressions for $A_\mathbf{k}$ and $B_\mathbf{k}$ have been obtained \cite{prx_anisotropic}  before \footnote{Note that there is a typo in Ref.\cite{prx_anisotropic}: wrong sign in front of $(1-\Delta)$ in $A_\mathbf{k}$.}, we also present the details of spin-wave calculations in Appendix~\ref{app:lswt}. If we define 
\begin{align}
B_\mathbf{k}=\left| B_\mathbf{k} \right| e^{i\phi_\mathbf{k}},
\end{align}
then the diagonalization is performed via transformation
\begin{align}
a^{\phantom \dagger}_\mathbf{k}=u^{\phantom \dagger}_\mathbf{k} e^{i\phi_\mathbf{k}/2} d^{\phantom \dagger}_\mathbf{k}+v^{\phantom \dagger}_\mathbf{k}e^{i\phi_\mathbf{k}/2} d^\dagger_\mathbf{-k}, 
\label{eq_bogolyubov}
\end{align}
where $u_\mathbf{k}^2-v_\mathbf{k}^2=1$ and
\begin{align}
u_\mathbf{k}^2+v_\mathbf{k}^2=\frac{A_\mathbf{k}}{\varepsilon_\mathbf{k}},~2 u^{\phantom \dagger}_\mathbf{k} v^{\phantom \dagger}_\mathbf{k}=\frac{\left| B_\mathbf{k} \right|}{\varepsilon_\mathbf{k}}
\end{align}
Note that here $u_\mathbf{k}$ and $v_\mathbf{k}$ are real functions. This transformation gives the quadratic Hamiltonian
\begin{align}
\mathcal{H}^{(2)}=\sum_\mathbf{k} \varepsilon_\mathbf{k} d^\dagger_\mathbf{k} d^{\phantom \dagger}_\mathbf{k} + \delta E^{(2)},
\end{align}
where the magnon spectrum is
\begin{align}
\varepsilon_\mathbf{k}&=\sqrt{A_\mathbf{k}^2-\left| B_\mathbf{k} \right|^2},
\label{eq_ek}
\end{align}
and the second term is zero-point quantum corrections
\begin{align}
\delta E^{(2)}=\frac{1}{2}\sum_\mathbf{k} \left(\varepsilon_\mathbf{k} -A_\mathbf{k} \right).
\label{eq_e2}
\end{align}
Now, since $A_\mathbf{k}$, $B_\mathbf{k}$ and subsequently, $\varepsilon_\mathbf{k}$ explicitly depend on $\varphi$ and $\theta$, $\delta E^{(2)}$ provides a contribution to the energy of the state that can be minimized to determine the magnetization direction. Minimization yields in-plane angle to be $\varphi=\pi/6+\pi n/3$, $n \in \mathbb{Z}$ for all values of $J_{\pm\pm}$ and $J_{z\pm}$, while canting angle $\theta$ becomes a function of anisotropic exchanges. Calculation for a representative set  $J_{\pm\pm}=0.2|J|$, $J_{z\pm}=0.6|J|$, $S=1/2$ are shown in Fig.~\ref{fig_delta1}(a).

The resulting phase diagram for $\Delta=1$ is shown in Fig.~\ref{fig_delta1}(b) as an intensity plot of the canting angle $\theta$ as a function of bond-dependent terms $J_{\pm\pm}$ and $J_{z\pm}$. There are two states that are stabilized in this phase diagram: FM-yz with non-zero canting angle and, unexpectedly, FM-z state with magnetic moments perpendicular to the lattice plane. It is somewhat expected that $J_{z\pm}$ interaction between in-plane and out-of-plane components of spin promotes a canted state. However, as one can see, for smaller  $J_{z\pm}$ and large $J_{\pm\pm}$ an Ising-like state is energetically preferred. This can be explained by the fact that in the FM-z state $J_{\pm\pm}$ interaction connects fluctuating components, and that is why quantum fluctuations prefer this state.

A clearer picture can be obtained using real-space perturbation theory \cite{Long_RSPT,Bergman_RSPT,Zhitomirsky_RSPT} that shows that virtual hoppings prefer FM-z state in the presence of $J_{\pm\pm}$ interaction. According to Ref.\cite{Zhitomirsky_RSPT}, the lowest-order corrections come from double spin-flip applied to the vacuum state $|0\rangle$ on a bond:
\begin{align}
|00\rangle \xrightarrow{S^-_i S^-_j} |11\rangle \xrightarrow{S^+_i S^+_j} |00\rangle .
\end{align}
The second-order perturbation is given by
\begin{align}
    \delta E_2 \simeq \frac{1}{E_0-E_2} \sum_{\alpha=1}^6 \left|V^{++}_\alpha \right|^2\langle 0| S^+_iS^+_j S^-_i S^-_j|0\rangle,
\end{align}
where $E_0$ and $E_2$ are vacuum and pair spin-flip energies, and the $S^+_i S^+_j$ flip term in the Hamiltonian \eqref{HJpm} in the local reference frame is
\begin{align}
    V^{++}_\alpha=-\frac{J_{\pm\pm}}{2} \left( \sin^2 \theta \tilde{c}_\alpha +\tilde{c}_\alpha-i\sin \theta \tilde{s}_\alpha\right).
\end{align}
Up to a constant, the second-order energy correction can be approximated by
\begin{align}
    \delta E_2 \simeq -\frac{J_{\pm\pm}^2 S}{|J|} \left( 3\sin^2\theta+\sin^4 \theta\right),
\end{align}
which yields a minimum at $\theta=\pm \pi/2$ in agreement with Fig.~\ref{fig_delta1}.

\section{Phase diagram for $\Delta \neq 1$}
\label{sec:MAGSWT}
While naively, one may expect an in-plane magnetization for $\Delta<1$ and out-of-plane FM-z state for $\Delta>1$, the appearance of FM-yz state in the $\Delta=1$ phase diagram suggests that the situation might be more complex. In fact, one can argue that due to continuity, it is very unlikely that FM-yz is stabilized only at $\Delta=1$ point. Instead, one should expect a finite window of its stability, and hence modification of $\Delta\neq 1$ phase diagrams from the simple in-plane and out-of-plane magnetization states.
\begin{figure}
\centering
\includegraphics[width=0.8\columnwidth]{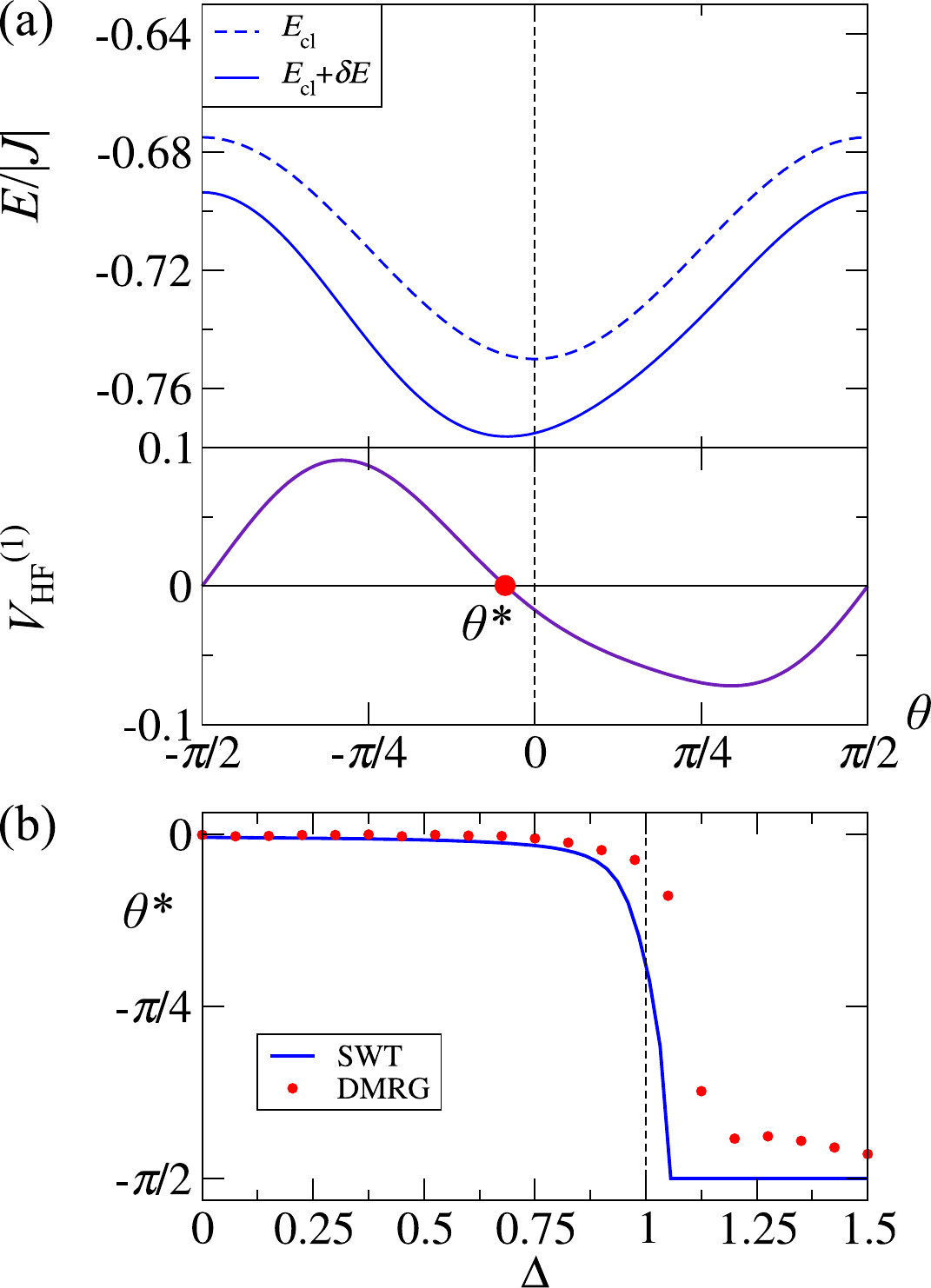}
\caption{(a) Classical energy and energy with quantum fluctuations included for $\Delta=0.9$, $J_{\pm\pm}=0.2|J|$, $J_{z\pm}=0.6|J|$ as a function of out-of-plane angle $\theta$, $\varphi=\pi/2$. Minimum of energy with quantum corrections at $\theta^*$ corresponds to the one-magnon term set to zero according to Eq.~\eqref{eq_V1HF}. (b) $\theta^*$  as a function of $\Delta$, calculated using spin-wave theory (solid line) and with DMRG method (dots).}
\label{fig_thetaHF}
\end{figure}

In order to calculate $\Delta \neq 1$ phase diagrams, one needs to obtain energies of FM-x, FM-y, FM-z and FM-yz states with quantum corrections \eqref{eq_e2} included. Calculations of energy corrections to FM-x or FM-y state for $\Delta<1$  are straightforward and analogous to previous section, since these states are minima of classical energy and their magnon spectrum is positive definite. However, the energies of other states cannot be calculated since their spectrum is imaginary. The same is true for $\Delta>1$ where only the spectrum of FM-z state is well-defined.

\begin{figure*}
\centering
\includegraphics[width=2.0\columnwidth]{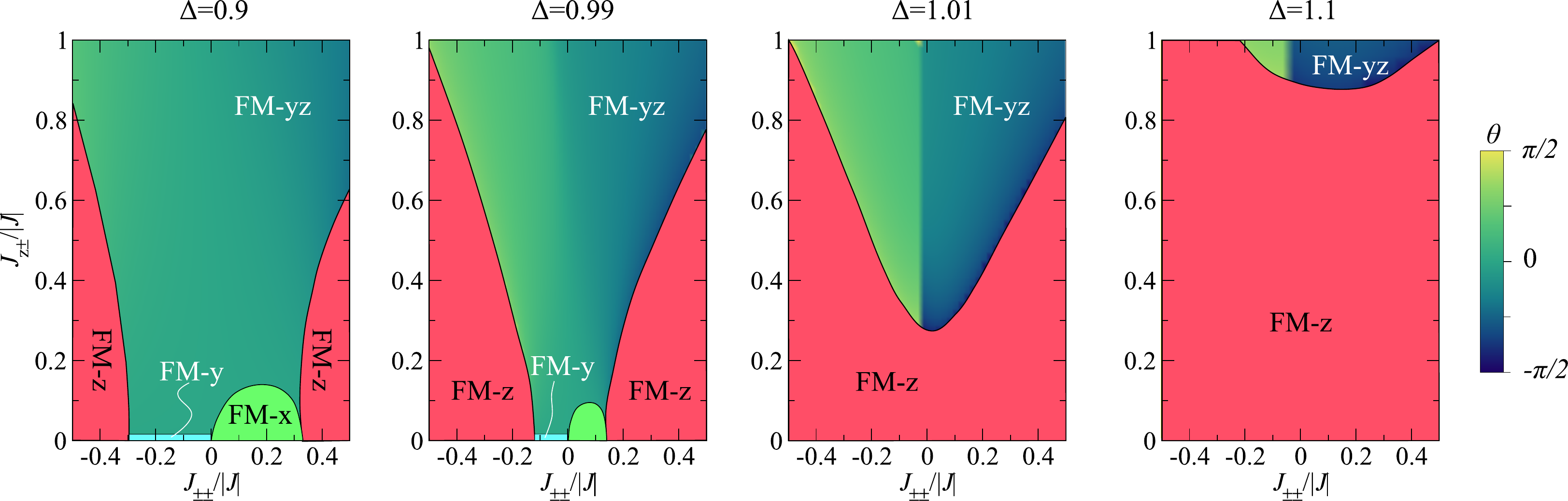}
\caption{Spin-wave theory phase diagram for various $\Delta$, calculated using energies of the states with quantum corrections \eqref{eq_e2} and MAGSWT procedure \eqref{eq_coletta}. Intensity plot indicates the canting angle of the FM-yz state $\theta^*$, which is calculated from Eq.\eqref{eq_V1HF}.}
\label{fig_deltaPD}
\end{figure*}

This artifact of semiclassical expansion can be mitigated by stabilizing the states first by utilizing MAGSWT procedure, which consists of adding a staggering field term \cite{Coletta1,Coletta2} to the Hamiltonian
\begin{align}
\mathcal{H} \rightarrow \mathcal{H}+\mu \sum_i (S-S^z).
\label{eq_coletta}
\end{align}
Using Holstein-Primakoff transformation, one can show that it acts as an additional chemical potential and it renormalizes $A_\mathbf{k}$:
\begin{align}
    \varepsilon_\mathbf{k} \rightarrow \sqrt{\left(A_\mathbf{k}+\mu\right)^2-\left| B_\mathbf{k} \right|^2}.
\end{align}
If $A_\mathbf{k}<\left| B_\mathbf{k} \right|$, the energy eigenvalue is not real, and positive $\mu$ is required to stabilized magnon spectrum.  Since $\mu$ is positive and  $S^z<S$, the additional term in Eq.~\eqref{eq_coletta} is positive and calculation of quantum corrections yields an upper bound of the energy of the state. This method was previously used to study plateau states in frustrated magnets \cite{Coletta3} and unique collinear states of the honeycomb $J_1-J_3$ model \cite{shengtao_j1j3}. One issue can be a search over the Brillouin zone for minimum value of $\mu$ that stabilizes magnon  spectrum in order to avoid overestimating the energy of the state. However, in the case of ferromagnetic states in this paper, instability occurs only at $\mathbf{k}=0$ and $\mu$ can be calculated as
\begin{align}
    \mu=\left| B_\mathbf{k=0} \right|-A_\mathbf{k=0}.
\end{align}

Another issue is that while FM-x, FM-y and FM-z are extrema of the classical energy and one can perform spin-wave expansion around these saddle points, FM-yz is not. It is equivalent to the fact that for the FM-yz state there is a non-zero linear term in the Holstein-Primakoff expansion \eqref{eq_Hexp}:
\begin{align}
\mathcal{H}^{(1)}=\sum_{\langle ij \rangle_\alpha} \left(a_i^{\phantom \dagger} +a_j^{\phantom \dagger} \right) V^{(1)}_\alpha +\text{H.c.},
\end{align}
where the sum is over nearest-neighbor bonds, and
\begin{align}
V^{(1)}_\alpha \frac{\sqrt{2}}{S^{3/2}}&=\frac{J}{2} \sin 2\theta (1-\Delta) \nonumber\\
&+J_{\pm\pm}\left[ \sin 2 \theta \cos(2\varphi+\varphi_\alpha)+2i \cos \theta \sin(2\varphi+\varphi_\alpha)\right]\nonumber\\
&+J_{z\pm}\left[\cos 2\theta \sin(\varphi_\alpha-\varphi)-i\sin\theta \cos(\varphi_\alpha-\varphi)\right].
\label{eq_V1}
\end{align}

Contribution from bond-dependent terms is zero on linear level since $\sum_{\alpha=1}^6 \cos \varphi_\alpha=\sum_{\alpha=1}^6 \sin \varphi_\alpha=0$:
\begin{align}
\sum_{\alpha=1}^6 V^{(1)}_\alpha=\frac{JS^{3/2}}{2\sqrt{2}} \sin 2\theta (1-\Delta).
\end{align}
Thus, $\sum_\alpha V^{(1)}_\alpha$ is zero only for $\theta = 0$ (FM-x and FM-y states), or $\theta=\pi/2$ (FM-z state). Therefore, harmonic expansion for these states is well-defined, but it is not the case for the FM-yz state.

Since the harmonic expansion for the FM-yz state is not well-defined on a linear level, we propose that quantum fluctuations help stabilize this state. In order to make FM-yz state a saddle point, we include higher-order corrections, which can make one-boson terms zero. We do that by including lowest-level corrections from Hartree-Fock decoupling of three-magnon terms
\begin{align}
\mathcal{H}^{(3)}&=\sum_{\langle ij \rangle_\alpha} \frac{V^{(3)}_\alpha}{4} \left( a^\dagger_i a^{\phantom \dagger}_i a^{\phantom \dagger}_i+(i\rightarrow j)\right)\nonumber\\
&+V^{(3)}_\alpha \left( a^{\phantom \dagger}_i a^\dagger_j  a^{\phantom \dagger}_j+(i\rightarrow j)\right)+\text{H.c.},
\label{eq_H3}
\end{align}
where three-magnon interaction vertex is given by
\begin{align}
V^{(3)}_\alpha=- V^{(1)}_\alpha/S.
\label{eq_V3V1}
\end{align}

A straightforward calculation yields renormalized one-boson term in the Hamiltonian:
\begin{align}
\mathcal{H}^{(1)}_\text{HF}=\sum_{\langle ij \rangle_\alpha} \left(a_i^{\phantom \dagger} +a_j^{\phantom \dagger} \right)V_\text{$\alpha$,HF}^{(1)} +\text{H.c.},
\end{align}
where renormalized vertex is given by
\begin{align}
V_\text{$\alpha$,HF}^{(1)}=V^{(1)}_\alpha -\frac{1}{S}\left[V^{(1)}_\alpha\left( \frac{3n}{2}+m_\alpha \right) - V^{(1)*}_\alpha \frac{\delta^*}{4} - V^{(1)*}_\alpha\Delta^*_\alpha\right],
\end{align}
and $n$, $m_\alpha$, $\delta$, $\Delta_\alpha$ are Hartree-Fock averages, whose definition is given in the Appendix~\ref{app:3HF}. The condition for the one-boson term to become zero is given by the following equation:
\begin{align}
V_\text{HF}^{(1)}\equiv\sum_{\alpha=1}^6 V_\text{$\alpha$,HF}^{(1)}=0.
\label{eq_V1HF}
\end{align}

By solving equation \eqref{eq_V1HF}, one can find $\theta^*$ where spin-wave expansion is valid and magnon spectrum is properly defined. Equivalently, it corresponds to minimization of the energy of magnetic state with quantum corrections included -- $E_\text{cl} +\delta E^{(2)}$. This procedure is analogous to the field-induced canting angle renormalization in frustrated systems \cite{Nikuni_1998,Shannon_2008,Mourigal10,umbrella}.

An example of calculation is shown in Fig.~\ref{fig_thetaHF}(a) where we present classical energy and quantum corrections as a function of the canting angle $\theta$ for a representative point $J_{\pm\pm}=0.2|J|$, $J_{z\pm}=0.6|J|$, $\varphi=\pi/2$. One can see that a new minimum appears at  $\theta^*$, which also corresponds to the root of real part of $V_\text{HF}^{(1)}$ (the imaginary part is zero for  $\varphi=\pi/6+\pi n/3$, $n \in \mathbb{Z}$), shown in lower panel.  It means that FM-yz state with canting angle $\theta^*$ becomes a saddle-point, its spectrum is well-defined and can be used to calculate its energy with quantum corrections \eqref{eq_e2} included. 

We also perform DMRG calculations to compare with spin-wave theory, the results are presented in Fig.~\ref{fig_thetaHF}(b) as canting angle $\theta^*$ as a function $XXZ$ anisotropy $\Delta$ for the same parameters as in Fig.~\ref{fig_thetaHF}(a). Solid line represents spin-wave theory, dots represent DMRG calculation. Numerical calculations were performed on a 100-site cluster with periodic-open boundary conditions with 8 sweeps by keeping up to $m=400$ states and truncation error up to $10^{-8}$, using ITensor package \cite{itensor}. One can see solid agreement between SWT and DMRG, which also predicts survival of FM-yz state for significant values of easy-plane anisotropy. Note that DMRG does not exhibit fully saturated FM-z state, most likely due to quantum fluctuations.

Using the procedure outlined above we calculate energies of all ferromagnetic states with quantum corrections included. The results are shown in Fig.~\ref{fig_deltaPD} for various values of $\Delta$ in both easy-plane and easy-axis regimes. The canting angle of the FM-yz state is presented as an intensity plot. Due to quantum fluctuations, finite $J_{z\pm}$ stabilizes FM-yz state in a wide range of $XXZ$ anisotropy for both $\Delta<1$ where simple easy-plane state is expected naively, and for $\Delta>1$ where one expects easy-axis out-of-plane state. Calculations for other $\Delta$ are also shown in Fig.~\ref{fig_states} where one can see that for smaller values of $\Delta$ the FM-x state becomes more stable and FM-z disappears. For $\Delta>1$, the region of stability of FM-yz state shrinks and Ising FM-z state encompasses most of the phase diagram.

\section{Non-linear spin-wave theory: dynamical structure factor}
\label{sec:spectrum}
\begin{figure}
\centering
\includegraphics[width=\columnwidth]{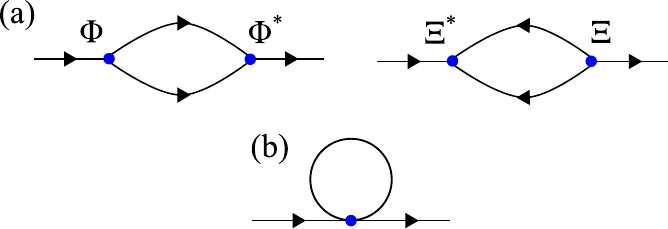}
\caption{(a) Decay and source diagrams for the $1/S$ contribution
to the self-energy \eqref{eq_sigma} from the three-magnon interactions \eqref{Hsource} and \eqref{Hdecay}. (b) The Hartree-Fock diagram from the four-magnon interactions \eqref{eq_H4}.}
\label{fig_diagrams}
\end{figure}
\begin{figure*}
\centering
\includegraphics[width=1.8\columnwidth]{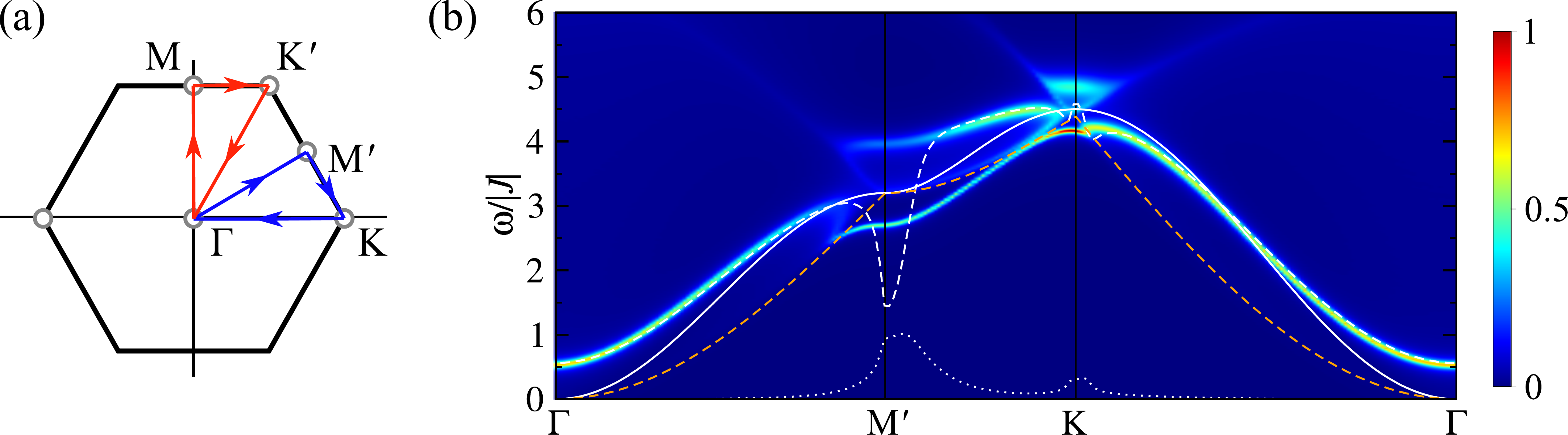}
\caption{(a) First Brillouin zone and one-dimensional contour used in (b) and Figs.~\ref{fig_fmz} and \ref{fig_fmyz}. (b) Magnetic spectrum of the FM-z state for $J_{\pm\pm}=0.6|J|$ and $J_{z\pm}=0.4|J|$ at $\Delta=1$. Dynamical structure factor $\mathcal{S}(\mathbf{k},\omega)$ \eqref{eq_Sqw} is displayed as intensity plot. Linear spin-wave spectrum \eqref{eq_ek} is shown with the solid line, magnon spectrum with $1/S$ on-shell corrections \eqref{eq_ek_onshell} is shown with the dashed line. Magnon decay rate $\Gamma_\mathbf{k}$ is indicated by dotted line. Dashed orange line shows the bottom of the two-magnon continuum. Units of $J^{-1}$ and artificial broadening $0.05|J|$ are used.}
\label{fig_spectrum}
\end{figure*}
While ferromagnetic state of the Heisenberg model conserves magnon number, and its excitations can only be affected by magnon-magnon interactions at high temperatures, ferromagnets with anisotropic terms in the Hamiltonian can exhibit renormalization of their spectrum even at lowest temperatures. This is the case for both spontaneously broken ferromagnetic state on various lattices with anisotropic exchanges \cite{us_kagome,rau_pyrochlore,Mook_honeycomb_2021,Hickey_pyrochlore_2025} and high-field polarized state, which is nominally free from quantum fluctuations \cite{McClarty2018,Vojta2020_NLSWT,Koyama_2023,Gallegos_CoNb2O6,us_jpp}. Here we show that bond-dependent terms of Hamiltonian \eqref{HJpm} can induce quantum effects in the excitation spectrum due to three- and four-magnon interactions, such as magnon energy renormalization \cite{triangle2} and magnon decays \cite{decay_review}.

The terms in the Holstein-Primakoff expansion \eqref{eq_Hexp} beyond quadratic represent magnon interactions that contribute to quantum corrections to the spin-wave spectrum \eqref{eq_ek}. We calculate corrections in the lowest $1/S$ order represented by the diagrams in Fig.~\ref{fig_diagrams}, which consist of three-magnon interactions and four-magnon Hartree-Fock corrections. Standard non-linear spin-wave theory yields self-energy
\begin{equation}
\Sigma(\mathbf{k},\omega)=\varepsilon^{(4)}_{\bf k}+\Sigma_{11}^{(a)} \left( \omega, {\bf k} \right)+\Sigma_{11}^{(b)} \left( \omega, {\bf k} \right),
\label{eq_sigma}
\end{equation}
which provides corrections to the magnon Green's function
\begin{equation}
G^{-1}(\mathbf{k},\omega)=\omega-\varepsilon_\mathbf{k}-\Sigma(\mathbf{k},\omega).
\end{equation}
First contribution to self-energy comes from three-magnon interaction (also shown in Fig.~\ref{fig_diagrams}(a)), which after Bogolyubov transformation \eqref{eq_bogolyubov} is given by
\begin{eqnarray}
&&\hat{\cal H}^{(3)}=\frac{1}{3!}\sum_{-{\bf p}={\bf k}+{\bf q}} \left(
\Xi_{{\bf q}{\bf k}{\bf p}} 
d^{\dagger}_{{\bf q}} d^{\dagger}_{{\bf k}} d^{\dagger}_{{\bf p}}+{\rm H.c.}\right)
\label{Hsource}
\\
&&\phantom{\hat{\cal H}^{(3)}}
+\frac{1}{2!}\sum_{-{\bf p}={\bf k}+{\bf q}} \left(
\Phi_{{\bf q}{\bf k}{\bf p}} 
d_{{\bf q}} d_{{\bf k}} d^\dagger_{-{\bf p}}+{\rm H.c.}\right).
\label{Hdecay}
\end{eqnarray}
Standard diagrammatic technique yields contributions
\begin{equation}
\Sigma_{11}^{(a)} \left( \omega, \bf k \right)=\frac{1}{2}  \sum_{\bf q} \frac{\left| {\Phi}_{{\bf q},{\bf k-q};{\bf k}}\right|^2}{\omega - \varepsilon_{\bf q} - \varepsilon_{{\bf k}- \bf q}+i0},
\end{equation}
\begin{equation}
\Sigma_{11}^{(b)} \left( \omega, \bf k \right)=-\frac{1}{2} \sum_{\bf q} \frac{\left| {\Xi}_{{\bf q},{\bf -k-q},{\bf k}}\right|^2}{\omega + \varepsilon_{\bf q} + \varepsilon_{-{\bf k}- \bf q}-i0}.
\end{equation}

A second contribution to the self-energy at the same level comes from Hartree-Fock decoupling of four-magnon interaction, shown in Fig.~\ref{fig_diagrams}(b). Straightforward calculation yields
\begin{eqnarray}
\delta {\cal{H}}_2^{(4)} =\sum_{{\bf k}} 
\Big[\varepsilon^{(4)}_{\bf k} d^\dagger_{{\bf k}} d^{\phantom{\dag}}_{{\bf k}}
-\frac{1}{2} \left(  B^{(4)}_{\bf k} d^\dagger_{{\bf k}} d^{\dagger}_{-{\bf k}}
+{\rm H.c.}\right)\Big],
\label{eq_H4}
\end{eqnarray}
where the correction to the spectrum is given by
\begin{equation}
\varepsilon^{(4)}_{\bf k}=\frac{A_{\bf k} \delta A^{(4)}_{\bf k} - \text{Re}\left( \delta B^{(4)}_{\bf k}B^*_\mathbf{k} \right)} {\sqrt{A_{\bf k}^2 - \left|B_{\bf k}\right|^2}}.
\label{eq_HF4ek}
\end{equation}
The expressions for three-magnon vertices $\Xi_{{\bf q}{\bf k}{\bf p}} $, $\Phi_{{\bf q}{\bf k}{\bf p}} $, and four-magnon Hartree-Fock terms $\delta A^{(4)}_{\bf k}$, $\delta B^{(4)}_{\bf k}$ are given in the Appendices~\ref{app:3mag} and \ref{app:4mag}.

Using self-energy \eqref{eq_sigma} we calculate dynamical structure factor
\begin{align}
\mathcal{S}(\mathbf{k},\omega)=\sum_{\alpha \beta} \left( \delta_{\alpha \beta} -\frac{k_\alpha k_\beta}{k^2}\right)\mathcal{S}^{\alpha \beta} (\mathbf{k},\omega),
\label{eq_Sqw}
\end{align}
where the sum is over $\alpha,\beta=\{x,y,z\}$ in the local reference frame, and the spin-spin correlation function 
\begin{align}
\mathcal{S}^{\alpha \beta}(\mathbf{k},\omega)=\frac{1}{\pi} \text{Im} \int_{-\infty}^\infty dt  e^{i\omega t} \  i\langle \mathcal{T} S^\alpha_\mathbf{k}(t) S^\beta_{-\mathbf{k}}(0)\rangle
\end{align}
can be related to spectral function
\begin{equation}
A(\mathbf{k},\omega)=-\frac{1}{\pi}\text{Im}\left[G(\mathbf{k},\omega)\right],
\end{equation}
see also Appendix~\ref{app:sqw} for details.

We calculate dynamical structure factor and on-shell self-energy for various states in the phase diagram. First, we explore excitations of the FM-z state for a representative set $J_{\pm\pm}=0.6|J|$ and $J_{z\pm}=0.4|J|$ at $\Delta=1$. Figure \ref{fig_spectrum} illustrates magnetic spectrum of the FM-z state with $1/S$ corrections from Eq.~\eqref{eq_sigma} included. Intensity plot is the dynamical structure factor $\mathcal{S}(\mathbf{k},\omega)$, solid line is non-interacting linear spin-wave theory result, dashed line is the on-shell calculation:
\begin{align}
\tilde{\varepsilon}_\mathbf{k}=\varepsilon_\mathbf{k}+\text{Re}\Sigma(\mathbf{k},\varepsilon_\mathbf{k}).
\label{eq_ek_onshell}
\end{align}
From this result one can clearly see one of the main effects of quantum order-by-disorder selection - opening of a gap in the spectrum at the $\Gamma$ point. Since the Hamiltonian \eqref{HJpm} does not have continuous symmetry, gapless linear spin-wave spectrum is an artifact, and we obtain physical result that higher-order corrections render the spectrum gapped. There are also corrections to the spectrum at higher energies due to interaction with two-magnon continuum, whose minimum is shown with the dashed orange line.

The shift of the real part of the spectrum is not the only effect, we also observe broadening of the magnon peaks at high energies from spontaneous magnon decays \cite{decay_review}, as shown in diagram in Fig.~\ref{fig_diagrams}(a). In order to estimate the strength of this broadening we calculate on-shell magnon decay rate
\begin{align}
    \Gamma_\mathbf{k}=-\text{Im}\Sigma(\mathbf{k},\varepsilon_\mathbf{k}),
\end{align}
which is shown with the dotted white line in Fig.~\ref{fig_spectrum}. One can see that even though the one-magnon mode overlaps with the continuum for all $\mathbf{k}$, the interaction vertex yields decays negligible almost everywhere except for the Brillouin zone boundary.

The results for other parameter sets of FM-z and FM-yz states are shown in Appendix~\ref{app:sqw} in Figs.~\ref{fig_fmz} and \ref{fig_fmyz}, where similar effects are observed.

Overall, we show that in ferromagnets in the presence of SOC-induced bond-dependent terms magnetic spectrum is not free from fluctuations, and corrections beyond linear spin-wave theory should be expected in future neutron scattering experiments in materials, such as NaRuO$_2$.

\section{Conclusions}
\label{sec:conclusions}
By studying the phase diagram of anisotropic-exchange (extended Kitaev-Heisenberg) model on the triangular lattice using spin-wave theory, we show that bond-dependent terms, induced by spin-orbit coupling, play important role in selection of ground state of the ferromagnetic regime of the model. These terms do not contribute to the classical energy and can only be accounted for by including quantum fluctuations. Therefore, a quantum order-by-disorder effect is realized in this model.

In the case of $\Delta=1$ - isotropic limit of bond-independent part of the Hamiltonian - where all spin directions have positive defined magnon spectrum, we found two stable states. We show that $J_{z\pm}$ exchange promotes out-of-plane canting of magnetization with spins in the $y-z$ plane, as shown in Fig.~\ref{fig_states}. This is somewhat expected since this interaction couples in-plane and out-of-plane spin components. More surprising is that $J_{\pm\pm}$ interaction, which couples in-plane $S^x$ and $S^y$ components, stabilizes out-of-plane magnetization with moments along $z$-axis. We argue that in the FM-z state additional interaction of in-plane components from $J_{\pm\pm}$ contribution enhances fluctuations and lowers its energy.

As anticipated, the states in the $\Delta=1$ phase diagram have some region of stability when $\Delta$ is tuned. For instance, we show that FM-z state survives in the easy-plane limit even when classical energy calculations predict in-plane states. Similarly, FM-yz state is stable in a wide region of $XXZ$ anisotropy. We calculate these $\Delta \neq 1$ phase diagrams using the extension of MAGSWT by searching for the minimum of energy with included quantum corrections, $E_\text{cl}+\delta E^{(2)}$. This is equivalent to stabilizing correct harmonic expansion with quantum fluctuations making linear boson terms zero. Our spin-wave theory results are also supported by DMRG calculations.

Moreover, anisotropic-exchange terms contribute to magnon-magnon interactions that introduce quantum corrections to the magnetic spectrum. One of the most important effects is the opening of a gap in the spectrum due to quantum fluctuations, while the spectrum is gapless on the linear spin-wave theory level due to accidental degeneracy. Spectrum renormalization is also pronounced at higher energies due to stronger interaction of the one-magnon mode with the two-magnon continuum, with larger bond-dependent terms yielding larger spectral redistribution.

Overall, we show that even in the simplest ferromagnetic state the presence of SOC-induced bond-dependent Kitaev-like exchange terms in the Hamiltonian can be important for both ground state and magnetic spectrum. These results may be useful for the identification of the ground state of recently studied triangular lattice ferromagnets \cite{Shikano_2004,Ortiz2023,Razpopov2023,Bhattacharyya2023,Moller_2012,Rawl_2017}.

\section{Acknowledgments}
We would like to thank Sasha Chernyshev for impactful discussions. P.A.M. acknowledges support from BASIS grant 24-1-3-11-1.
\appendix
\section{The Hamiltonian}
\label{app:ham}
The rotation between crystallographic and cubic reference frames ${\bf S}_{\rm cryst}\!=\!\hat{\mathbf{R}}_c{\bf S}_{\rm cubic}$, shown in Fig.~\ref{fig_axes}, is given by
\begin{align}
\hat{\mathbf{R}}_c=\left(
\begin{array}{ccc}
 \frac{1}{\sqrt{2}} & -\frac{1}{\sqrt{2}} &  0\\
 \frac{1}{\sqrt{6}} & \frac{1}{\sqrt{6}} & -\frac{2}{\sqrt{6}} \\
 \frac{1}{\sqrt{3}} & \frac{1}{\sqrt{3}} & \frac{1}{\sqrt{3}} \\
\end{array} 
\right).
\label{eq_cubic_transform}
\end{align}
This transformation yields the realtion between two forms of the Hamiltonian \eqref{HJpm} and \eqref{eq_H_JKG}:
\begin{align}
{\rm J}&=\frac{1}{3}\left( 2J+\Delta J+2J_{\pm\pm}+\sqrt{2} J_{z\pm}\right),\nonumber\\
\label{eq_jkg_transform}
K&=-2J_{\pm\pm}-\sqrt{2}J_{z\pm},\\
\Gamma&=\frac{1}{3} \left( -J+\Delta J-4J_{\pm\pm}+\sqrt{2} J_{z\pm}\right),\nonumber\\
\Gamma'&=\frac{1}{6} \left( -2J+2\Delta J+4J_{\pm\pm}-\sqrt{2} J_{z\pm}\right).\nonumber
\end{align}
\section{Spin-wave theory details}
\subsection{Linear spin-wave theory}
\label{app:lswt}
Holstein-Primakoff transformation \eqref{eq_HP} of the Hamiltonian \eqref{HJpm} with terms up to quadratic order yields expression \eqref{eq_H2}, whose elements are given by
\begin{align}
A_\mathbf{k}=&-6JS\left(\cos^2 \theta+\Delta \sin^2 \theta\right) +3JS \gamma_\mathbf{k} \left( 1+\sin^2 \theta+\Delta\cos^2 \theta \right)\nonumber\\
&-6J_{\pm\pm} S \cos^2 \theta \gamma_{F,\bf k} -3J_{z\pm}S \sin 2\theta \overline{\gamma}_{F,\bf k}, 
\end{align}
\begin{align}
&B_{\bf k}=3JS\left(1-\Delta\right)\cos^2\theta\,\gamma_{\bf k}
-6J_{\pm\pm}S\left(1+\sin^2\theta\right)\gamma_{F,\bf k}\nonumber\\ 
&+3J_{z\pm}
S\sin 2 \theta\,\overline{\gamma}_{F,\bf k}
+6iS\Big(2J_{\pm\pm}\sin\theta\,\widetilde{\gamma}_{F,\bf k}+J_{z\pm}\cos \theta\,\hat{\gamma}_{F,\bf k}\Big),
\end{align}
where
\begin{align}
    \gamma_\mathbf{k}&=\frac{1}{3}\sum_{\alpha=1}^3 \cos {\bf k}\bm{\delta}_\alpha,\nonumber\\
    \gamma_{F,\bf k}&=\frac{1}{3}\sum_{\alpha=1}^3 \cos {\bf k}\bm{\delta}_\alpha \cos(2\varphi +\varphi_\alpha),\nonumber\\
    \overline{\gamma}_{F,\bf k}&=\frac{1}{3}\sum_{\alpha=1}^3  \cos {\bf k}\bm{\delta}_\alpha
\sin(\varphi-\varphi_\alpha),\nonumber\\
    \widetilde{\gamma}_{F,\bf k}&=\frac{1}{3}\sum_{\alpha=1}^3 \cos {\bf k}\bm{\delta}_\alpha
\sin(2\varphi+\varphi_\alpha),\nonumber\\
\hat{\gamma}_{F,\bf k}&=\frac{1}{3}\sum_{\alpha=1}^3  \cos {\bf k}\bm{\delta}_\alpha \cos(\varphi-\varphi_\alpha),
\label{gammaFM}
\end{align} 
in agreement (up to a typo in $A_\mathbf{k}$) with Ref.~\cite{prx_anisotropic}.
\subsection{Corrections to one-magnon term}
\label{app:3HF}
Hartree-Fock decoupling of three-magnon interaction \eqref{eq_H3}
\begin{align}
a^\dagger_i a^{\phantom \dagger}_i a^{\phantom \dagger}_i &\rightarrow 2 n a^{\phantom \dagger}_i +\delta a^\dagger_i\nonumber\\
a^{\phantom \dagger}_i a^\dagger_j  a^{\phantom \dagger}_j &\rightarrow m_\alpha  a^{\phantom \dagger}_j + n a^{\phantom \dagger}_i + \Delta_\alpha a^\dagger_j\nonumber\\
a^{\phantom \dagger}_j a^\dagger_i  a^{\phantom \dagger}_i &\rightarrow m_\alpha  a^{\phantom \dagger}_i + n a^{\phantom \dagger}_j + \Delta_\alpha a^\dagger_i
\end{align}
yields quantum corrections to the one magnon term in the Holstein-Primakoff expansion:
\begin{align}
\mathcal{H}^{(3)}_\text{HF}=\sum_{\langle ij \rangle_\alpha}\left( a^{\phantom \dagger}_i +a^{\phantom \dagger}_j \right) &\left[ V^{(3)}_\alpha\left( \frac{3n}{2}+m_\alpha \right) \right.\nonumber\\
&\left. + \delta^*\frac{V_\alpha^{(3)*}}{4} +\Delta^*_\alpha V_\alpha^{(3)*} \right]+\text{H.c.},
\end{align}
where the Hartree-Fock averages are given by
\begin{align}
n&=\frac{1}{N}\sum_i \langle  a^\dagger_i  a^{\phantom \dagger}_i\rangle=\sum_\mathbf{k} v_\mathbf{k}^2,\\
m_\alpha &= \frac{1}{N}\sum_{\mathbf{r}_j=\mathbf{r}_i+\bm{\delta}_\alpha} \langle  a^\dagger_i  a^{\phantom \dagger}_j\rangle=\sum_\mathbf{k} v_\mathbf{k}^2 \cos \mathbf{k} \bm{\delta}_\alpha,\\
\delta &= \frac{1}{N}\sum_{i} \langle  a^{\phantom \dagger}_i  a^{\phantom \dagger}_i \rangle=\sum_\mathbf{k}u_\mathbf{k}  v_\mathbf{k} e^{i\phi_\mathbf{k}},\\
\Delta_\alpha &= \frac{1}{N}\sum_{\mathbf{r}_j=\mathbf{r}_i+\bm{\delta}\alpha} \langle  a^{\phantom \dagger}_i  a^{\phantom \dagger}_j\rangle=\sum_\mathbf{k} u_\mathbf{k} v_\mathbf{k}  \cos \mathbf{k} \bm{\delta}_\alpha e^{i\phi_\mathbf{k}}.
\end{align}
\begin{figure}
\centering
\includegraphics[width=\columnwidth]{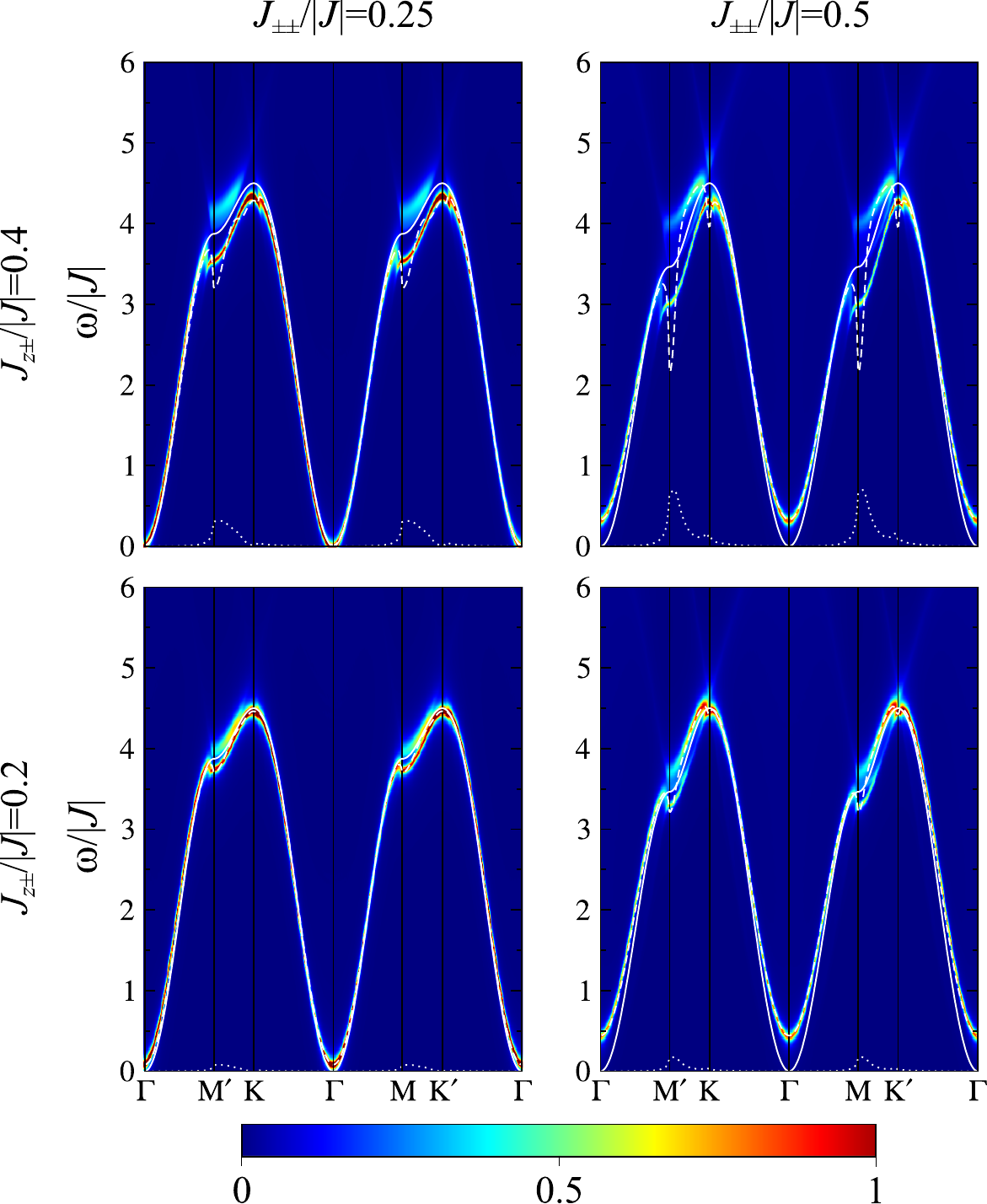}
\caption{Magnetic spectrum of the FM-z state for various $J_{\pm\pm}$ and $J_{z\pm}$ at $\Delta=1$. Dynamical structure factor $\mathcal{S}(\mathbf{k},\omega)$ \eqref{eq_Sqw} is displayed as intensity plot. Linear spin-wave spectrum \eqref{eq_ek} is shown with the solid line, magnon spectrum with $1/S$ on-shell corrections \eqref{eq_ek_onshell} is shown with the dashed line. Magnon decay rate $\Gamma_\mathbf{k}$ is indicated by dotted line. Units of $J^{-1}$ and artificial broadening $0.05|J|$ are used.}
\label{fig_fmz}
\end{figure}
\subsection{Three-magnon interaction}
\label{app:3mag}
Interaction vertices of three-magnon terms in Eqs.~\eqref{Hdecay} and \eqref{Hsource} are given by
\begin{align}
\Phi_{{\bf q}{\bf k}{\bf p}} =& F_\mathbf{qkp} u_\mathbf{q}u_\mathbf{k}u_\mathbf{p}+F_\mathbf{kqp} u_\mathbf{q}u_\mathbf{k}u_\mathbf{p}\nonumber\\
+&F_\mathbf{qpk} u_\mathbf{q}v_\mathbf{k}v_\mathbf{p}+F_\mathbf{kpq} v_\mathbf{q}u_\mathbf{k}v_\mathbf{p}\nonumber\\
+&F_\mathbf{pkq} v_\mathbf{q}u_\mathbf{k}v_\mathbf{p}+F_\mathbf{pqk} u_\mathbf{q}v_\mathbf{k}v_\mathbf{p}\nonumber\\
+&F^*_\mathbf{pkq} u_\mathbf{q}v_\mathbf{k}u_\mathbf{p}+F^*_\mathbf{pqk} v_\mathbf{q}u_\mathbf{k}u_\mathbf{p}\nonumber\\
+&F^*_\mathbf{qpk} v_\mathbf{q}u_\mathbf{k}u_\mathbf{p}+F^*_\mathbf{kpq} u_\mathbf{q}v_\mathbf{k}u_\mathbf{p}\nonumber\\
+&F^*_\mathbf{qkp} v_\mathbf{q}v_\mathbf{k}v_\mathbf{p}+F^*_\mathbf{kqp} v_\mathbf{q}v_\mathbf{k}v_\mathbf{p}\nonumber,\\
\end{align}
\begin{align}
\Xi_{{\bf q}{\bf k}{\bf p}} = &F_\mathbf{qkp} v_\mathbf{q}v_\mathbf{k}u_\mathbf{p}+F_\mathbf{kqp} v_\mathbf{q}v_\mathbf{k}u_\mathbf{p}\nonumber\\
+&F_\mathbf{qpk} v_\mathbf{q}u_\mathbf{k}v_\mathbf{p}+F_\mathbf{kpq} u_\mathbf{q}v_\mathbf{k}v_\mathbf{p}\nonumber\\
+&F_\mathbf{pqk} v_\mathbf{q}u_\mathbf{k}v_\mathbf{p}+F_\mathbf{pkq} u_\mathbf{q}v_\mathbf{k}v_\mathbf{p}\nonumber\\
+&F^*_\mathbf{qkp} u_\mathbf{q}u_\mathbf{k}v_\mathbf{p}+F^*_\mathbf{qpk} u_\mathbf{q}v_\mathbf{k}u_\mathbf{p}\nonumber\\
+&F^*_\mathbf{kqp} u_\mathbf{q}u_\mathbf{k}v_\mathbf{p}+F^*_\mathbf{kpq} v_\mathbf{q}u_\mathbf{k}u_\mathbf{p}\nonumber\\
+&F^*_\mathbf{pkq} v_\mathbf{q}u_\mathbf{k}u_\mathbf{p}+F^*_\mathbf{pqk} u_\mathbf{q}v_\mathbf{k}u_\mathbf{p}\nonumber,\\
\end{align}
where 
\begin{align}
F_\mathbf{qkp}= \frac{1}{2} e^{i\frac{\phi_\mathbf{q}+\phi_\mathbf{k}-\phi_\mathbf{p}}{2}} \sum_{\alpha=1}^3 \left(1+4\cos \mathbf{q}\delta_\alpha \right)V^{(3)}_\alpha.
\end{align}
\subsection{Four-magnon interaction}
\label{app:4mag}
Hartree-Fock decoupling of four-magnon interaction yields corrections to the magnon spectrum
\begin{align} 
\delta {\cal{H}}_2^{(4)} =\sum_{{\bf k}} 
\Big[\delta A^{(4)}_{\bf k} a^\dagger_{{\bf k}} a^{\phantom{\dag}}_{{\bf k}}-\frac{1}{2} \left(  \delta B^{(4)}_{\bf k}a^\dagger_{{\bf k}} a^{\dagger}_{-{\bf k}}
+{\rm H.c.}\right)\Big],
\end{align}
where coefficients are given by
\begin{align}
\delta A^{(4)}_{\bf k} &= \sum_{\alpha=1}^3 J^{xx} \left[ -(m_\alpha+\text{Re}~\Delta_\alpha)-\frac{1}{2} (2n+\text{Re}~\delta) \cos \mathbf{k} \delta_\alpha\right]\nonumber\\
&+J^{yy}\left[ (\text{Re}~\Delta_\alpha-m_\alpha)+\frac{1}{2} (\text{Re}~\delta-2n) \cos \mathbf{k} \delta_\alpha\right]\nonumber\\
&+J^{zz} \left[2n+2m_\alpha \cos \mathbf{k} \delta_\alpha  \right] +J^{xy} \left[-2\text{Im}~\Delta_\alpha-\text{Im}~\delta \cos \mathbf{k} \delta_\alpha \right],
\end{align}
\begin{align}
\delta B^{(4)}_{\bf k} &=\sum_{\alpha=1}^3 \frac{J^{xx}}{2} \left[ m_\alpha +\Delta_\alpha +\left(2n+\delta\right) \cos \mathbf{k} \delta_\alpha\right]\nonumber\\
&-\frac{J^{yy}}{2}\left[ m_\alpha -\Delta_\alpha +\left(2n-\delta\right)\cos \mathbf{k} \delta_\alpha \right]\nonumber\\
&+i J^{xy} \left( m_\alpha+2n \cos \mathbf{k} \delta_\alpha\right)-2 J^{zz} \Delta_\alpha \cos \mathbf{k} \delta_\alpha.
\end{align}
Here $J^{\alpha \beta}$ are the elements of the Hamiltonian matrix in the local reference frame:
\begin{align}
J^{xx}&=\left[ J \left(\sin^2 \theta +\Delta \cos^2 \theta \right) +2 J_{\pm\pm} \sin^2 \theta \cos(2\varphi+\varphi_\alpha)\right.\nonumber\\&\left. - J_{z\pm}\sin 2\theta \sin(\varphi-\varphi_\alpha)\right],\\
J^{yy}&= \left[ J- 2 J_{\pm\pm} \cos(2\varphi+\varphi_\alpha) \right],\\
J^{zz}&= \left[J\left( \Delta \sin^2 \theta+\cos^2 \theta \right) +2 J_{\pm\pm}\cos^2 \theta \cos(2\varphi+\varphi_\alpha)\right.\nonumber\\&\left.+ J_{z\pm}\sin 2\theta \sin(\varphi-\varphi_\alpha) \right],\\
J^{xy}&=-\left[2J_{\pm\pm}\sin\theta \sin(2\varphi+\varphi_\alpha) +J_{z\pm} \cos \theta \cos(\varphi-\varphi_\alpha)\right].
\end{align}
Bogolyubov transformation \eqref{eq_bogolyubov} yields
\begin{eqnarray}
\delta {\cal{H}}_2^{(4)} =\sum_{{\bf k}} 
\Big[\varepsilon^{(4)}_{\bf k} d^\dagger_{{\bf k}} d^{\phantom{\dag}}_{{\bf k}}
-\frac{1}{2} \left(  B^{(4)}_{\bf k} d^\dagger_{{\bf k}} d^{\dagger}_{-{\bf k}}
+{\rm H.c.}\right)\Big]. \nonumber
\end{eqnarray}
Thus, four-magnon Hartree-Fock corrections to energy spectrum are shown in Eq.~\eqref{eq_HF4ek},
and  $B_{\bf k}$ is given by
\begin{equation}
B^{(4)}_{\bf k} = \frac{-\left|B_{\bf k}\right| \delta A^{(4)}_{\bf k} +A_{\bf k} \text{Re}\left( \delta B^{(4)}_{\bf k}e^{-i\phi_\mathbf{k}} \right)}{\sqrt{A_{\bf k}^2 -\left|B_{\bf k}\right|^2}}+i\text{Im} \left( \delta B^{(4)}_{\bf k}e^{-i\phi_\mathbf{k}} \right).
\end{equation}

\begin{figure}
\centering
\includegraphics[width=\columnwidth]{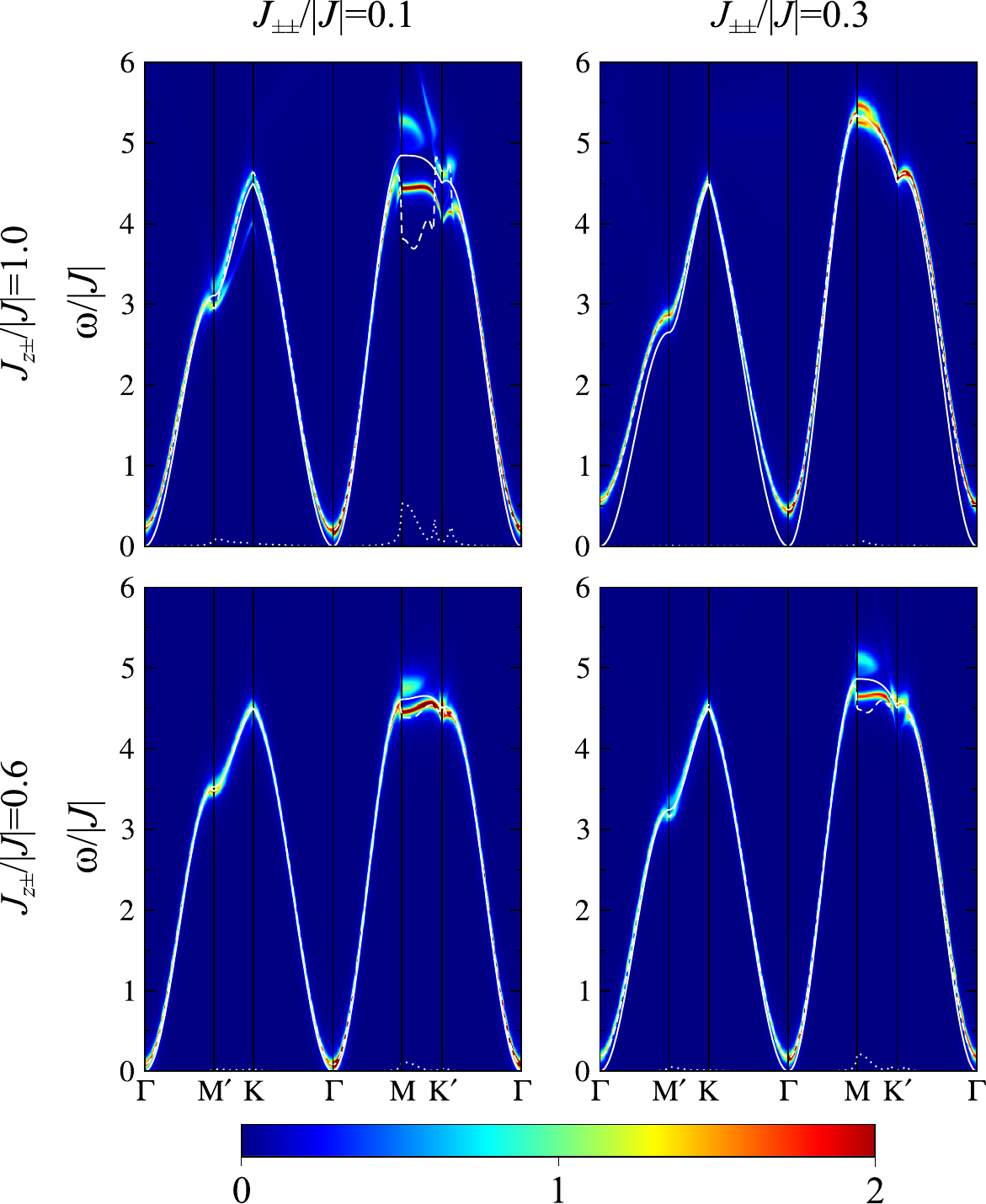}
\caption{Magnetic spectrum of the FM-yz state for various $J_{\pm\pm}$ and $J_{z\pm}$ at $\Delta=1$. Dynamical structure factor $\mathcal{S}(\mathbf{k},\omega)$ \eqref{eq_Sqw} is displayed as intensity plot. Linear spin-wave spectrum \eqref{eq_ek} is shown with the solid line, magnon spectrum with $1/S$ on-shell corrections \eqref{eq_ek_onshell} is shown with the dashed line. Magnon decay rate $\Gamma_\mathbf{k}$ is indicated by dotted line. Units of $J^{-1}$ and artificial broadening $0.05|J|$ are used.}
\label{fig_fmyz}
\end{figure}

\subsection{Dynamical structure factor intensity}
\label{app:sqw}
Rotating to the local reference frame and summing over $\alpha$ and $\beta$ in Eq.~\eqref{eq_Sqw} yields (for $k_z=0$)
\begin{align}
\mathcal{S}(\mathbf{k},\omega)=\frac{k_y^2}{k^2}S^{x_0 x_0}_{\mathbf{k},\omega}+\frac{k_x^2}{k^2}S^{y_0 y_0}_{\mathbf{k},\omega}-\frac{k_x k_y}{k^2}\left(S^{x_0 y_0}_{\mathbf{k},\omega}+S^{y_0 x_0}_{\mathbf{k},\omega}\right)+S^{z_0 z_0}_{\mathbf{k},\omega},
\end{align}
where components of dynamical structure factor are given by
\begin{align}
    S^{x_0 x_0}_{\mathbf{k},\omega}&= \mathcal{S}^{xx}_{\mathbf{k},\omega} \sin^2 \theta \cos^2\varphi+\mathcal{S}^{yy}_{\mathbf{k},\omega} \sin^2\varphi \nonumber\\&-\frac{\mathcal{S}^{x y}_{\mathbf{k},\omega}+\mathcal{S}^{yx}_{\mathbf{k},\omega}}{2}\sin\theta \sin 2\varphi+\mathcal{S}^{zz}_{\mathbf{k},\omega} \cos^2 \theta \cos^2\varphi,
\end{align}
\begin{align}
    S^{y_0 y_0}_{\mathbf{k},\omega}&= \mathcal{S}^{xx}_{\mathbf{k},\omega} \sin^2 \theta \sin^2\varphi+\mathcal{S}^{yy}_{\mathbf{k},\omega} \cos^2\varphi \nonumber\\&+\frac{\mathcal{S}^{x y}_{\mathbf{k},\omega}+\mathcal{S}^{yx}_{\mathbf{k},\omega}}{2}\sin\theta \sin 2\varphi+\mathcal{S}^{zz}_{\mathbf{k},\omega} \cos^2 \theta \sin^2\varphi,
\end{align}
\begin{align}
    S^{z_0 z_0}_{\mathbf{k},\omega}= \mathcal{S}^{xx}_{\mathbf{k},\omega} \cos^2 \theta +\mathcal{S}^{zz}_{\mathbf{k},\omega} \sin^2 \theta
\end{align}
\begin{align}
    S^{x_0 y_0}_{\mathbf{k},\omega}+S^{y_0 x_0}_{\mathbf{k},\omega}&= \mathcal{S}^{xx}_{\mathbf{k},\omega} \sin^2 \theta \sin 2\varphi-\mathcal{S}^{yy}_{\mathbf{k},\omega} \sin 2\varphi \nonumber\\&+\left(\mathcal{S}^{x y}_{\mathbf{k},\omega}+\mathcal{S}^{yx}_{\mathbf{k},\omega}\right)\sin\theta \cos 2\varphi+\mathcal{S}^{zz}_{\mathbf{k},\omega} \cos^2 \theta \sin 2\varphi.
\end{align}
Transverse components of $\mathcal{S}({\bf q},\omega)$ are given by
\begin{eqnarray}
&&\mathcal{S}^{xx(yy)}_{\mathbf{k},\omega}=\frac{S}{2}\left|\Lambda_{\pm}\right|^2 
\left[u_{\mathbf k}^2 + v_{\mathbf k}^2 \pm 2 u_{\mathbf k} v_{\mathbf k} \cos \left(\phi_\mathbf{k} -2\phi_{\pm}\right)\right] A(\mathbf{k},\omega),\nonumber\\
&&\mathcal{S}^{x y (yx)}_{\mathbf{k},\omega}=\frac{S}{2i}\left| \Lambda_+\right| \left|\Lambda_-\right| 
\left[\mp (u_{\mathbf k}^2  e^{\pm i(\phi_{-}-\phi_{+})} - v_{\mathbf k}^2 e^{\pm i(\phi_{+}-\phi_{-})}) \right.\nonumber\\ &&\left.+ 2i u_{\mathbf k} v_{\mathbf k} \sin \left(\phi_\mathbf{k}-\phi_{+}-\phi_{-}\right)\right] A(\mathbf{k},\omega),
\end{eqnarray}
\begin{align}
\mathcal{S}^{x y}_{\mathbf{k},\omega}+\mathcal{S}^{yx}_{\mathbf{k},\omega}&=S \left| \Lambda_+\right| \left|\Lambda_-\right| \left[ 2 u_{\mathbf k} v_{\mathbf k}\sin \left(\phi_\mathbf{k}-\phi_{+}-\phi_{-}\right)\right.\nonumber\\&\left.+ \left(u_{\mathbf k}^2 + v_{\mathbf k}^2 \right) \sin \left(\phi_{+}-\phi_{-}\right)\right]A(\mathbf{k},\omega)
\end{align}
where $\Lambda_\pm=1-(2n\pm\delta)/4S=\left| \Lambda_\pm\right| e^{i\phi_{\pm}}$.

The two-magnon (longitudinal) component is given by \cite{us_jpp}
\begin{align}
\mathcal{S}^{zz}\left( \mathbf{k},\omega \right)=\sum_{\mathbf{q}} \mathcal{F}_{\mathbf{k}\mathbf{q}} \delta\left(\omega - \varepsilon_{\bf q} - \varepsilon_{\bf k-q} \right).
\label{eq_twomag}
\end{align}
The expression for the intensity of two-magnon contribution to the dynamical structure factor \eqref{eq_twomag} can be seen as a sum over two-magnon density of states with each component having different intensities. The expression for these intensities is generally given by
\begin{align}
\mathcal{F}_{\mathbf{k}\mathbf{q}}=\left( \tilde{u}_{\mathbf{q}} \tilde{u}^{*}_{\mathbf{q}} \tilde{v}_{\mathbf{q-k}} \tilde{v}^{*}_{\mathbf{q-k}}+\tilde{u}_{\mathbf{q}} \tilde{v}_{\mathbf{-q}} \tilde{v}^{*}_{\mathbf{q-k}} \tilde{u}^{*}_{\mathbf{-q+k}}\right)
\end{align}
The elements of transformation matrix are given by Eq.\eqref{eq_bogolyubov}:
\begin{align}
\tilde{u}_\mathbf{k}=e^{i\phi_\mathbf{k}/2} u_\mathbf{k},~\tilde{v}_\mathbf{k}=e^{i\phi_\mathbf{k}/2} v_\mathbf{k},
\end{align}
which yields
\begin{align}
\mathcal{F}_{\mathbf{k}\mathbf{q}}=\frac{1}{2} &\left( u_\mathbf{q}^2 v_\mathbf{q}^2+u_\mathbf{k-q}^2 v_\mathbf{k-q}^2\right)\nonumber\\&+u_\mathbf{q} v_\mathbf{q} u_\mathbf{k-q} v_\mathbf{k-q} \cos \left(\phi_\mathbf{q}- \phi_\mathbf{k-q}\right).
\end{align}
\bibliography{FM_Kitaev}

\begin{thebibliography}{111}%
\makeatletter
\providecommand \@ifxundefined [1]{%
 \@ifx{#1\undefined}
}%
\providecommand \@ifnum [1]{%
 \ifnum #1\expandafter \@firstoftwo
 \else \expandafter \@secondoftwo
 \fi
}%
\providecommand \@ifx [1]{%
 \ifx #1\expandafter \@firstoftwo
 \else \expandafter \@secondoftwo
 \fi
}%
\providecommand \natexlab [1]{#1}%
\providecommand \enquote  [1]{``#1''}%
\providecommand \bibnamefont  [1]{#1}%
\providecommand \bibfnamefont [1]{#1}%
\providecommand \citenamefont [1]{#1}%
\providecommand \href@noop [0]{\@secondoftwo}%
\providecommand \href [0]{\begingroup \@sanitize@url \@href}%
\providecommand \@href[1]{\@@startlink{#1}\@@href}%
\providecommand \@@href[1]{\endgroup#1\@@endlink}%
\providecommand \@sanitize@url [0]{\catcode `\\12\catcode `\$12\catcode `\&12\catcode `\#12\catcode `\^12\catcode `\_12\catcode `\%12\relax}%
\providecommand \@@startlink[1]{}%
\providecommand \@@endlink[0]{}%
\providecommand \url  [0]{\begingroup\@sanitize@url \@url }%
\providecommand \@url [1]{\endgroup\@href {#1}{\urlprefix }}%
\providecommand \urlprefix  [0]{URL }%
\providecommand \Eprint [0]{\href }%
\providecommand \doibase [0]{https://doi.org/}%
\providecommand \selectlanguage [0]{\@gobble}%
\providecommand \bibinfo  [0]{\@secondoftwo}%
\providecommand \bibfield  [0]{\@secondoftwo}%
\providecommand \translation [1]{[#1]}%
\providecommand \BibitemOpen [0]{}%
\providecommand \bibitemStop [0]{}%
\providecommand \bibitemNoStop [0]{.\EOS\space}%
\providecommand \EOS [0]{\spacefactor3000\relax}%
\providecommand \BibitemShut  [1]{\csname bibitem#1\endcsname}%
\let\auto@bib@innerbib\@empty
\bibitem [{\citenamefont {Gong}\ \emph {et~al.}(2017)\citenamefont {Gong}, \citenamefont {Li}, \citenamefont {Li}, \citenamefont {Ji}, \citenamefont {Stern}, \citenamefont {Xia}, \citenamefont {Cao}, \citenamefont {Bao}, \citenamefont {Wang}, \citenamefont {Wang}, \citenamefont {Qiu}, \citenamefont {Cava}, \citenamefont {Louie}, \citenamefont {Xia},\ and\ \citenamefont {Zhang}}]{Gong2017}%
  \BibitemOpen
  \bibfield  {author} {\bibinfo {author} {\bibfnamefont {C.}~\bibnamefont {Gong}}, \bibinfo {author} {\bibfnamefont {L.}~\bibnamefont {Li}}, \bibinfo {author} {\bibfnamefont {Z.}~\bibnamefont {Li}}, \bibinfo {author} {\bibfnamefont {H.}~\bibnamefont {Ji}}, \bibinfo {author} {\bibfnamefont {A.}~\bibnamefont {Stern}}, \bibinfo {author} {\bibfnamefont {Y.}~\bibnamefont {Xia}}, \bibinfo {author} {\bibfnamefont {T.}~\bibnamefont {Cao}}, \bibinfo {author} {\bibfnamefont {W.}~\bibnamefont {Bao}}, \bibinfo {author} {\bibfnamefont {C.}~\bibnamefont {Wang}}, \bibinfo {author} {\bibfnamefont {Y.}~\bibnamefont {Wang}}, \bibinfo {author} {\bibfnamefont {Z.~Q.}\ \bibnamefont {Qiu}}, \bibinfo {author} {\bibfnamefont {R.~J.}\ \bibnamefont {Cava}}, \bibinfo {author} {\bibfnamefont {S.~G.}\ \bibnamefont {Louie}}, \bibinfo {author} {\bibfnamefont {J.}~\bibnamefont {Xia}},\ and\ \bibinfo {author} {\bibfnamefont {X.}~\bibnamefont {Zhang}},\ }\bibfield  {title} {\bibinfo {title} {{Discovery of intrinsic ferromagnetism in
  two-dimensional van der Waals crystals}},\ }\href {https://doi.org/10.1038/nature22060} {\bibfield  {journal} {\bibinfo  {journal} {Nature}\ }\textbf {\bibinfo {volume} {546}},\ \bibinfo {pages} {265} (\bibinfo {year} {2017})}\BibitemShut {NoStop}%
\bibitem [{\citenamefont {Fei}\ \emph {et~al.}(2018)\citenamefont {Fei}, \citenamefont {Huang}, \citenamefont {Malinowski}, \citenamefont {Wang}, \citenamefont {Song}, \citenamefont {Sanchez}, \citenamefont {Yao}, \citenamefont {Xiao}, \citenamefont {Zhu}, \citenamefont {May}, \citenamefont {Wu}, \citenamefont {Cobden}, \citenamefont {Chu},\ and\ \citenamefont {Xu}}]{Fei2018}%
  \BibitemOpen
  \bibfield  {author} {\bibinfo {author} {\bibfnamefont {Z.}~\bibnamefont {Fei}}, \bibinfo {author} {\bibfnamefont {B.}~\bibnamefont {Huang}}, \bibinfo {author} {\bibfnamefont {P.}~\bibnamefont {Malinowski}}, \bibinfo {author} {\bibfnamefont {W.}~\bibnamefont {Wang}}, \bibinfo {author} {\bibfnamefont {T.}~\bibnamefont {Song}}, \bibinfo {author} {\bibfnamefont {J.}~\bibnamefont {Sanchez}}, \bibinfo {author} {\bibfnamefont {W.}~\bibnamefont {Yao}}, \bibinfo {author} {\bibfnamefont {D.}~\bibnamefont {Xiao}}, \bibinfo {author} {\bibfnamefont {X.}~\bibnamefont {Zhu}}, \bibinfo {author} {\bibfnamefont {A.~F.}\ \bibnamefont {May}}, \bibinfo {author} {\bibfnamefont {W.}~\bibnamefont {Wu}}, \bibinfo {author} {\bibfnamefont {D.~H.}\ \bibnamefont {Cobden}}, \bibinfo {author} {\bibfnamefont {J.-H.}\ \bibnamefont {Chu}},\ and\ \bibinfo {author} {\bibfnamefont {X.}~\bibnamefont {Xu}},\ }\bibfield  {title} {\bibinfo {title} {{Two-dimensional itinerant ferromagnetism in atomically thin Fe$_3$GeTe$_2$}},\ }\href
  {https://doi.org/10.1038/s41563-018-0149-7} {\bibfield  {journal} {\bibinfo  {journal} {Nat. Mater.}\ }\textbf {\bibinfo {volume} {17}},\ \bibinfo {pages} {778} (\bibinfo {year} {2018})}\BibitemShut {NoStop}%
\bibitem [{\citenamefont {Deng}\ \emph {et~al.}(2018)\citenamefont {Deng}, \citenamefont {Yu}, \citenamefont {Song}, \citenamefont {Zhang}, \citenamefont {Wang}, \citenamefont {Sun}, \citenamefont {Yi}, \citenamefont {Wu}, \citenamefont {Wu}, \citenamefont {Zhu}, \citenamefont {Wang}, \citenamefont {Chen},\ and\ \citenamefont {Zhang}}]{Deng2018}%
  \BibitemOpen
  \bibfield  {author} {\bibinfo {author} {\bibfnamefont {Y.}~\bibnamefont {Deng}}, \bibinfo {author} {\bibfnamefont {Y.}~\bibnamefont {Yu}}, \bibinfo {author} {\bibfnamefont {Y.}~\bibnamefont {Song}}, \bibinfo {author} {\bibfnamefont {J.}~\bibnamefont {Zhang}}, \bibinfo {author} {\bibfnamefont {N.~Z.}\ \bibnamefont {Wang}}, \bibinfo {author} {\bibfnamefont {Z.}~\bibnamefont {Sun}}, \bibinfo {author} {\bibfnamefont {Y.}~\bibnamefont {Yi}}, \bibinfo {author} {\bibfnamefont {Y.~Z.}\ \bibnamefont {Wu}}, \bibinfo {author} {\bibfnamefont {S.}~\bibnamefont {Wu}}, \bibinfo {author} {\bibfnamefont {J.}~\bibnamefont {Zhu}}, \bibinfo {author} {\bibfnamefont {J.}~\bibnamefont {Wang}}, \bibinfo {author} {\bibfnamefont {X.~H.}\ \bibnamefont {Chen}},\ and\ \bibinfo {author} {\bibfnamefont {Y.}~\bibnamefont {Zhang}},\ }\bibfield  {title} {\bibinfo {title} {{Gate-tunable room-temperature ferromagnetism in two-dimensional Fe$_3$GeTe$_2$}},\ }\href {https://doi.org/10.1038/s41586-018-0626-9} {\bibfield  {journal} {\bibinfo
  {journal} {Nature}\ }\textbf {\bibinfo {volume} {563}},\ \bibinfo {pages} {94} (\bibinfo {year} {2018})}\BibitemShut {NoStop}%
\bibitem [{\citenamefont {Zhang}\ \emph {et~al.}(2019)\citenamefont {Zhang}, \citenamefont {Shang}, \citenamefont {Jiang}, \citenamefont {Rasmita}, \citenamefont {Gao},\ and\ \citenamefont {Yu}}]{Zhang_2019}%
  \BibitemOpen
  \bibfield  {author} {\bibinfo {author} {\bibfnamefont {Z.}~\bibnamefont {Zhang}}, \bibinfo {author} {\bibfnamefont {J.}~\bibnamefont {Shang}}, \bibinfo {author} {\bibfnamefont {C.}~\bibnamefont {Jiang}}, \bibinfo {author} {\bibfnamefont {A.}~\bibnamefont {Rasmita}}, \bibinfo {author} {\bibfnamefont {W.}~\bibnamefont {Gao}},\ and\ \bibinfo {author} {\bibfnamefont {T.}~\bibnamefont {Yu}},\ }\bibfield  {title} {\bibinfo {title} {{Direct Photoluminescence Probing of Ferromagnetism in Monolayer Two-Dimensional CrBr$_3$}},\ }\href {https://doi.org/10.1021/acs.nanolett.9b00553} {\bibfield  {journal} {\bibinfo  {journal} {Nano Letters}\ }\textbf {\bibinfo {volume} {19}},\ \bibinfo {pages} {3138} (\bibinfo {year} {2019})}\BibitemShut {NoStop}%
\bibitem [{\citenamefont {Bedoya-Pinto}\ \emph {et~al.}(2021)\citenamefont {Bedoya-Pinto}, \citenamefont {Ji}, \citenamefont {Pandeya}, \citenamefont {Gargiani}, \citenamefont {Valvidares}, \citenamefont {Sessi}, \citenamefont {Taylor}, \citenamefont {Radu}, \citenamefont {Chang},\ and\ \citenamefont {Parkin}}]{Parkin_2021}%
  \BibitemOpen
  \bibfield  {author} {\bibinfo {author} {\bibfnamefont {A.}~\bibnamefont {Bedoya-Pinto}}, \bibinfo {author} {\bibfnamefont {J.-R.}\ \bibnamefont {Ji}}, \bibinfo {author} {\bibfnamefont {A.~K.}\ \bibnamefont {Pandeya}}, \bibinfo {author} {\bibfnamefont {P.}~\bibnamefont {Gargiani}}, \bibinfo {author} {\bibfnamefont {M.}~\bibnamefont {Valvidares}}, \bibinfo {author} {\bibfnamefont {P.}~\bibnamefont {Sessi}}, \bibinfo {author} {\bibfnamefont {J.~M.}\ \bibnamefont {Taylor}}, \bibinfo {author} {\bibfnamefont {F.}~\bibnamefont {Radu}}, \bibinfo {author} {\bibfnamefont {K.}~\bibnamefont {Chang}},\ and\ \bibinfo {author} {\bibfnamefont {S.~S.~P.}\ \bibnamefont {Parkin}},\ }\bibfield  {title} {\bibinfo {title} {{Intrinsic 2D-XY ferromagnetism in a van der Waals monolayer}},\ }\href {https://doi.org/10.1126/science.abd5146} {\bibfield  {journal} {\bibinfo  {journal} {Science}\ }\textbf {\bibinfo {volume} {374}},\ \bibinfo {pages} {616} (\bibinfo {year} {2021})}\BibitemShut {NoStop}%
\bibitem [{\citenamefont {Gong}\ and\ \citenamefont {Zhang}(2019)}]{Gong_2019_review}%
  \BibitemOpen
  \bibfield  {author} {\bibinfo {author} {\bibfnamefont {C.}~\bibnamefont {Gong}}\ and\ \bibinfo {author} {\bibfnamefont {X.}~\bibnamefont {Zhang}},\ }\bibfield  {title} {\bibinfo {title} {{Two-dimensional magnetic crystals and emergent heterostructure devices}},\ }\href {https://doi.org/10.1126/science.aav4450} {\bibfield  {journal} {\bibinfo  {journal} {Science}\ }\textbf {\bibinfo {volume} {363}},\ \bibinfo {pages} {eaav4450} (\bibinfo {year} {2019})}\BibitemShut {NoStop}%
\bibitem [{\citenamefont {Soriano}\ \emph {et~al.}(2020)\citenamefont {Soriano}, \citenamefont {Katsnelson},\ and\ \citenamefont {Fernández-Rossier}}]{Soriano_2020_review}%
  \BibitemOpen
  \bibfield  {author} {\bibinfo {author} {\bibfnamefont {D.}~\bibnamefont {Soriano}}, \bibinfo {author} {\bibfnamefont {M.~I.}\ \bibnamefont {Katsnelson}},\ and\ \bibinfo {author} {\bibfnamefont {J.}~\bibnamefont {Fernández-Rossier}},\ }\bibfield  {title} {\bibinfo {title} {{Magnetic Two-Dimensional Chromium Trihalides: A Theoretical Perspective}},\ }\href {https://doi.org/10.1021/acs.nanolett.0c02381} {\bibfield  {journal} {\bibinfo  {journal} {Nano Lett.}\ }\textbf {\bibinfo {volume} {20}},\ \bibinfo {pages} {6225} (\bibinfo {year} {2020})}\BibitemShut {NoStop}%
\bibitem [{\citenamefont {Mermin}\ and\ \citenamefont {Wagner}(1966)}]{MerminWagner}%
  \BibitemOpen
  \bibfield  {author} {\bibinfo {author} {\bibfnamefont {N.~D.}\ \bibnamefont {Mermin}}\ and\ \bibinfo {author} {\bibfnamefont {H.}~\bibnamefont {Wagner}},\ }\bibfield  {title} {\bibinfo {title} {{Absence of Ferromagnetism or Antiferromagnetism in One- or Two-Dimensional Isotropic Heisenberg Models}},\ }\href {https://doi.org/10.1103/PhysRevLett.17.1133} {\bibfield  {journal} {\bibinfo  {journal} {Phys. Rev. Lett.}\ }\textbf {\bibinfo {volume} {17}},\ \bibinfo {pages} {1133} (\bibinfo {year} {1966})}\BibitemShut {NoStop}%
\bibitem [{\citenamefont {Huang}\ \emph {et~al.}(2017)\citenamefont {Huang}, \citenamefont {Clark}, \citenamefont {Navarro-Moratalla}, \citenamefont {Klein}, \citenamefont {Cheng}, \citenamefont {Seyler}, \citenamefont {Zhong}, \citenamefont {Schmidgall}, \citenamefont {McGuire}, \citenamefont {Cobden}, \citenamefont {Yao}, \citenamefont {Xiao}, \citenamefont {Jarillo-Herrero},\ and\ \citenamefont {Xu}}]{Huang2017}%
  \BibitemOpen
  \bibfield  {author} {\bibinfo {author} {\bibfnamefont {B.}~\bibnamefont {Huang}}, \bibinfo {author} {\bibfnamefont {G.}~\bibnamefont {Clark}}, \bibinfo {author} {\bibfnamefont {E.}~\bibnamefont {Navarro-Moratalla}}, \bibinfo {author} {\bibfnamefont {D.~R.}\ \bibnamefont {Klein}}, \bibinfo {author} {\bibfnamefont {R.}~\bibnamefont {Cheng}}, \bibinfo {author} {\bibfnamefont {K.~L.}\ \bibnamefont {Seyler}}, \bibinfo {author} {\bibfnamefont {D.}~\bibnamefont {Zhong}}, \bibinfo {author} {\bibfnamefont {E.}~\bibnamefont {Schmidgall}}, \bibinfo {author} {\bibfnamefont {M.~A.}\ \bibnamefont {McGuire}}, \bibinfo {author} {\bibfnamefont {D.~H.}\ \bibnamefont {Cobden}}, \bibinfo {author} {\bibfnamefont {W.}~\bibnamefont {Yao}}, \bibinfo {author} {\bibfnamefont {D.}~\bibnamefont {Xiao}}, \bibinfo {author} {\bibfnamefont {P.}~\bibnamefont {Jarillo-Herrero}},\ and\ \bibinfo {author} {\bibfnamefont {X.}~\bibnamefont {Xu}},\ }\bibfield  {title} {\bibinfo {title} {{Layer-dependent ferromagnetism in a van der Waals crystal
  down to the monolayer limit}},\ }\href {https://doi.org/10.1038/nature22391} {\bibfield  {journal} {\bibinfo  {journal} {Nature}\ }\textbf {\bibinfo {volume} {546}},\ \bibinfo {pages} {270} (\bibinfo {year} {2017})}\BibitemShut {NoStop}%
\bibitem [{\citenamefont {Stavropoulos}\ \emph {et~al.}(2021)\citenamefont {Stavropoulos}, \citenamefont {Liu},\ and\ \citenamefont {Kee}}]{HYKee_2021}%
  \BibitemOpen
  \bibfield  {author} {\bibinfo {author} {\bibfnamefont {P.~P.}\ \bibnamefont {Stavropoulos}}, \bibinfo {author} {\bibfnamefont {X.}~\bibnamefont {Liu}},\ and\ \bibinfo {author} {\bibfnamefont {H.-Y.}\ \bibnamefont {Kee}},\ }\bibfield  {title} {\bibinfo {title} {{Magnetic anisotropy in spin-3/2 with heavy ligand in honeycomb Mott insulators: Application to ${\mathrm{CrI}}_{3}$}},\ }\href {https://doi.org/10.1103/PhysRevResearch.3.013216} {\bibfield  {journal} {\bibinfo  {journal} {Phys. Rev. Res.}\ }\textbf {\bibinfo {volume} {3}},\ \bibinfo {pages} {013216} (\bibinfo {year} {2021})}\BibitemShut {NoStop}%
\bibitem [{\citenamefont {Witczak-Krempa}\ \emph {et~al.}(2014)\citenamefont {Witczak-Krempa}, \citenamefont {Chen}, \citenamefont {Kim},\ and\ \citenamefont {Balents}}]{Witczak_Krempa14}%
  \BibitemOpen
  \bibfield  {author} {\bibinfo {author} {\bibfnamefont {W.}~\bibnamefont {Witczak-Krempa}}, \bibinfo {author} {\bibfnamefont {G.}~\bibnamefont {Chen}}, \bibinfo {author} {\bibfnamefont {Y.~B.}\ \bibnamefont {Kim}},\ and\ \bibinfo {author} {\bibfnamefont {L.}~\bibnamefont {Balents}},\ }\bibfield  {title} {\bibinfo {title} {{Correlated {Quantum} {Phenomena} in the {Strong} {Spin}-{Orbit} {Regime}}},\ }\href {https://doi.org/10.1146/annurev-conmatphys-020911-125138} {\bibfield  {journal} {\bibinfo  {journal} {Annu. Rev. Condens. Matter Phys.}\ }\textbf {\bibinfo {volume} {5}},\ \bibinfo {pages} {57} (\bibinfo {year} {2014})}\BibitemShut {NoStop}%
\bibitem [{\citenamefont {Toulouse}()}]{Toulouse_frustration}%
  \BibitemOpen
  \bibfield  {author} {\bibinfo {author} {\bibfnamefont {G.}~\bibnamefont {Toulouse}},\ }\bibinfo {title} {{Theory of the frustration effect in spin glasses: I}},\ in\ \href {https://doi.org/10.1142/9789812799371_0009} {\emph {\bibinfo {booktitle} {Spin Glass Theory and Beyond}}},\ pp.\ \bibinfo {pages} {99--103}\BibitemShut {NoStop}%
\bibitem [{\citenamefont {Vannimenus}\ and\ \citenamefont {Toulouse}(1977)}]{Vannimenus_1977}%
  \BibitemOpen
  \bibfield  {author} {\bibinfo {author} {\bibfnamefont {J.}~\bibnamefont {Vannimenus}}\ and\ \bibinfo {author} {\bibfnamefont {G.}~\bibnamefont {Toulouse}},\ }\bibfield  {title} {\bibinfo {title} {{Theory of the frustration effect. II. Ising spins on a square lattice}},\ }\href {https://doi.org/10.1088/0022-3719/10/18/008} {\bibfield  {journal} {\bibinfo  {journal} {J. Phys. C}\ }\textbf {\bibinfo {volume} {10}},\ \bibinfo {pages} {L537} (\bibinfo {year} {1977})}\BibitemShut {NoStop}%
\bibitem [{\citenamefont {Ferreiro}\ \emph {et~al.}(2018)\citenamefont {Ferreiro}, \citenamefont {Komives},\ and\ \citenamefont {Wolynes}}]{Ferreiro_2018}%
  \BibitemOpen
  \bibfield  {author} {\bibinfo {author} {\bibfnamefont {D.~U.}\ \bibnamefont {Ferreiro}}, \bibinfo {author} {\bibfnamefont {E.~A.}\ \bibnamefont {Komives}},\ and\ \bibinfo {author} {\bibfnamefont {P.~G.}\ \bibnamefont {Wolynes}},\ }\bibfield  {title} {\bibinfo {title} {Frustration, function and folding},\ }\href {https://doi.org/https://doi.org/10.1016/j.sbi.2017.09.006} {\bibfield  {journal} {\bibinfo  {journal} {Curr. Opin. Struct. Biol.}\ }\textbf {\bibinfo {volume} {48}},\ \bibinfo {pages} {68} (\bibinfo {year} {2018})}\BibitemShut {NoStop}%
\bibitem [{\citenamefont {Shintani}\ and\ \citenamefont {Tanaka}(2006)}]{Shintani2006}%
  \BibitemOpen
  \bibfield  {author} {\bibinfo {author} {\bibfnamefont {H.}~\bibnamefont {Shintani}}\ and\ \bibinfo {author} {\bibfnamefont {H.}~\bibnamefont {Tanaka}},\ }\bibfield  {title} {\bibinfo {title} {Frustration on the way to crystallization in glass},\ }\href {https://doi.org/10.1038/nphys235} {\bibfield  {journal} {\bibinfo  {journal} {Nat. Phys.}\ }\textbf {\bibinfo {volume} {2}},\ \bibinfo {pages} {200} (\bibinfo {year} {2006})}\BibitemShut {NoStop}%
\bibitem [{\citenamefont {Ramirez}(1994)}]{ramirez_strongly_1994}%
  \BibitemOpen
  \bibfield  {author} {\bibinfo {author} {\bibfnamefont {A.~P.}\ \bibnamefont {Ramirez}},\ }\bibfield  {title} {\bibinfo {title} {Strongly {Geometrically} {Frustrated} {Magnets}},\ }\href {https://doi.org/10.1146/annurev.ms.24.080194.002321} {\bibfield  {journal} {\bibinfo  {journal} {Annu. Rev. Mater. Sci.}\ }\textbf {\bibinfo {volume} {24}},\ \bibinfo {pages} {453} (\bibinfo {year} {1994})}\BibitemShut {NoStop}%
\bibitem [{\citenamefont {Wannier}(1950)}]{wannier_antiferromagnetism._1950}%
  \BibitemOpen
  \bibfield  {author} {\bibinfo {author} {\bibfnamefont {G.~H.}\ \bibnamefont {Wannier}},\ }\bibfield  {title} {\bibinfo {title} {{Antiferromagnetism. The Triangular Ising Net}},\ }\href {https://doi.org/10.1103/PhysRev.79.357} {\bibfield  {journal} {\bibinfo  {journal} {Phys. Rev.}\ }\textbf {\bibinfo {volume} {79}},\ \bibinfo {pages} {357} (\bibinfo {year} {1950})}\BibitemShut {NoStop}%
\bibitem [{\citenamefont {{Sriram Shastry}}\ and\ \citenamefont {Sutherland}(1981)}]{ShastrySutherland}%
  \BibitemOpen
  \bibfield  {author} {\bibinfo {author} {\bibfnamefont {B.}~\bibnamefont {{Sriram Shastry}}}\ and\ \bibinfo {author} {\bibfnamefont {B.}~\bibnamefont {Sutherland}},\ }\bibfield  {title} {\bibinfo {title} {Exact ground state of a quantum mechanical antiferromagnet},\ }\href {https://doi.org/https://doi.org/10.1016/0378-4363(81)90838-X} {\bibfield  {journal} {\bibinfo  {journal} {Physica B+C}\ }\textbf {\bibinfo {volume} {108}},\ \bibinfo {pages} {1069} (\bibinfo {year} {1981})}\BibitemShut {NoStop}%
\bibitem [{\citenamefont {Balents}(2010)}]{balents_review}%
  \BibitemOpen
  \bibfield  {author} {\bibinfo {author} {\bibfnamefont {L.}~\bibnamefont {Balents}},\ }\bibfield  {title} {\bibinfo {title} {Spin liquids in frustrated magnets},\ }\href {http://dx.doi.org/10.1038/nature08917} {\bibfield  {journal} {\bibinfo  {journal} {Nature}\ }\textbf {\bibinfo {volume} {464}},\ \bibinfo {pages} {199 EP } (\bibinfo {year} {2010})}\BibitemShut {NoStop}%
\bibitem [{\citenamefont {Savary}\ and\ \citenamefont {Balents}(2017)}]{savary_balents}%
  \BibitemOpen
  \bibfield  {author} {\bibinfo {author} {\bibfnamefont {L.}~\bibnamefont {Savary}}\ and\ \bibinfo {author} {\bibfnamefont {L.}~\bibnamefont {Balents}},\ }\bibfield  {title} {\bibinfo {title} {Quantum spin liquids: a review},\ }\href {http://stacks.iop.org/0034-4885/80/i=1/a=016502} {\bibfield  {journal} {\bibinfo  {journal} {Rep. Prog. Phys.}\ }\textbf {\bibinfo {volume} {80}},\ \bibinfo {pages} {016502} (\bibinfo {year} {2017})}\BibitemShut {NoStop}%
\bibitem [{\citenamefont {Yan}\ \emph {et~al.}(2011)\citenamefont {Yan}, \citenamefont {Huse},\ and\ \citenamefont {White}}]{White_kagome_2011}%
  \BibitemOpen
  \bibfield  {author} {\bibinfo {author} {\bibfnamefont {S.}~\bibnamefont {Yan}}, \bibinfo {author} {\bibfnamefont {D.~A.}\ \bibnamefont {Huse}},\ and\ \bibinfo {author} {\bibfnamefont {S.~R.}\ \bibnamefont {White}},\ }\bibfield  {title} {\bibinfo {title} {{Spin-Liquid Ground State of the $S$ = 1/2 Kagome Heisenberg Antiferromagnet}},\ }\href {https://doi.org/10.1126/science.1201080} {\bibfield  {journal} {\bibinfo  {journal} {Science}\ }\textbf {\bibinfo {volume} {332}},\ \bibinfo {pages} {1173} (\bibinfo {year} {2011})}\BibitemShut {NoStop}%
\bibitem [{\citenamefont {Zhu}\ and\ \citenamefont {White}(2015)}]{zhu_white15}%
  \BibitemOpen
  \bibfield  {author} {\bibinfo {author} {\bibfnamefont {Z.}~\bibnamefont {Zhu}}\ and\ \bibinfo {author} {\bibfnamefont {S.~R.}\ \bibnamefont {White}},\ }\bibfield  {title} {\bibinfo {title} {{Spin liquid phase of the $S=\frac{1}{2}\phantom{\rule{4.pt}{0ex}}{J}_{1}\ensuremath{-}{J}_{2}$ Heisenberg model on the triangular lattice}},\ }\href {https://doi.org/10.1103/PhysRevB.92.041105} {\bibfield  {journal} {\bibinfo  {journal} {Phys. Rev. B}\ }\textbf {\bibinfo {volume} {92}},\ \bibinfo {pages} {041105} (\bibinfo {year} {2015})}\BibitemShut {NoStop}%
\bibitem [{\citenamefont {Iqbal}\ \emph {et~al.}(2016)\citenamefont {Iqbal}, \citenamefont {Hu}, \citenamefont {Thomale}, \citenamefont {Poilblanc},\ and\ \citenamefont {Becca}}]{iqbal16_j1j2}%
  \BibitemOpen
  \bibfield  {author} {\bibinfo {author} {\bibfnamefont {Y.}~\bibnamefont {Iqbal}}, \bibinfo {author} {\bibfnamefont {W.-J.}\ \bibnamefont {Hu}}, \bibinfo {author} {\bibfnamefont {R.}~\bibnamefont {Thomale}}, \bibinfo {author} {\bibfnamefont {D.}~\bibnamefont {Poilblanc}},\ and\ \bibinfo {author} {\bibfnamefont {F.}~\bibnamefont {Becca}},\ }\bibfield  {title} {\bibinfo {title} {{Spin liquid nature in the Heisenberg ${J}_{1}\ensuremath{-}{J}_{2}$ triangular antiferromagnet}},\ }\href {https://doi.org/10.1103/PhysRevB.93.144411} {\bibfield  {journal} {\bibinfo  {journal} {Phys. Rev. B}\ }\textbf {\bibinfo {volume} {93}},\ \bibinfo {pages} {144411} (\bibinfo {year} {2016})}\BibitemShut {NoStop}%
\bibitem [{\citenamefont {Villain}\ \emph {et~al.}(1980)\citenamefont {Villain}, \citenamefont {Bidaux}, \citenamefont {Carton},\ and\ \citenamefont {Conte}}]{villain1980-Ordereffect}%
  \BibitemOpen
  \bibfield  {author} {\bibinfo {author} {\bibfnamefont {J.}~\bibnamefont {Villain}}, \bibinfo {author} {\bibfnamefont {R.}~\bibnamefont {Bidaux}}, \bibinfo {author} {\bibfnamefont {J.-P.}\ \bibnamefont {Carton}},\ and\ \bibinfo {author} {\bibfnamefont {R.}~\bibnamefont {Conte}},\ }\bibfield  {title} {\bibinfo {title} {Order as an effect of disorder},\ }\href@noop {} {\bibfield  {journal} {\bibinfo  {journal} {Journal de Physique}\ }\textbf {\bibinfo {volume} {41}},\ \bibinfo {pages} {1263} (\bibinfo {year} {1980})}\BibitemShut {NoStop}%
\bibitem [{\citenamefont {Shender}(1982)}]{shender1982-Antiferromagneticgarnets}%
  \BibitemOpen
  \bibfield  {author} {\bibinfo {author} {\bibfnamefont {E.~F.}\ \bibnamefont {Shender}},\ }\bibfield  {title} {\bibinfo {title} {Antiferromagnetic garnets with fluctuationally interacting sublattices},\ }\href@noop {} {\bibfield  {journal} {\bibinfo  {journal} {Soviet Journal of Experimental and Theoretical Physics}\ }\textbf {\bibinfo {volume} {56}},\ \bibinfo {pages} {178} (\bibinfo {year} {1982})}\BibitemShut {NoStop}%
\bibitem [{\citenamefont {Henley}(1987)}]{henley1987-Orderingdisorder}%
  \BibitemOpen
  \bibfield  {author} {\bibinfo {author} {\bibfnamefont {C.~L.}\ \bibnamefont {Henley}},\ }\bibfield  {title} {\bibinfo {title} {Ordering by disorder: {{Ground}}-state selection in fcc vector antiferromagnets},\ }\href {https://doi.org/10.1063/1.338570} {\bibfield  {journal} {\bibinfo  {journal} {J. Appl. Phys.}\ }\textbf {\bibinfo {volume} {61}},\ \bibinfo {pages} {3962} (\bibinfo {year} {1987})}\BibitemShut {NoStop}%
\bibitem [{\citenamefont {Henley}(1989)}]{henley1989-Orderingdue}%
  \BibitemOpen
  \bibfield  {author} {\bibinfo {author} {\bibfnamefont {C.~L.}\ \bibnamefont {Henley}},\ }\bibfield  {title} {\bibinfo {title} {Ordering due to disorder in a frustrated vector antiferromagnet},\ }\href {https://doi.org/10.1103/PhysRevLett.62.2056} {\bibfield  {journal} {\bibinfo  {journal} {Phys. Rev. Lett.}\ }\textbf {\bibinfo {volume} {62}},\ \bibinfo {pages} {2056} (\bibinfo {year} {1989})}\BibitemShut {NoStop}%
\bibitem [{\citenamefont {Chandra}\ \emph {et~al.}(1990)\citenamefont {Chandra}, \citenamefont {Coleman},\ and\ \citenamefont {Larkin}}]{chandra_coleman_larkin}%
  \BibitemOpen
  \bibfield  {author} {\bibinfo {author} {\bibfnamefont {P.}~\bibnamefont {Chandra}}, \bibinfo {author} {\bibfnamefont {P.}~\bibnamefont {Coleman}},\ and\ \bibinfo {author} {\bibfnamefont {A.~I.}\ \bibnamefont {Larkin}},\ }\bibfield  {title} {\bibinfo {title} {{Ising transition in frustrated Heisenberg models}},\ }\href {https://doi.org/10.1103/PhysRevLett.64.88} {\bibfield  {journal} {\bibinfo  {journal} {Phys. Rev. Lett.}\ }\textbf {\bibinfo {volume} {64}},\ \bibinfo {pages} {88} (\bibinfo {year} {1990})}\BibitemShut {NoStop}%
\bibitem [{\citenamefont {Chubukov}\ and\ \citenamefont {Jolicoeur}(1992)}]{chubukov_j1j2}%
  \BibitemOpen
  \bibfield  {author} {\bibinfo {author} {\bibfnamefont {A.~V.}\ \bibnamefont {Chubukov}}\ and\ \bibinfo {author} {\bibfnamefont {T.}~\bibnamefont {Jolicoeur}},\ }\bibfield  {title} {\bibinfo {title} {{Order-from-disorder phenomena in Heisenberg antiferromagnets on a triangular lattice}},\ }\href {https://doi.org/10.1103/PhysRevB.46.11137} {\bibfield  {journal} {\bibinfo  {journal} {Phys. Rev. B}\ }\textbf {\bibinfo {volume} {46}},\ \bibinfo {pages} {11137} (\bibinfo {year} {1992})}\BibitemShut {NoStop}%
\bibitem [{\citenamefont {Chubukov}\ and\ \citenamefont {Golosov}(1991)}]{chubukov91}%
  \BibitemOpen
  \bibfield  {author} {\bibinfo {author} {\bibfnamefont {A.~V.}\ \bibnamefont {Chubukov}}\ and\ \bibinfo {author} {\bibfnamefont {D.~I.}\ \bibnamefont {Golosov}},\ }\bibfield  {title} {\bibinfo {title} {Quantum theory of an antiferromagnet on a triangular lattice in a magnetic field},\ }\href {https://doi.org/10.1088/0953-8984/3/1/005} {\bibfield  {journal} {\bibinfo  {journal} {J. Phys.: Condens. Matter}\ }\textbf {\bibinfo {volume} {3}},\ \bibinfo {pages} {69} (\bibinfo {year} {1991})}\BibitemShut {NoStop}%
\bibitem [{\citenamefont {Alicea}\ \emph {et~al.}(2009)\citenamefont {Alicea}, \citenamefont {Chubukov},\ and\ \citenamefont {Starykh}}]{alicea_plateau}%
  \BibitemOpen
  \bibfield  {author} {\bibinfo {author} {\bibfnamefont {J.}~\bibnamefont {Alicea}}, \bibinfo {author} {\bibfnamefont {A.~V.}\ \bibnamefont {Chubukov}},\ and\ \bibinfo {author} {\bibfnamefont {O.~A.}\ \bibnamefont {Starykh}},\ }\bibfield  {title} {\bibinfo {title} {{Quantum Stabilization of the $1/3$-Magnetization Plateau in ${\mathrm{Cs}}_{2}{\mathrm{CuBr}}_{4}$}},\ }\href {https://doi.org/10.1103/PhysRevLett.102.137201} {\bibfield  {journal} {\bibinfo  {journal} {Phys. Rev. Lett.}\ }\textbf {\bibinfo {volume} {102}},\ \bibinfo {pages} {137201} (\bibinfo {year} {2009})}\BibitemShut {NoStop}%
\bibitem [{\citenamefont {Takano}\ \emph {et~al.}(2011)\citenamefont {Takano}, \citenamefont {Tsunetsugu},\ and\ \citenamefont {Zhitomirsky}}]{Takano_2011}%
  \BibitemOpen
  \bibfield  {author} {\bibinfo {author} {\bibfnamefont {J.}~\bibnamefont {Takano}}, \bibinfo {author} {\bibfnamefont {H.}~\bibnamefont {Tsunetsugu}},\ and\ \bibinfo {author} {\bibfnamefont {M.~E.}\ \bibnamefont {Zhitomirsky}},\ }\bibfield  {title} {\bibinfo {title} {Self-consistent spin wave analysis of the magnetization plateau in triangular antiferromagnet},\ }\href {https://doi.org/10.1088/1742-6596/320/1/012011} {\bibfield  {journal} {\bibinfo  {journal} {J. Phys.: Conf. Ser.}\ }\textbf {\bibinfo {volume} {320}},\ \bibinfo {pages} {012011} (\bibinfo {year} {2011})}\BibitemShut {NoStop}%
\bibitem [{\citenamefont {Chubukov}(1992)}]{chubukov_kagome}%
  \BibitemOpen
  \bibfield  {author} {\bibinfo {author} {\bibfnamefont {A.}~\bibnamefont {Chubukov}},\ }\bibfield  {title} {\bibinfo {title} {{Order from disorder in a kagom\'e antiferromagnet}},\ }\href {https://doi.org/10.1103/PhysRevLett.69.832} {\bibfield  {journal} {\bibinfo  {journal} {Phys. Rev. Lett.}\ }\textbf {\bibinfo {volume} {69}},\ \bibinfo {pages} {832} (\bibinfo {year} {1992})}\BibitemShut {NoStop}%
\bibitem [{\citenamefont {Chern}\ and\ \citenamefont {Moessner}(2013)}]{Chern_kagome_2013}%
  \BibitemOpen
  \bibfield  {author} {\bibinfo {author} {\bibfnamefont {G.-W.}\ \bibnamefont {Chern}}\ and\ \bibinfo {author} {\bibfnamefont {R.}~\bibnamefont {Moessner}},\ }\bibfield  {title} {\bibinfo {title} {{Dipolar Order by Disorder in the Classical Heisenberg Antiferromagnet on the Kagome Lattice}},\ }\href {https://doi.org/10.1103/PhysRevLett.110.077201} {\bibfield  {journal} {\bibinfo  {journal} {Phys. Rev. Lett.}\ }\textbf {\bibinfo {volume} {110}},\ \bibinfo {pages} {077201} (\bibinfo {year} {2013})}\BibitemShut {NoStop}%
\bibitem [{\citenamefont {Chernyshev}\ and\ \citenamefont {Zhitomirsky}(2014)}]{AFkagome1}%
  \BibitemOpen
  \bibfield  {author} {\bibinfo {author} {\bibfnamefont {A.~L.}\ \bibnamefont {Chernyshev}}\ and\ \bibinfo {author} {\bibfnamefont {M.~E.}\ \bibnamefont {Zhitomirsky}},\ }\bibfield  {title} {\bibinfo {title} {{Quantum Selection of Order in an $XXZ$ Antiferromagnet on a Kagome Lattice}},\ }\href {https://doi.org/10.1103/PhysRevLett.113.237202} {\bibfield  {journal} {\bibinfo  {journal} {Phys. Rev. Lett.}\ }\textbf {\bibinfo {volume} {113}},\ \bibinfo {pages} {237202} (\bibinfo {year} {2014})}\BibitemShut {NoStop}%
\bibitem [{\citenamefont {Zhitomirsky}\ \emph {et~al.}(2012)\citenamefont {Zhitomirsky}, \citenamefont {Gvozdikova}, \citenamefont {Holdsworth},\ and\ \citenamefont {Moessner}}]{Zh_pyrochlore}%
  \BibitemOpen
  \bibfield  {author} {\bibinfo {author} {\bibfnamefont {M.~E.}\ \bibnamefont {Zhitomirsky}}, \bibinfo {author} {\bibfnamefont {M.~V.}\ \bibnamefont {Gvozdikova}}, \bibinfo {author} {\bibfnamefont {P.~C.~W.}\ \bibnamefont {Holdsworth}},\ and\ \bibinfo {author} {\bibfnamefont {R.}~\bibnamefont {Moessner}},\ }\bibfield  {title} {\bibinfo {title} {{Quantum Order by Disorder and Accidental Soft Mode in ${\mathrm{Er}}_{2}{\mathrm{Ti}}_{2}{\mathbf{O}}_{7}$}},\ }\href {https://doi.org/10.1103/PhysRevLett.109.077204} {\bibfield  {journal} {\bibinfo  {journal} {Phys. Rev. Lett.}\ }\textbf {\bibinfo {volume} {109}},\ \bibinfo {pages} {077204} (\bibinfo {year} {2012})}\BibitemShut {NoStop}%
\bibitem [{\citenamefont {McClarty}\ \emph {et~al.}(2014)\citenamefont {McClarty}, \citenamefont {Stasiak},\ and\ \citenamefont {Gingras}}]{McClarty_pyrochlore_2014}%
  \BibitemOpen
  \bibfield  {author} {\bibinfo {author} {\bibfnamefont {P.~A.}\ \bibnamefont {McClarty}}, \bibinfo {author} {\bibfnamefont {P.}~\bibnamefont {Stasiak}},\ and\ \bibinfo {author} {\bibfnamefont {M.~J.~P.}\ \bibnamefont {Gingras}},\ }\bibfield  {title} {\bibinfo {title} {{Order-by-disorder in the $XY$ pyrochlore antiferromagnet}},\ }\href {https://doi.org/10.1103/PhysRevB.89.024425} {\bibfield  {journal} {\bibinfo  {journal} {Phys. Rev. B}\ }\textbf {\bibinfo {volume} {89}},\ \bibinfo {pages} {024425} (\bibinfo {year} {2014})}\BibitemShut {NoStop}%
\bibitem [{\citenamefont {Jaubert}\ \emph {et~al.}(2015)\citenamefont {Jaubert}, \citenamefont {Benton}, \citenamefont {Rau}, \citenamefont {Oitmaa}, \citenamefont {Singh}, \citenamefont {Shannon},\ and\ \citenamefont {Gingras}}]{Jaubert_pyrochlore_2015}%
  \BibitemOpen
  \bibfield  {author} {\bibinfo {author} {\bibfnamefont {L.~D.~C.}\ \bibnamefont {Jaubert}}, \bibinfo {author} {\bibfnamefont {O.}~\bibnamefont {Benton}}, \bibinfo {author} {\bibfnamefont {J.~G.}\ \bibnamefont {Rau}}, \bibinfo {author} {\bibfnamefont {J.}~\bibnamefont {Oitmaa}}, \bibinfo {author} {\bibfnamefont {R.~R.~P.}\ \bibnamefont {Singh}}, \bibinfo {author} {\bibfnamefont {N.}~\bibnamefont {Shannon}},\ and\ \bibinfo {author} {\bibfnamefont {M.~J.~P.}\ \bibnamefont {Gingras}},\ }\bibfield  {title} {\bibinfo {title} {{Are Multiphase Competition and Order by Disorder the Keys to Understanding ${\mathrm{Yb}}_{2}{\mathrm{Ti}}_{2}{\mathrm{O}}_{7}$?}},\ }\href {https://doi.org/10.1103/PhysRevLett.115.267208} {\bibfield  {journal} {\bibinfo  {journal} {Phys. Rev. Lett.}\ }\textbf {\bibinfo {volume} {115}},\ \bibinfo {pages} {267208} (\bibinfo {year} {2015})}\BibitemShut {NoStop}%
\bibitem [{\citenamefont {Ter~Haar}\ \emph {et~al.}(1962)\citenamefont {Ter~Haar}, \citenamefont {Lines},\ and\ \citenamefont {Bleaney}}]{Lines_fcc}%
  \BibitemOpen
  \bibfield  {author} {\bibinfo {author} {\bibfnamefont {D.}~\bibnamefont {Ter~Haar}}, \bibinfo {author} {\bibfnamefont {M.~E.}\ \bibnamefont {Lines}},\ and\ \bibinfo {author} {\bibfnamefont {B.}~\bibnamefont {Bleaney}},\ }\bibfield  {title} {\bibinfo {title} {A spin-wave theory of anisotropic antiferromagnetica},\ }\href {https://doi.org/10.1098/rsta.1962.0008} {\bibfield  {journal} {\bibinfo  {journal} {Phil. Trans. R. Soc., A}\ }\textbf {\bibinfo {volume} {255}},\ \bibinfo {pages} {1} (\bibinfo {year} {1962})}\BibitemShut {NoStop}%
\bibitem [{\citenamefont {Schick}\ \emph {et~al.}(2020)\citenamefont {Schick}, \citenamefont {Ziman},\ and\ \citenamefont {Zhitomirsky}}]{Zh_fcc1}%
  \BibitemOpen
  \bibfield  {author} {\bibinfo {author} {\bibfnamefont {R.}~\bibnamefont {Schick}}, \bibinfo {author} {\bibfnamefont {T.}~\bibnamefont {Ziman}},\ and\ \bibinfo {author} {\bibfnamefont {M.~E.}\ \bibnamefont {Zhitomirsky}},\ }\bibfield  {title} {\bibinfo {title} {{Quantum versus thermal fluctuations in the fcc antiferromagnet: Alternative routes to order by disorder}},\ }\href {https://doi.org/10.1103/PhysRevB.102.220405} {\bibfield  {journal} {\bibinfo  {journal} {Phys. Rev. B}\ }\textbf {\bibinfo {volume} {102}},\ \bibinfo {pages} {220405} (\bibinfo {year} {2020})}\BibitemShut {NoStop}%
\bibitem [{\citenamefont {Schick}\ \emph {et~al.}(2022)\citenamefont {Schick}, \citenamefont {G\"otze}, \citenamefont {Ziman}, \citenamefont {Zinke}, \citenamefont {Richter},\ and\ \citenamefont {Zhitomirsky}}]{Zh_fcc2}%
  \BibitemOpen
  \bibfield  {author} {\bibinfo {author} {\bibfnamefont {R.}~\bibnamefont {Schick}}, \bibinfo {author} {\bibfnamefont {O.}~\bibnamefont {G\"otze}}, \bibinfo {author} {\bibfnamefont {T.}~\bibnamefont {Ziman}}, \bibinfo {author} {\bibfnamefont {R.}~\bibnamefont {Zinke}}, \bibinfo {author} {\bibfnamefont {J.}~\bibnamefont {Richter}},\ and\ \bibinfo {author} {\bibfnamefont {M.~E.}\ \bibnamefont {Zhitomirsky}},\ }\bibfield  {title} {\bibinfo {title} {Ground-state selection by magnon interactions in a fcc antiferromagnet},\ }\href {https://doi.org/10.1103/PhysRevB.106.094431} {\bibfield  {journal} {\bibinfo  {journal} {Phys. Rev. B}\ }\textbf {\bibinfo {volume} {106}},\ \bibinfo {pages} {094431} (\bibinfo {year} {2022})}\BibitemShut {NoStop}%
\bibitem [{\citenamefont {Nussinov}\ and\ \citenamefont {{van den Brink}}(2015)}]{nussinov2015-Compassmodels}%
  \BibitemOpen
  \bibfield  {author} {\bibinfo {author} {\bibfnamefont {Z.}~\bibnamefont {Nussinov}}\ and\ \bibinfo {author} {\bibfnamefont {J.}~\bibnamefont {{van den Brink}}},\ }\bibfield  {title} {\bibinfo {title} {Compass models: {{Theory}} and physical motivations},\ }\href {https://doi.org/10.1103/RevModPhys.87.1} {\bibfield  {journal} {\bibinfo  {journal} {Rev. Mod. Phys.}\ }\textbf {\bibinfo {volume} {87}},\ \bibinfo {pages} {1} (\bibinfo {year} {2015})}\BibitemShut {NoStop}%
\bibitem [{\citenamefont {Khaliullin}(2001)}]{Giniyat_ObD_2001}%
  \BibitemOpen
  \bibfield  {author} {\bibinfo {author} {\bibfnamefont {G.}~\bibnamefont {Khaliullin}},\ }\bibfield  {title} {\bibinfo {title} {{Order from disorder: Quantum spin gap in magnon spectra of ${\mathrm{LaTiO}}_{3}$}},\ }\href {https://doi.org/10.1103/PhysRevB.64.212405} {\bibfield  {journal} {\bibinfo  {journal} {Phys. Rev. B}\ }\textbf {\bibinfo {volume} {64}},\ \bibinfo {pages} {212405} (\bibinfo {year} {2001})}\BibitemShut {NoStop}%
\bibitem [{\citenamefont {Kitaev}(2006)}]{KITAEV2006}%
  \BibitemOpen
  \bibfield  {author} {\bibinfo {author} {\bibfnamefont {A.}~\bibnamefont {Kitaev}},\ }\bibfield  {title} {\bibinfo {title} {Anyons in an exactly solved model and beyond},\ }\href {https://doi.org/https://doi.org/10.1016/j.aop.2005.10.005} {\bibfield  {journal} {\bibinfo  {journal} {Annals of Physics}\ }\textbf {\bibinfo {volume} {321}},\ \bibinfo {pages} {2} (\bibinfo {year} {2006})},\ \bibinfo {note} {january Special Issue}\BibitemShut {NoStop}%
\bibitem [{\citenamefont {Chaloupka}\ \emph {et~al.}(2010)\citenamefont {Chaloupka}, \citenamefont {Jackeli},\ and\ \citenamefont {Khaliullin}}]{chaloupka2010}%
  \BibitemOpen
  \bibfield  {author} {\bibinfo {author} {\bibfnamefont {J.}~\bibnamefont {Chaloupka}}, \bibinfo {author} {\bibfnamefont {G.}~\bibnamefont {Jackeli}},\ and\ \bibinfo {author} {\bibfnamefont {G.}~\bibnamefont {Khaliullin}},\ }\bibfield  {title} {\bibinfo {title} {{Kitaev-Heisenberg Model on a Honeycomb Lattice: Possible Exotic Phases in Iridium Oxides ${A}_{2}{\mathrm{IrO}}_{3}$}},\ }\href {https://doi.org/10.1103/PhysRevLett.105.027204} {\bibfield  {journal} {\bibinfo  {journal} {Phys. Rev. Lett.}\ }\textbf {\bibinfo {volume} {105}},\ \bibinfo {pages} {027204} (\bibinfo {year} {2010})}\BibitemShut {NoStop}%
\bibitem [{\citenamefont {Sizyuk}\ \emph {et~al.}(2016)\citenamefont {Sizyuk}, \citenamefont {W\"olfle},\ and\ \citenamefont {Perkins}}]{Perkins_KH_ObD_2016}%
  \BibitemOpen
  \bibfield  {author} {\bibinfo {author} {\bibfnamefont {Y.}~\bibnamefont {Sizyuk}}, \bibinfo {author} {\bibfnamefont {P.}~\bibnamefont {W\"olfle}},\ and\ \bibinfo {author} {\bibfnamefont {N.~B.}\ \bibnamefont {Perkins}},\ }\bibfield  {title} {\bibinfo {title} {{Selection of direction of the ordered moments in ${\text{Na}}_{2}{\text{IrO}}_{3}$ and $\ensuremath{\alpha}\ensuremath{-}{\text{RuCl}}_{3}$}},\ }\href {https://doi.org/10.1103/PhysRevB.94.085109} {\bibfield  {journal} {\bibinfo  {journal} {Phys. Rev. B}\ }\textbf {\bibinfo {volume} {94}},\ \bibinfo {pages} {085109} (\bibinfo {year} {2016})}\BibitemShut {NoStop}%
\bibitem [{\citenamefont {Rousochatzakis}\ \emph {et~al.}(2015)\citenamefont {Rousochatzakis}, \citenamefont {Reuther}, \citenamefont {Thomale}, \citenamefont {Rachel},\ and\ \citenamefont {Perkins}}]{rousochatzakis2015-PhaseDiagram}%
  \BibitemOpen
  \bibfield  {author} {\bibinfo {author} {\bibfnamefont {I.}~\bibnamefont {Rousochatzakis}}, \bibinfo {author} {\bibfnamefont {J.}~\bibnamefont {Reuther}}, \bibinfo {author} {\bibfnamefont {R.}~\bibnamefont {Thomale}}, \bibinfo {author} {\bibfnamefont {S.}~\bibnamefont {Rachel}},\ and\ \bibinfo {author} {\bibfnamefont {N.~B.}\ \bibnamefont {Perkins}},\ }\bibfield  {title} {\bibinfo {title} {{Phase {{Diagram}} and {{Quantum Order}} by {{Disorder}} in the {{Kitaev}} \$\{\vphantom\}{{K}}\vphantom\{\}\_\{1\}{\textbackslash}ensuremath\{-\}\{\vphantom\}{{K}}\vphantom\{\}\_\{2\}\$ {{Honeycomb Magnet}}}},\ }\href {https://doi.org/10.1103/PhysRevX.5.041035} {\bibfield  {journal} {\bibinfo  {journal} {Phys. Rev. X}\ }\textbf {\bibinfo {volume} {5}},\ \bibinfo {pages} {041035} (\bibinfo {year} {2015})}\BibitemShut {NoStop}%
\bibitem [{\citenamefont {Elliot}\ \emph {et~al.}(2021)\citenamefont {Elliot}, \citenamefont {McClarty}, \citenamefont {Prabhakaran}, \citenamefont {Johnson}, \citenamefont {Walker}, \citenamefont {Manuel},\ and\ \citenamefont {Coldea}}]{Coldea_Co_2020}%
  \BibitemOpen
  \bibfield  {author} {\bibinfo {author} {\bibfnamefont {M.}~\bibnamefont {Elliot}}, \bibinfo {author} {\bibfnamefont {P.~A.}\ \bibnamefont {McClarty}}, \bibinfo {author} {\bibfnamefont {D.}~\bibnamefont {Prabhakaran}}, \bibinfo {author} {\bibfnamefont {R.~D.}\ \bibnamefont {Johnson}}, \bibinfo {author} {\bibfnamefont {H.~C.}\ \bibnamefont {Walker}}, \bibinfo {author} {\bibfnamefont {P.}~\bibnamefont {Manuel}},\ and\ \bibinfo {author} {\bibfnamefont {R.}~\bibnamefont {Coldea}},\ }\bibfield  {title} {\bibinfo {title} {Order-by-disorder from bond-dependent exchange and intensity signature of nodal quasiparticles in a honeycomb cobaltate},\ }\href {https://doi.org/10.1038/s41467-021-23851-0} {\bibfield  {journal} {\bibinfo  {journal} {Nature Communications}\ }\textbf {\bibinfo {volume} {12}},\ \bibinfo {pages} {3936} (\bibinfo {year} {2021})}\BibitemShut {NoStop}%
\bibitem [{\citenamefont {Kimchi}\ and\ \citenamefont {Vishwanath}(2014)}]{Kimchi14}%
  \BibitemOpen
  \bibfield  {author} {\bibinfo {author} {\bibfnamefont {I.}~\bibnamefont {Kimchi}}\ and\ \bibinfo {author} {\bibfnamefont {A.}~\bibnamefont {Vishwanath}},\ }\bibfield  {title} {\bibinfo {title} {{Kitaev-Heisenberg models for iridates on the triangular, hyperkagome, kagome, fcc, and pyrochlore lattices}},\ }\href {https://doi.org/10.1103/PhysRevB.89.014414} {\bibfield  {journal} {\bibinfo  {journal} {Phys. Rev. B}\ }\textbf {\bibinfo {volume} {89}},\ \bibinfo {pages} {014414} (\bibinfo {year} {2014})}\BibitemShut {NoStop}%
\bibitem [{\citenamefont {Rousochatzakis}\ \emph {et~al.}(2016)\citenamefont {Rousochatzakis}, \citenamefont {R\"ossler}, \citenamefont {van~den Brink},\ and\ \citenamefont {Daghofer}}]{Ioannis_Z2}%
  \BibitemOpen
  \bibfield  {author} {\bibinfo {author} {\bibfnamefont {I.}~\bibnamefont {Rousochatzakis}}, \bibinfo {author} {\bibfnamefont {U.~K.}\ \bibnamefont {R\"ossler}}, \bibinfo {author} {\bibfnamefont {J.}~\bibnamefont {van~den Brink}},\ and\ \bibinfo {author} {\bibfnamefont {M.}~\bibnamefont {Daghofer}},\ }\bibfield  {title} {\bibinfo {title} {{Kitaev anisotropy induces mesoscopic ${\mathbb{Z}}_{2}$ vortex crystals in frustrated hexagonal antiferromagnets}},\ }\href {https://doi.org/10.1103/PhysRevB.93.104417} {\bibfield  {journal} {\bibinfo  {journal} {Phys. Rev. B}\ }\textbf {\bibinfo {volume} {93}},\ \bibinfo {pages} {104417} (\bibinfo {year} {2016})}\BibitemShut {NoStop}%
\bibitem [{\citenamefont {Luo}\ \emph {et~al.}(2017)\citenamefont {Luo}, \citenamefont {Hu}, \citenamefont {Xi}, \citenamefont {Zhao},\ and\ \citenamefont {Wang}}]{Wang17}%
  \BibitemOpen
  \bibfield  {author} {\bibinfo {author} {\bibfnamefont {Q.}~\bibnamefont {Luo}}, \bibinfo {author} {\bibfnamefont {S.}~\bibnamefont {Hu}}, \bibinfo {author} {\bibfnamefont {B.}~\bibnamefont {Xi}}, \bibinfo {author} {\bibfnamefont {J.}~\bibnamefont {Zhao}},\ and\ \bibinfo {author} {\bibfnamefont {X.}~\bibnamefont {Wang}},\ }\bibfield  {title} {\bibinfo {title} {Ground-state phase diagram of an anisotropic spin-$\frac{1}{2}$ model on the triangular lattice},\ }\href {https://doi.org/10.1103/PhysRevB.95.165110} {\bibfield  {journal} {\bibinfo  {journal} {Phys. Rev. B}\ }\textbf {\bibinfo {volume} {95}},\ \bibinfo {pages} {165110} (\bibinfo {year} {2017})}\BibitemShut {NoStop}%
\bibitem [{\citenamefont {Li}\ \emph {et~al.}(2015{\natexlab{a}})\citenamefont {Li}, \citenamefont {Yu},\ and\ \citenamefont {Li}}]{Li_CSL}%
  \BibitemOpen
  \bibfield  {author} {\bibinfo {author} {\bibfnamefont {K.}~\bibnamefont {Li}}, \bibinfo {author} {\bibfnamefont {S.-L.}\ \bibnamefont {Yu}},\ and\ \bibinfo {author} {\bibfnamefont {J.-X.}\ \bibnamefont {Li}},\ }\bibfield  {title} {\bibinfo {title} {{Global phase diagram, possible chiral spin liquid, and topological superconductivity in the triangular Kitaev–Heisenberg model}},\ }\href {http://stacks.iop.org/1367-2630/17/i=4/a=043032} {\bibfield  {journal} {\bibinfo  {journal} {New J. Phys.}\ }\textbf {\bibinfo {volume} {17}},\ \bibinfo {pages} {043032} (\bibinfo {year} {2015}{\natexlab{a}})}\BibitemShut {NoStop}%
\bibitem [{\citenamefont {Kos}\ and\ \citenamefont {Punk}(2017)}]{Punk}%
  \BibitemOpen
  \bibfield  {author} {\bibinfo {author} {\bibfnamefont {P.}~\bibnamefont {Kos}}\ and\ \bibinfo {author} {\bibfnamefont {M.}~\bibnamefont {Punk}},\ }\bibfield  {title} {\bibinfo {title} {{Quantum spin liquid ground states of the Heisenberg-Kitaev model on the triangular lattice}},\ }\href {https://doi.org/10.1103/PhysRevB.95.024421} {\bibfield  {journal} {\bibinfo  {journal} {Phys. Rev. B}\ }\textbf {\bibinfo {volume} {95}},\ \bibinfo {pages} {024421} (\bibinfo {year} {2017})}\BibitemShut {NoStop}%
\bibitem [{\citenamefont {Maksimov}\ \emph {et~al.}(2019)\citenamefont {Maksimov}, \citenamefont {Zhu}, \citenamefont {White},\ and\ \citenamefont {Chernyshev}}]{prx_anisotropic}%
  \BibitemOpen
  \bibfield  {author} {\bibinfo {author} {\bibfnamefont {P.~A.}\ \bibnamefont {Maksimov}}, \bibinfo {author} {\bibfnamefont {Z.}~\bibnamefont {Zhu}}, \bibinfo {author} {\bibfnamefont {S.~R.}\ \bibnamefont {White}},\ and\ \bibinfo {author} {\bibfnamefont {A.~L.}\ \bibnamefont {Chernyshev}},\ }\bibfield  {title} {\bibinfo {title} {{Anisotropic-Exchange Magnets on a Triangular Lattice: Spin Waves, Accidental Degeneracies, and Dual Spin Liquids}},\ }\href {https://doi.org/10.1103/PhysRevX.9.021017} {\bibfield  {journal} {\bibinfo  {journal} {Phys. Rev. X}\ }\textbf {\bibinfo {volume} {9}},\ \bibinfo {pages} {021017} (\bibinfo {year} {2019})}\BibitemShut {NoStop}%
\bibitem [{\citenamefont {Shinjo}\ \emph {et~al.}(2016)\citenamefont {Shinjo}, \citenamefont {Sota}, \citenamefont {Yunoki}, \citenamefont {Totsuka},\ and\ \citenamefont {Tohyama}}]{Tohyama}%
  \BibitemOpen
  \bibfield  {author} {\bibinfo {author} {\bibfnamefont {K.}~\bibnamefont {Shinjo}}, \bibinfo {author} {\bibfnamefont {S.}~\bibnamefont {Sota}}, \bibinfo {author} {\bibfnamefont {S.}~\bibnamefont {Yunoki}}, \bibinfo {author} {\bibfnamefont {K.}~\bibnamefont {Totsuka}},\ and\ \bibinfo {author} {\bibfnamefont {T.}~\bibnamefont {Tohyama}},\ }\bibfield  {title} {\bibinfo {title} {{Density-Matrix Renormalization Group Study of Kitaev–Heisenberg Model on a Triangular Lattice}},\ }\href {https://doi.org/10.7566/JPSJ.85.114710} {\bibfield  {journal} {\bibinfo  {journal} {J. Phys. Soc. Jpn.}\ }\textbf {\bibinfo {volume} {85}},\ \bibinfo {pages} {114710} (\bibinfo {year} {2016})}\BibitemShut {NoStop}%
\bibitem [{\citenamefont {Jackeli}\ and\ \citenamefont {Khaliullin}(2009)}]{Jackeli}%
  \BibitemOpen
  \bibfield  {author} {\bibinfo {author} {\bibfnamefont {G.}~\bibnamefont {Jackeli}}\ and\ \bibinfo {author} {\bibfnamefont {G.}~\bibnamefont {Khaliullin}},\ }\bibfield  {title} {\bibinfo {title} {{Mott Insulators in the Strong Spin-Orbit Coupling Limit: From Heisenberg to a Quantum Compass and Kitaev Models}},\ }\href {https://doi.org/10.1103/PhysRevLett.102.017205} {\bibfield  {journal} {\bibinfo  {journal} {Phys. Rev. Lett.}\ }\textbf {\bibinfo {volume} {102}},\ \bibinfo {pages} {017205} (\bibinfo {year} {2009})}\BibitemShut {NoStop}%
\bibitem [{\citenamefont {Rau}\ \emph {et~al.}(2014)\citenamefont {Rau}, \citenamefont {Lee},\ and\ \citenamefont {Kee}}]{rau_jkg}%
  \BibitemOpen
  \bibfield  {author} {\bibinfo {author} {\bibfnamefont {J.~G.}\ \bibnamefont {Rau}}, \bibinfo {author} {\bibfnamefont {E.~K.-H.}\ \bibnamefont {Lee}},\ and\ \bibinfo {author} {\bibfnamefont {H.-Y.}\ \bibnamefont {Kee}},\ }\bibfield  {title} {\bibinfo {title} {{Generic Spin Model for the Honeycomb Iridates beyond the Kitaev Limit}},\ }\href {https://doi.org/10.1103/PhysRevLett.112.077204} {\bibfield  {journal} {\bibinfo  {journal} {Phys. Rev. Lett.}\ }\textbf {\bibinfo {volume} {112}},\ \bibinfo {pages} {077204} (\bibinfo {year} {2014})}\BibitemShut {NoStop}%
\bibitem [{\citenamefont {{Rau}}\ and\ \citenamefont {{Kee}}(2014)}]{rau2014trigonal}%
  \BibitemOpen
  \bibfield  {author} {\bibinfo {author} {\bibfnamefont {J.~G.}\ \bibnamefont {{Rau}}}\ and\ \bibinfo {author} {\bibfnamefont {H.-Y.}\ \bibnamefont {{Kee}}},\ }\href@noop {} {\bibinfo {title} {{Trigonal distortion in the honeycomb iridates: Proximity of zigzag and spiral phases in Na2IrO3}}} (\bibinfo {year} {2014}),\ \Eprint {https://arxiv.org/abs/1408.4811} {arXiv:1408.4811} \BibitemShut {NoStop}%
\bibitem [{\citenamefont {Li}\ \emph {et~al.}(2016)\citenamefont {Li}, \citenamefont {Wang},\ and\ \citenamefont {Chen}}]{Chen3}%
  \BibitemOpen
  \bibfield  {author} {\bibinfo {author} {\bibfnamefont {Y.-D.}\ \bibnamefont {Li}}, \bibinfo {author} {\bibfnamefont {X.}~\bibnamefont {Wang}},\ and\ \bibinfo {author} {\bibfnamefont {G.}~\bibnamefont {Chen}},\ }\bibfield  {title} {\bibinfo {title} {Anisotropic spin model of strong spin-orbit-coupled triangular antiferromagnets},\ }\href {https://doi.org/10.1103/PhysRevB.94.035107} {\bibfield  {journal} {\bibinfo  {journal} {Phys. Rev. B}\ }\textbf {\bibinfo {volume} {94}},\ \bibinfo {pages} {035107} (\bibinfo {year} {2016})}\BibitemShut {NoStop}%
\bibitem [{\citenamefont {Catuneanu}\ \emph {et~al.}(2015)\citenamefont {Catuneanu}, \citenamefont {Rau}, \citenamefont {Kim},\ and\ \citenamefont {Kee}}]{Rau_tr}%
  \BibitemOpen
  \bibfield  {author} {\bibinfo {author} {\bibfnamefont {A.}~\bibnamefont {Catuneanu}}, \bibinfo {author} {\bibfnamefont {J.~G.}\ \bibnamefont {Rau}}, \bibinfo {author} {\bibfnamefont {H.-S.}\ \bibnamefont {Kim}},\ and\ \bibinfo {author} {\bibfnamefont {H.-Y.}\ \bibnamefont {Kee}},\ }\bibfield  {title} {\bibinfo {title} {{Magnetic orders proximal to the Kitaev limit in frustrated triangular systems: Application to ${\mathrm{Ba}}_{3}{\mathrm{IrTi}}_{2}{\mathrm{O}}_{9}$}},\ }\href {https://doi.org/10.1103/PhysRevB.92.165108} {\bibfield  {journal} {\bibinfo  {journal} {Phys. Rev. B}\ }\textbf {\bibinfo {volume} {92}},\ \bibinfo {pages} {165108} (\bibinfo {year} {2015})}\BibitemShut {NoStop}%
\bibitem [{\citenamefont {Li}\ \emph {et~al.}(2015{\natexlab{b}})\citenamefont {Li}, \citenamefont {Liao}, \citenamefont {Zhang}, \citenamefont {Li}, \citenamefont {Jin}, \citenamefont {Ling}, \citenamefont {Zhang}, \citenamefont {Zou}, \citenamefont {Pi}, \citenamefont {Yang}, \citenamefont {Wang}, \citenamefont {Wu},\ and\ \citenamefont {Zhang}}]{SciRep}%
  \BibitemOpen
  \bibfield  {author} {\bibinfo {author} {\bibfnamefont {Y.}~\bibnamefont {Li}}, \bibinfo {author} {\bibfnamefont {H.}~\bibnamefont {Liao}}, \bibinfo {author} {\bibfnamefont {Z.}~\bibnamefont {Zhang}}, \bibinfo {author} {\bibfnamefont {S.}~\bibnamefont {Li}}, \bibinfo {author} {\bibfnamefont {F.}~\bibnamefont {Jin}}, \bibinfo {author} {\bibfnamefont {L.}~\bibnamefont {Ling}}, \bibinfo {author} {\bibfnamefont {L.}~\bibnamefont {Zhang}}, \bibinfo {author} {\bibfnamefont {Y.}~\bibnamefont {Zou}}, \bibinfo {author} {\bibfnamefont {L.}~\bibnamefont {Pi}}, \bibinfo {author} {\bibfnamefont {Z.}~\bibnamefont {Yang}}, \bibinfo {author} {\bibfnamefont {J.}~\bibnamefont {Wang}}, \bibinfo {author} {\bibfnamefont {Z.}~\bibnamefont {Wu}},\ and\ \bibinfo {author} {\bibfnamefont {Q.}~\bibnamefont {Zhang}},\ }\bibfield  {title} {\bibinfo {title} {{Gapless quantum spin liquid ground state in the two-dimensional spin-1/2 triangular antiferromagnet YbMgGaO$_4$}},\ }\href {https://doi.org/10.1038/srep16419} {\bibfield  {journal}
  {\bibinfo  {journal} {Sci. Rep.}\ }\textbf {\bibinfo {volume} {5}},\ \bibinfo {pages} {16419} (\bibinfo {year} {2015}{\natexlab{b}})}\BibitemShut {NoStop}%
\bibitem [{\citenamefont {Li}\ \emph {et~al.}(2015{\natexlab{c}})\citenamefont {Li}, \citenamefont {Chen}, \citenamefont {Tong}, \citenamefont {Pi}, \citenamefont {Liu}, \citenamefont {Yang}, \citenamefont {Wang},\ and\ \citenamefont {Zhang}}]{Chen1}%
  \BibitemOpen
  \bibfield  {author} {\bibinfo {author} {\bibfnamefont {Y.}~\bibnamefont {Li}}, \bibinfo {author} {\bibfnamefont {G.}~\bibnamefont {Chen}}, \bibinfo {author} {\bibfnamefont {W.}~\bibnamefont {Tong}}, \bibinfo {author} {\bibfnamefont {L.}~\bibnamefont {Pi}}, \bibinfo {author} {\bibfnamefont {J.}~\bibnamefont {Liu}}, \bibinfo {author} {\bibfnamefont {Z.}~\bibnamefont {Yang}}, \bibinfo {author} {\bibfnamefont {X.}~\bibnamefont {Wang}},\ and\ \bibinfo {author} {\bibfnamefont {Q.}~\bibnamefont {Zhang}},\ }\bibfield  {title} {\bibinfo {title} {{Rare-Earth Triangular Lattice Spin Liquid: A Single-Crystal Study of ${\mathrm{YbMgGaO}}_{4}$}},\ }\href {https://doi.org/10.1103/PhysRevLett.115.167203} {\bibfield  {journal} {\bibinfo  {journal} {Phys. Rev. Lett.}\ }\textbf {\bibinfo {volume} {115}},\ \bibinfo {pages} {167203} (\bibinfo {year} {2015}{\natexlab{c}})}\BibitemShut {NoStop}%
\bibitem [{\citenamefont {Paddison}\ \emph {et~al.}(2017)\citenamefont {Paddison}, \citenamefont {Daum}, \citenamefont {Dun}, \citenamefont {Ehlers}, \citenamefont {Liu}, \citenamefont {Stone}, \citenamefont {Zhou},\ and\ \citenamefont {Mourigal}}]{MM}%
  \BibitemOpen
  \bibfield  {author} {\bibinfo {author} {\bibfnamefont {J.~A.}\ \bibnamefont {Paddison}}, \bibinfo {author} {\bibfnamefont {M.}~\bibnamefont {Daum}}, \bibinfo {author} {\bibfnamefont {Z.}~\bibnamefont {Dun}}, \bibinfo {author} {\bibfnamefont {G.}~\bibnamefont {Ehlers}}, \bibinfo {author} {\bibfnamefont {Y.}~\bibnamefont {Liu}}, \bibinfo {author} {\bibfnamefont {M.}~\bibnamefont {Stone}}, \bibinfo {author} {\bibfnamefont {H.}~\bibnamefont {Zhou}},\ and\ \bibinfo {author} {\bibfnamefont {M.}~\bibnamefont {Mourigal}},\ }\bibfield  {title} {\bibinfo {title} {{Continuous excitations of the triangular-lattice quantum spin liquid YbMgGaO$_4$}},\ }\href {https://doi.org/10.1038/nphys3971} {\bibfield  {journal} {\bibinfo  {journal} {Nat. Phys.}\ }\textbf {\bibinfo {volume} {13}},\ \bibinfo {pages} {117} (\bibinfo {year} {2017})}\BibitemShut {NoStop}%
\bibitem [{\citenamefont {Zhang}\ \emph {et~al.}(2018)\citenamefont {Zhang}, \citenamefont {Mahmood}, \citenamefont {Daum}, \citenamefont {Dun}, \citenamefont {Paddison}, \citenamefont {Laurita}, \citenamefont {Hong}, \citenamefont {Zhou}, \citenamefont {Armitage},\ and\ \citenamefont {Mourigal}}]{MM2}%
  \BibitemOpen
  \bibfield  {author} {\bibinfo {author} {\bibfnamefont {X.}~\bibnamefont {Zhang}}, \bibinfo {author} {\bibfnamefont {F.}~\bibnamefont {Mahmood}}, \bibinfo {author} {\bibfnamefont {M.}~\bibnamefont {Daum}}, \bibinfo {author} {\bibfnamefont {Z.}~\bibnamefont {Dun}}, \bibinfo {author} {\bibfnamefont {J.~A.~M.}\ \bibnamefont {Paddison}}, \bibinfo {author} {\bibfnamefont {N.~J.}\ \bibnamefont {Laurita}}, \bibinfo {author} {\bibfnamefont {T.}~\bibnamefont {Hong}}, \bibinfo {author} {\bibfnamefont {H.}~\bibnamefont {Zhou}}, \bibinfo {author} {\bibfnamefont {N.~P.}\ \bibnamefont {Armitage}},\ and\ \bibinfo {author} {\bibfnamefont {M.}~\bibnamefont {Mourigal}},\ }\bibfield  {title} {\bibinfo {title} {{Hierarchy of Exchange Interactions in the Triangular-Lattice Spin Liquid ${\mathrm{YbMgGaO}}_{4}$}},\ }\href {https://doi.org/10.1103/PhysRevX.8.031001} {\bibfield  {journal} {\bibinfo  {journal} {Phys. Rev. X}\ }\textbf {\bibinfo {volume} {8}},\ \bibinfo {pages} {031001} (\bibinfo {year} {2018})}\BibitemShut {NoStop}%
\bibitem [{\citenamefont {Zhu}\ \emph {et~al.}(2017)\citenamefont {Zhu}, \citenamefont {Maksimov}, \citenamefont {White},\ and\ \citenamefont {Chernyshev}}]{us_mimicry}%
  \BibitemOpen
  \bibfield  {author} {\bibinfo {author} {\bibfnamefont {Z.}~\bibnamefont {Zhu}}, \bibinfo {author} {\bibfnamefont {P.~A.}\ \bibnamefont {Maksimov}}, \bibinfo {author} {\bibfnamefont {S.~R.}\ \bibnamefont {White}},\ and\ \bibinfo {author} {\bibfnamefont {A.~L.}\ \bibnamefont {Chernyshev}},\ }\bibfield  {title} {\bibinfo {title} {{Disorder-Induced Mimicry of a Spin Liquid in ${\mathrm{YbMgGaO}}_{4}$}},\ }\href {https://doi.org/10.1103/PhysRevLett.119.157201} {\bibfield  {journal} {\bibinfo  {journal} {Phys. Rev. Lett.}\ }\textbf {\bibinfo {volume} {119}},\ \bibinfo {pages} {157201} (\bibinfo {year} {2017})}\BibitemShut {NoStop}%
\bibitem [{\citenamefont {Ma}\ \emph {et~al.}(2018)\citenamefont {Ma}, \citenamefont {Wang}, \citenamefont {Dong}, \citenamefont {Zhang}, \citenamefont {Li}, \citenamefont {Zheng}, \citenamefont {Yu}, \citenamefont {Wang}, \citenamefont {Che}, \citenamefont {Ran}, \citenamefont {Bao}, \citenamefont {Cai}, \citenamefont {\ifmmode~\check{C}\else \v{C}\fi{}erm\'ak}, \citenamefont {Schneidewind}, \citenamefont {Yano}, \citenamefont {Gardner}, \citenamefont {Lu}, \citenamefont {Yu}, \citenamefont {Liu}, \citenamefont {Li}, \citenamefont {Li},\ and\ \citenamefont {Wen}}]{Wen_freeze18}%
  \BibitemOpen
  \bibfield  {author} {\bibinfo {author} {\bibfnamefont {Z.}~\bibnamefont {Ma}}, \bibinfo {author} {\bibfnamefont {J.}~\bibnamefont {Wang}}, \bibinfo {author} {\bibfnamefont {Z.-Y.}\ \bibnamefont {Dong}}, \bibinfo {author} {\bibfnamefont {J.}~\bibnamefont {Zhang}}, \bibinfo {author} {\bibfnamefont {S.}~\bibnamefont {Li}}, \bibinfo {author} {\bibfnamefont {S.-H.}\ \bibnamefont {Zheng}}, \bibinfo {author} {\bibfnamefont {Y.}~\bibnamefont {Yu}}, \bibinfo {author} {\bibfnamefont {W.}~\bibnamefont {Wang}}, \bibinfo {author} {\bibfnamefont {L.}~\bibnamefont {Che}}, \bibinfo {author} {\bibfnamefont {K.}~\bibnamefont {Ran}}, \bibinfo {author} {\bibfnamefont {S.}~\bibnamefont {Bao}}, \bibinfo {author} {\bibfnamefont {Z.}~\bibnamefont {Cai}}, \bibinfo {author} {\bibfnamefont {P.}~\bibnamefont {\ifmmode~\check{C}\else \v{C}\fi{}erm\'ak}}, \bibinfo {author} {\bibfnamefont {A.}~\bibnamefont {Schneidewind}}, \bibinfo {author} {\bibfnamefont {S.}~\bibnamefont {Yano}}, \bibinfo {author} {\bibfnamefont {J.~S.}\ \bibnamefont
  {Gardner}}, \bibinfo {author} {\bibfnamefont {X.}~\bibnamefont {Lu}}, \bibinfo {author} {\bibfnamefont {S.-L.}\ \bibnamefont {Yu}}, \bibinfo {author} {\bibfnamefont {J.-M.}\ \bibnamefont {Liu}}, \bibinfo {author} {\bibfnamefont {S.}~\bibnamefont {Li}}, \bibinfo {author} {\bibfnamefont {J.-X.}\ \bibnamefont {Li}},\ and\ \bibinfo {author} {\bibfnamefont {J.}~\bibnamefont {Wen}},\ }\bibfield  {title} {\bibinfo {title} {{Spin-Glass Ground State in a Triangular-Lattice Compound ${\mathrm{YbZnGaO}}_{4}$}},\ }\href {https://doi.org/10.1103/PhysRevLett.120.087201} {\bibfield  {journal} {\bibinfo  {journal} {Phys. Rev. Lett.}\ }\textbf {\bibinfo {volume} {120}},\ \bibinfo {pages} {087201} (\bibinfo {year} {2018})}\BibitemShut {NoStop}%
\bibitem [{\citenamefont {Steinhardt}\ \emph {et~al.}(2021)\citenamefont {Steinhardt}, \citenamefont {Maksimov}, \citenamefont {Dissanayake}, \citenamefont {Shi}, \citenamefont {Butch}, \citenamefont {Graf}, \citenamefont {Podlesnyak}, \citenamefont {Liu}, \citenamefont {Zhao}, \citenamefont {Xu}, \citenamefont {Lynn}, \citenamefont {Marjerrison}, \citenamefont {Chernyshev},\ and\ \citenamefont {Haravifard}}]{YZGO_2021}%
  \BibitemOpen
  \bibfield  {author} {\bibinfo {author} {\bibfnamefont {W.}~\bibnamefont {Steinhardt}}, \bibinfo {author} {\bibfnamefont {P.~A.}\ \bibnamefont {Maksimov}}, \bibinfo {author} {\bibfnamefont {S.}~\bibnamefont {Dissanayake}}, \bibinfo {author} {\bibfnamefont {Z.}~\bibnamefont {Shi}}, \bibinfo {author} {\bibfnamefont {N.~P.}\ \bibnamefont {Butch}}, \bibinfo {author} {\bibfnamefont {D.}~\bibnamefont {Graf}}, \bibinfo {author} {\bibfnamefont {A.}~\bibnamefont {Podlesnyak}}, \bibinfo {author} {\bibfnamefont {Y.}~\bibnamefont {Liu}}, \bibinfo {author} {\bibfnamefont {Y.}~\bibnamefont {Zhao}}, \bibinfo {author} {\bibfnamefont {G.}~\bibnamefont {Xu}}, \bibinfo {author} {\bibfnamefont {J.~W.}\ \bibnamefont {Lynn}}, \bibinfo {author} {\bibfnamefont {C.}~\bibnamefont {Marjerrison}}, \bibinfo {author} {\bibfnamefont {A.~L.}\ \bibnamefont {Chernyshev}},\ and\ \bibinfo {author} {\bibfnamefont {S.}~\bibnamefont {Haravifard}},\ }\bibfield  {title} {\bibinfo {title} {{Phase diagram of YbZnGaO$_4$ in applied magnetic field}},\
  }\href {https://doi.org/10.1038/s41535-021-00380-z} {\bibfield  {journal} {\bibinfo  {journal} {npj Quantum Mater.}\ }\textbf {\bibinfo {volume} {6}},\ \bibinfo {pages} {78} (\bibinfo {year} {2021})}\BibitemShut {NoStop}%
\bibitem [{\citenamefont {Ma}\ \emph {et~al.}(2021)\citenamefont {Ma}, \citenamefont {Dong}, \citenamefont {Wang}, \citenamefont {Zheng}, \citenamefont {Ran}, \citenamefont {Bao}, \citenamefont {Cai}, \citenamefont {Shangguan}, \citenamefont {Wang}, \citenamefont {Boehm}, \citenamefont {Steffens}, \citenamefont {Regnault}, \citenamefont {Wang}, \citenamefont {Su}, \citenamefont {Yu}, \citenamefont {Liu}, \citenamefont {Li},\ and\ \citenamefont {Wen}}]{Wen_YZGO_2021}%
  \BibitemOpen
  \bibfield  {author} {\bibinfo {author} {\bibfnamefont {Z.}~\bibnamefont {Ma}}, \bibinfo {author} {\bibfnamefont {Z.-Y.}\ \bibnamefont {Dong}}, \bibinfo {author} {\bibfnamefont {J.}~\bibnamefont {Wang}}, \bibinfo {author} {\bibfnamefont {S.}~\bibnamefont {Zheng}}, \bibinfo {author} {\bibfnamefont {K.}~\bibnamefont {Ran}}, \bibinfo {author} {\bibfnamefont {S.}~\bibnamefont {Bao}}, \bibinfo {author} {\bibfnamefont {Z.}~\bibnamefont {Cai}}, \bibinfo {author} {\bibfnamefont {Y.}~\bibnamefont {Shangguan}}, \bibinfo {author} {\bibfnamefont {W.}~\bibnamefont {Wang}}, \bibinfo {author} {\bibfnamefont {M.}~\bibnamefont {Boehm}}, \bibinfo {author} {\bibfnamefont {P.}~\bibnamefont {Steffens}}, \bibinfo {author} {\bibfnamefont {L.-P.}\ \bibnamefont {Regnault}}, \bibinfo {author} {\bibfnamefont {X.}~\bibnamefont {Wang}}, \bibinfo {author} {\bibfnamefont {Y.}~\bibnamefont {Su}}, \bibinfo {author} {\bibfnamefont {S.-L.}\ \bibnamefont {Yu}}, \bibinfo {author} {\bibfnamefont {J.-M.}\ \bibnamefont {Liu}}, \bibinfo {author}
  {\bibfnamefont {J.-X.}\ \bibnamefont {Li}},\ and\ \bibinfo {author} {\bibfnamefont {J.}~\bibnamefont {Wen}},\ }\bibfield  {title} {\bibinfo {title} {{Disorder-induced broadening of the spin waves in the triangular-lattice quantum spin liquid candidate ${\mathrm{YbZnGaO}}_{4}$}},\ }\href {https://doi.org/10.1103/PhysRevB.104.224433} {\bibfield  {journal} {\bibinfo  {journal} {Phys. Rev. B}\ }\textbf {\bibinfo {volume} {104}},\ \bibinfo {pages} {224433} (\bibinfo {year} {2021})}\BibitemShut {NoStop}%
\bibitem [{\citenamefont {Pratt}\ \emph {et~al.}(2022)\citenamefont {Pratt}, \citenamefont {Lang}, \citenamefont {Steinhardt}, \citenamefont {Haravifard},\ and\ \citenamefont {Blundell}}]{Blundell_YZGO_2022}%
  \BibitemOpen
  \bibfield  {author} {\bibinfo {author} {\bibfnamefont {F.~L.}\ \bibnamefont {Pratt}}, \bibinfo {author} {\bibfnamefont {F.}~\bibnamefont {Lang}}, \bibinfo {author} {\bibfnamefont {W.}~\bibnamefont {Steinhardt}}, \bibinfo {author} {\bibfnamefont {S.}~\bibnamefont {Haravifard}},\ and\ \bibinfo {author} {\bibfnamefont {S.~J.}\ \bibnamefont {Blundell}},\ }\bibfield  {title} {\bibinfo {title} {{Spin dynamics, entanglement, and the nature of the spin liquid state in ${\mathrm{YbZnGaO}}_{4}$}},\ }\href {https://doi.org/10.1103/PhysRevB.106.L060401} {\bibfield  {journal} {\bibinfo  {journal} {Phys. Rev. B}\ }\textbf {\bibinfo {volume} {106}},\ \bibinfo {pages} {L060401} (\bibinfo {year} {2022})}\BibitemShut {NoStop}%
\bibitem [{\citenamefont {Liu}\ \emph {et~al.}(2018)\citenamefont {Liu}, \citenamefont {Zhang}, \citenamefont {Ji}, \citenamefont {Liu}, \citenamefont {Li}, \citenamefont {Wang}, \citenamefont {Lei}, \citenamefont {Chen},\ and\ \citenamefont {Zhang}}]{Liu_2018}%
  \BibitemOpen
  \bibfield  {author} {\bibinfo {author} {\bibfnamefont {W.}~\bibnamefont {Liu}}, \bibinfo {author} {\bibfnamefont {Z.}~\bibnamefont {Zhang}}, \bibinfo {author} {\bibfnamefont {J.}~\bibnamefont {Ji}}, \bibinfo {author} {\bibfnamefont {Y.}~\bibnamefont {Liu}}, \bibinfo {author} {\bibfnamefont {J.}~\bibnamefont {Li}}, \bibinfo {author} {\bibfnamefont {X.}~\bibnamefont {Wang}}, \bibinfo {author} {\bibfnamefont {H.}~\bibnamefont {Lei}}, \bibinfo {author} {\bibfnamefont {G.}~\bibnamefont {Chen}},\ and\ \bibinfo {author} {\bibfnamefont {Q.}~\bibnamefont {Zhang}},\ }\bibfield  {title} {\bibinfo {title} {{Rare-Earth Chalcogenides: A Large Family of Triangular Lattice Spin Liquid Candidates}},\ }\href {https://doi.org/10.1088/0256-307X/35/11/117501} {\bibfield  {journal} {\bibinfo  {journal} {Chin. Phys. Lett.}\ }\textbf {\bibinfo {volume} {35}},\ \bibinfo {pages} {117501} (\bibinfo {year} {2018})}\BibitemShut {NoStop}%
\bibitem [{\citenamefont {Bordelon}\ \emph {et~al.}(2019)\citenamefont {Bordelon}, \citenamefont {Kenney}, \citenamefont {Liu}, \citenamefont {Hogan}, \citenamefont {Posthuma}, \citenamefont {Kavand}, \citenamefont {Lyu}, \citenamefont {Sherwin}, \citenamefont {Butch}, \citenamefont {Brown}, \citenamefont {Graf}, \citenamefont {Balents},\ and\ \citenamefont {Wilson}}]{Wilson_NaYbO2_2019}%
  \BibitemOpen
  \bibfield  {author} {\bibinfo {author} {\bibfnamefont {M.~M.}\ \bibnamefont {Bordelon}}, \bibinfo {author} {\bibfnamefont {E.}~\bibnamefont {Kenney}}, \bibinfo {author} {\bibfnamefont {C.}~\bibnamefont {Liu}}, \bibinfo {author} {\bibfnamefont {T.}~\bibnamefont {Hogan}}, \bibinfo {author} {\bibfnamefont {L.}~\bibnamefont {Posthuma}}, \bibinfo {author} {\bibfnamefont {M.}~\bibnamefont {Kavand}}, \bibinfo {author} {\bibfnamefont {Y.}~\bibnamefont {Lyu}}, \bibinfo {author} {\bibfnamefont {M.}~\bibnamefont {Sherwin}}, \bibinfo {author} {\bibfnamefont {N.~P.}\ \bibnamefont {Butch}}, \bibinfo {author} {\bibfnamefont {C.}~\bibnamefont {Brown}}, \bibinfo {author} {\bibfnamefont {M.~J.}\ \bibnamefont {Graf}}, \bibinfo {author} {\bibfnamefont {L.}~\bibnamefont {Balents}},\ and\ \bibinfo {author} {\bibfnamefont {S.~D.}\ \bibnamefont {Wilson}},\ }\bibfield  {title} {\bibinfo {title} {{Field-tunable quantum disordered ground state in the triangular-lattice antiferromagnet NaYbO$_2$}},\ }\href
  {https://doi.org/10.1038/s41567-019-0594-5} {\bibfield  {journal} {\bibinfo  {journal} {Nat. Phys.}\ }\textbf {\bibinfo {volume} {15}},\ \bibinfo {pages} {1058} (\bibinfo {year} {2019})}\BibitemShut {NoStop}%
\bibitem [{\citenamefont {Wu}\ \emph {et~al.}(2022)\citenamefont {Wu}, \citenamefont {Li}, \citenamefont {Zhang}, \citenamefont {Liu}, \citenamefont {Gao}, \citenamefont {Feng}, \citenamefont {Deng}, \citenamefont {Ren}, \citenamefont {Wang}, \citenamefont {Chen}, \citenamefont {Embs}, \citenamefont {Zhu}, \citenamefont {Huang}, \citenamefont {Xiang}, \citenamefont {Chen}, \citenamefont {Wu}, \citenamefont {Choi}, \citenamefont {Qu}, \citenamefont {Li}, \citenamefont {Wang}, \citenamefont {Zhou}, \citenamefont {Su}, \citenamefont {Wang}, \citenamefont {Chen}, \citenamefont {Zhang},\ and\ \citenamefont {Ma}}]{Wu2022}%
  \BibitemOpen
  \bibfield  {author} {\bibinfo {author} {\bibfnamefont {J.}~\bibnamefont {Wu}}, \bibinfo {author} {\bibfnamefont {J.}~\bibnamefont {Li}}, \bibinfo {author} {\bibfnamefont {Z.}~\bibnamefont {Zhang}}, \bibinfo {author} {\bibfnamefont {C.}~\bibnamefont {Liu}}, \bibinfo {author} {\bibfnamefont {Y.~H.}\ \bibnamefont {Gao}}, \bibinfo {author} {\bibfnamefont {E.}~\bibnamefont {Feng}}, \bibinfo {author} {\bibfnamefont {G.}~\bibnamefont {Deng}}, \bibinfo {author} {\bibfnamefont {Q.}~\bibnamefont {Ren}}, \bibinfo {author} {\bibfnamefont {Z.}~\bibnamefont {Wang}}, \bibinfo {author} {\bibfnamefont {R.}~\bibnamefont {Chen}}, \bibinfo {author} {\bibfnamefont {J.}~\bibnamefont {Embs}}, \bibinfo {author} {\bibfnamefont {F.}~\bibnamefont {Zhu}}, \bibinfo {author} {\bibfnamefont {Q.}~\bibnamefont {Huang}}, \bibinfo {author} {\bibfnamefont {Z.}~\bibnamefont {Xiang}}, \bibinfo {author} {\bibfnamefont {L.}~\bibnamefont {Chen}}, \bibinfo {author} {\bibfnamefont {Y.}~\bibnamefont {Wu}}, \bibinfo {author} {\bibfnamefont {E.~S.}\
  \bibnamefont {Choi}}, \bibinfo {author} {\bibfnamefont {Z.}~\bibnamefont {Qu}}, \bibinfo {author} {\bibfnamefont {L.}~\bibnamefont {Li}}, \bibinfo {author} {\bibfnamefont {J.}~\bibnamefont {Wang}}, \bibinfo {author} {\bibfnamefont {H.}~\bibnamefont {Zhou}}, \bibinfo {author} {\bibfnamefont {Y.}~\bibnamefont {Su}}, \bibinfo {author} {\bibfnamefont {X.}~\bibnamefont {Wang}}, \bibinfo {author} {\bibfnamefont {G.}~\bibnamefont {Chen}}, \bibinfo {author} {\bibfnamefont {Q.}~\bibnamefont {Zhang}},\ and\ \bibinfo {author} {\bibfnamefont {J.}~\bibnamefont {Ma}},\ }\bibfield  {title} {\bibinfo {title} {{Magnetic field effects on the quantum spin liquid behaviors of NaYbS$_2$}},\ }\href {https://doi.org/10.1007/s44214-022-00011-z} {\bibfield  {journal} {\bibinfo  {journal} {Quantum Front.}\ }\textbf {\bibinfo {volume} {1}},\ \bibinfo {pages} {13} (\bibinfo {year} {2022})}\BibitemShut {NoStop}%
\bibitem [{\citenamefont {Xie}\ \emph {et~al.}(2023)\citenamefont {Xie}, \citenamefont {Eberharter}, \citenamefont {Xing}, \citenamefont {Nishimoto}, \citenamefont {Brando}, \citenamefont {Khanenko}, \citenamefont {Sichelschmidt}, \citenamefont {Turrini}, \citenamefont {Mazzone}, \citenamefont {Naumov}, \citenamefont {Sanjeewa}, \citenamefont {Harrison}, \citenamefont {Sefat}, \citenamefont {Normand}, \citenamefont {L{\"a}uchli}, \citenamefont {Podlesnyak},\ and\ \citenamefont {Nikitin}}]{Xie2023}%
  \BibitemOpen
  \bibfield  {author} {\bibinfo {author} {\bibfnamefont {T.}~\bibnamefont {Xie}}, \bibinfo {author} {\bibfnamefont {A.~A.}\ \bibnamefont {Eberharter}}, \bibinfo {author} {\bibfnamefont {J.}~\bibnamefont {Xing}}, \bibinfo {author} {\bibfnamefont {S.}~\bibnamefont {Nishimoto}}, \bibinfo {author} {\bibfnamefont {M.}~\bibnamefont {Brando}}, \bibinfo {author} {\bibfnamefont {P.}~\bibnamefont {Khanenko}}, \bibinfo {author} {\bibfnamefont {J.}~\bibnamefont {Sichelschmidt}}, \bibinfo {author} {\bibfnamefont {A.~A.}\ \bibnamefont {Turrini}}, \bibinfo {author} {\bibfnamefont {D.~G.}\ \bibnamefont {Mazzone}}, \bibinfo {author} {\bibfnamefont {P.~G.}\ \bibnamefont {Naumov}}, \bibinfo {author} {\bibfnamefont {L.~D.}\ \bibnamefont {Sanjeewa}}, \bibinfo {author} {\bibfnamefont {N.}~\bibnamefont {Harrison}}, \bibinfo {author} {\bibfnamefont {A.~S.}\ \bibnamefont {Sefat}}, \bibinfo {author} {\bibfnamefont {B.}~\bibnamefont {Normand}}, \bibinfo {author} {\bibfnamefont {A.~M.}\ \bibnamefont {L{\"a}uchli}}, \bibinfo {author}
  {\bibfnamefont {A.}~\bibnamefont {Podlesnyak}},\ and\ \bibinfo {author} {\bibfnamefont {S.~E.}\ \bibnamefont {Nikitin}},\ }\bibfield  {title} {\bibinfo {title} {{Complete field-induced spectral response of the spin-1/2 triangular-lattice antiferromagnet CsYbSe$_2$}},\ }\href {https://doi.org/10.1038/s41535-023-00580-9} {\bibfield  {journal} {\bibinfo  {journal} {npj Quantum Mater.}\ }\textbf {\bibinfo {volume} {8}},\ \bibinfo {pages} {48} (\bibinfo {year} {2023})}\BibitemShut {NoStop}%
\bibitem [{\citenamefont {Ding}\ \emph {et~al.}(2023)\citenamefont {Ding}, \citenamefont {Wo}, \citenamefont {Luo}, \citenamefont {Gu}, \citenamefont {Gu}, \citenamefont {Bewley}, \citenamefont {Chen},\ and\ \citenamefont {Zhao}}]{KErSe2_INS_2023}%
  \BibitemOpen
  \bibfield  {author} {\bibinfo {author} {\bibfnamefont {G.}~\bibnamefont {Ding}}, \bibinfo {author} {\bibfnamefont {H.}~\bibnamefont {Wo}}, \bibinfo {author} {\bibfnamefont {R.~L.}\ \bibnamefont {Luo}}, \bibinfo {author} {\bibfnamefont {Y.}~\bibnamefont {Gu}}, \bibinfo {author} {\bibfnamefont {Y.}~\bibnamefont {Gu}}, \bibinfo {author} {\bibfnamefont {R.}~\bibnamefont {Bewley}}, \bibinfo {author} {\bibfnamefont {G.}~\bibnamefont {Chen}},\ and\ \bibinfo {author} {\bibfnamefont {J.}~\bibnamefont {Zhao}},\ }\bibfield  {title} {\bibinfo {title} {{Stripe order and spin dynamics in the triangular-lattice antiferromagnet ${\mathrm{KErSe}}_{2}$: A single-crystal study with a theoretical description}},\ }\href {https://doi.org/10.1103/PhysRevB.107.L100411} {\bibfield  {journal} {\bibinfo  {journal} {Phys. Rev. B}\ }\textbf {\bibinfo {volume} {107}},\ \bibinfo {pages} {L100411} (\bibinfo {year} {2023})}\BibitemShut {NoStop}%
\bibitem [{\citenamefont {Scheie}\ \emph {et~al.}(2024)\citenamefont {Scheie}, \citenamefont {Ghioldi}, \citenamefont {Xing}, \citenamefont {Paddison}, \citenamefont {Sherman}, \citenamefont {Dupont}, \citenamefont {Sanjeewa}, \citenamefont {Lee}, \citenamefont {Woods}, \citenamefont {Abernathy}, \citenamefont {Pajerowski}, \citenamefont {Williams}, \citenamefont {Zhang}, \citenamefont {Manuel}, \citenamefont {Trumper}, \citenamefont {Pemmaraju}, \citenamefont {Sefat}, \citenamefont {Parker}, \citenamefont {Devereaux}, \citenamefont {Movshovich}, \citenamefont {Moore}, \citenamefont {Batista},\ and\ \citenamefont {Tennant}}]{Scheie2024}%
  \BibitemOpen
  \bibfield  {author} {\bibinfo {author} {\bibfnamefont {A.~O.}\ \bibnamefont {Scheie}}, \bibinfo {author} {\bibfnamefont {E.~A.}\ \bibnamefont {Ghioldi}}, \bibinfo {author} {\bibfnamefont {J.}~\bibnamefont {Xing}}, \bibinfo {author} {\bibfnamefont {J.~A.~M.}\ \bibnamefont {Paddison}}, \bibinfo {author} {\bibfnamefont {N.~E.}\ \bibnamefont {Sherman}}, \bibinfo {author} {\bibfnamefont {M.}~\bibnamefont {Dupont}}, \bibinfo {author} {\bibfnamefont {L.~D.}\ \bibnamefont {Sanjeewa}}, \bibinfo {author} {\bibfnamefont {S.}~\bibnamefont {Lee}}, \bibinfo {author} {\bibfnamefont {A.~J.}\ \bibnamefont {Woods}}, \bibinfo {author} {\bibfnamefont {D.}~\bibnamefont {Abernathy}}, \bibinfo {author} {\bibfnamefont {D.~M.}\ \bibnamefont {Pajerowski}}, \bibinfo {author} {\bibfnamefont {T.~J.}\ \bibnamefont {Williams}}, \bibinfo {author} {\bibfnamefont {S.-S.}\ \bibnamefont {Zhang}}, \bibinfo {author} {\bibfnamefont {L.~O.}\ \bibnamefont {Manuel}}, \bibinfo {author} {\bibfnamefont {A.~E.}\ \bibnamefont {Trumper}}, \bibinfo
  {author} {\bibfnamefont {C.~D.}\ \bibnamefont {Pemmaraju}}, \bibinfo {author} {\bibfnamefont {A.~S.}\ \bibnamefont {Sefat}}, \bibinfo {author} {\bibfnamefont {D.~S.}\ \bibnamefont {Parker}}, \bibinfo {author} {\bibfnamefont {T.~P.}\ \bibnamefont {Devereaux}}, \bibinfo {author} {\bibfnamefont {R.}~\bibnamefont {Movshovich}}, \bibinfo {author} {\bibfnamefont {J.~E.}\ \bibnamefont {Moore}}, \bibinfo {author} {\bibfnamefont {C.~D.}\ \bibnamefont {Batista}},\ and\ \bibinfo {author} {\bibfnamefont {D.~A.}\ \bibnamefont {Tennant}},\ }\bibfield  {title} {\bibinfo {title} {{Proximate spin liquid and fractionalization in the triangular antiferromagnet KYbSe$_2$}},\ }\href {https://doi.org/10.1038/s41567-023-02259-1} {\bibfield  {journal} {\bibinfo  {journal} {Nat. Phys.}\ }\textbf {\bibinfo {volume} {20}},\ \bibinfo {pages} {74} (\bibinfo {year} {2024})}\BibitemShut {NoStop}%
\bibitem [{\citenamefont {Xie}\ \emph {et~al.}(2024{\natexlab{a}})\citenamefont {Xie}, \citenamefont {Gozel}, \citenamefont {Xing}, \citenamefont {Zhao}, \citenamefont {Avdoshenko}, \citenamefont {Wu}, \citenamefont {Sefat}, \citenamefont {Chernyshev}, \citenamefont {L\"auchli}, \citenamefont {Podlesnyak},\ and\ \citenamefont {Nikitin}}]{CsCeSe2_Nikitin_2024}%
  \BibitemOpen
  \bibfield  {author} {\bibinfo {author} {\bibfnamefont {T.}~\bibnamefont {Xie}}, \bibinfo {author} {\bibfnamefont {S.}~\bibnamefont {Gozel}}, \bibinfo {author} {\bibfnamefont {J.}~\bibnamefont {Xing}}, \bibinfo {author} {\bibfnamefont {N.}~\bibnamefont {Zhao}}, \bibinfo {author} {\bibfnamefont {S.~M.}\ \bibnamefont {Avdoshenko}}, \bibinfo {author} {\bibfnamefont {L.}~\bibnamefont {Wu}}, \bibinfo {author} {\bibfnamefont {A.~S.}\ \bibnamefont {Sefat}}, \bibinfo {author} {\bibfnamefont {A.~L.}\ \bibnamefont {Chernyshev}}, \bibinfo {author} {\bibfnamefont {A.~M.}\ \bibnamefont {L\"auchli}}, \bibinfo {author} {\bibfnamefont {A.}~\bibnamefont {Podlesnyak}},\ and\ \bibinfo {author} {\bibfnamefont {S.~E.}\ \bibnamefont {Nikitin}},\ }\bibfield  {title} {\bibinfo {title} {{Quantum Spin Dynamics Due to Strong Kitaev Interactions in the Triangular-Lattice Antiferromagnet ${\mathrm{CsCeSe}}_{2}$}},\ }\href {https://doi.org/10.1103/PhysRevLett.133.096703} {\bibfield  {journal} {\bibinfo  {journal} {Phys. Rev. Lett.}\
  }\textbf {\bibinfo {volume} {133}},\ \bibinfo {pages} {096703} (\bibinfo {year} {2024}{\natexlab{a}})}\BibitemShut {NoStop}%
\bibitem [{\citenamefont {Xie}\ \emph {et~al.}(2024{\natexlab{b}})\citenamefont {Xie}, \citenamefont {Zhuo}, \citenamefont {Cai}, \citenamefont {Zhang},\ and\ \citenamefont {Zhang}}]{Xie_2024}%
  \BibitemOpen
  \bibfield  {author} {\bibinfo {author} {\bibfnamefont {M.}~\bibnamefont {Xie}}, \bibinfo {author} {\bibfnamefont {W.}~\bibnamefont {Zhuo}}, \bibinfo {author} {\bibfnamefont {Y.}~\bibnamefont {Cai}}, \bibinfo {author} {\bibfnamefont {Z.}~\bibnamefont {Zhang}},\ and\ \bibinfo {author} {\bibfnamefont {Q.}~\bibnamefont {Zhang}},\ }\bibfield  {title} {\bibinfo {title} {{Rare-Earth Chalcogenides: An Inspiring Playground for Exploring Frustrated Magnetism}},\ }\href {https://doi.org/10.1088/0256-307X/41/11/117505} {\bibfield  {journal} {\bibinfo  {journal} {Chin. Phys. Lett.}\ }\textbf {\bibinfo {volume} {41}},\ \bibinfo {pages} {117505} (\bibinfo {year} {2024}{\natexlab{b}})}\BibitemShut {NoStop}%
\bibitem [{\citenamefont {Shikano}\ \emph {et~al.}(2004)\citenamefont {Shikano}, \citenamefont {Delmas},\ and\ \citenamefont {Darriet}}]{Shikano_2004}%
  \BibitemOpen
  \bibfield  {author} {\bibinfo {author} {\bibfnamefont {M.}~\bibnamefont {Shikano}}, \bibinfo {author} {\bibfnamefont {C.}~\bibnamefont {Delmas}},\ and\ \bibinfo {author} {\bibfnamefont {J.}~\bibnamefont {Darriet}},\ }\bibfield  {title} {\bibinfo {title} {{NaRuO$_2$ and Na$_x$RuO$_2$·$y$H2O: New Oxide and Oxyhydrate with Two Dimensional RuO$_2$ Layers}},\ }\href {https://doi.org/10.1021/ic035324d} {\bibfield  {journal} {\bibinfo  {journal} {Inorg. Chem.}\ }\textbf {\bibinfo {volume} {43}},\ \bibinfo {pages} {1214} (\bibinfo {year} {2004})}\BibitemShut {NoStop}%
\bibitem [{\citenamefont {Ortiz}\ \emph {et~al.}(2023)\citenamefont {Ortiz}, \citenamefont {Sarte}, \citenamefont {Avidor}, \citenamefont {Hay}, \citenamefont {Kenney}, \citenamefont {Kolesnikov}, \citenamefont {Pajerowski}, \citenamefont {Aczel}, \citenamefont {Taddei}, \citenamefont {Brown}, \citenamefont {Wang}, \citenamefont {Graf}, \citenamefont {Seshadri}, \citenamefont {Balents},\ and\ \citenamefont {Wilson}}]{Ortiz2023}%
  \BibitemOpen
  \bibfield  {author} {\bibinfo {author} {\bibfnamefont {B.~R.}\ \bibnamefont {Ortiz}}, \bibinfo {author} {\bibfnamefont {P.~M.}\ \bibnamefont {Sarte}}, \bibinfo {author} {\bibfnamefont {A.~H.}\ \bibnamefont {Avidor}}, \bibinfo {author} {\bibfnamefont {A.}~\bibnamefont {Hay}}, \bibinfo {author} {\bibfnamefont {E.}~\bibnamefont {Kenney}}, \bibinfo {author} {\bibfnamefont {A.~I.}\ \bibnamefont {Kolesnikov}}, \bibinfo {author} {\bibfnamefont {D.~M.}\ \bibnamefont {Pajerowski}}, \bibinfo {author} {\bibfnamefont {A.~A.}\ \bibnamefont {Aczel}}, \bibinfo {author} {\bibfnamefont {K.~M.}\ \bibnamefont {Taddei}}, \bibinfo {author} {\bibfnamefont {C.~M.}\ \bibnamefont {Brown}}, \bibinfo {author} {\bibfnamefont {C.}~\bibnamefont {Wang}}, \bibinfo {author} {\bibfnamefont {M.~J.}\ \bibnamefont {Graf}}, \bibinfo {author} {\bibfnamefont {R.}~\bibnamefont {Seshadri}}, \bibinfo {author} {\bibfnamefont {L.}~\bibnamefont {Balents}},\ and\ \bibinfo {author} {\bibfnamefont {S.~D.}\ \bibnamefont {Wilson}},\ }\bibfield  {title}
  {\bibinfo {title} {{Quantum disordered ground state in the triangular-lattice magnet NaRuO$_2$}},\ }\href {https://doi.org/10.1038/s41567-023-02039-x} {\bibfield  {journal} {\bibinfo  {journal} {Nat. Phys.}\ }\textbf {\bibinfo {volume} {19}},\ \bibinfo {pages} {943} (\bibinfo {year} {2023})}\BibitemShut {NoStop}%
\bibitem [{\citenamefont {Razpopov}\ \emph {et~al.}(2023)\citenamefont {Razpopov}, \citenamefont {Kaib}, \citenamefont {Backes}, \citenamefont {Balents}, \citenamefont {Wilson}, \citenamefont {Ferrari}, \citenamefont {Riedl},\ and\ \citenamefont {Valent{\'i}}}]{Razpopov2023}%
  \BibitemOpen
  \bibfield  {author} {\bibinfo {author} {\bibfnamefont {A.}~\bibnamefont {Razpopov}}, \bibinfo {author} {\bibfnamefont {D.~A.~S.}\ \bibnamefont {Kaib}}, \bibinfo {author} {\bibfnamefont {S.}~\bibnamefont {Backes}}, \bibinfo {author} {\bibfnamefont {L.}~\bibnamefont {Balents}}, \bibinfo {author} {\bibfnamefont {S.~D.}\ \bibnamefont {Wilson}}, \bibinfo {author} {\bibfnamefont {F.}~\bibnamefont {Ferrari}}, \bibinfo {author} {\bibfnamefont {K.}~\bibnamefont {Riedl}},\ and\ \bibinfo {author} {\bibfnamefont {R.}~\bibnamefont {Valent{\'i}}},\ }\bibfield  {title} {\bibinfo {title} {{A $j_\text{eff}{\thinspace}={\thinspace}1/2$ Kitaev material on the triangular lattice: the case of NaRuO$_2$}},\ }\href {https://doi.org/10.1038/s41535-023-00567-6} {\bibfield  {journal} {\bibinfo  {journal} {npj Quantum Mater.}\ }\textbf {\bibinfo {volume} {8}},\ \bibinfo {pages} {36} (\bibinfo {year} {2023})}\BibitemShut {NoStop}%
\bibitem [{\citenamefont {Bhattacharyya}\ \emph {et~al.}(2023)\citenamefont {Bhattacharyya}, \citenamefont {Bogdanov}, \citenamefont {Nishimoto}, \citenamefont {Wilson},\ and\ \citenamefont {Hozoi}}]{Bhattacharyya2023}%
  \BibitemOpen
  \bibfield  {author} {\bibinfo {author} {\bibfnamefont {P.}~\bibnamefont {Bhattacharyya}}, \bibinfo {author} {\bibfnamefont {N.~A.}\ \bibnamefont {Bogdanov}}, \bibinfo {author} {\bibfnamefont {S.}~\bibnamefont {Nishimoto}}, \bibinfo {author} {\bibfnamefont {S.~D.}\ \bibnamefont {Wilson}},\ and\ \bibinfo {author} {\bibfnamefont {L.}~\bibnamefont {Hozoi}},\ }\bibfield  {title} {\bibinfo {title} {{NaRuO$_2$: Kitaev-Heisenberg exchange in triangular-lattice setting}},\ }\href {https://doi.org/10.1038/s41535-023-00582-7} {\bibfield  {journal} {\bibinfo  {journal} {npj Quantum Mater.}\ }\textbf {\bibinfo {volume} {8}},\ \bibinfo {pages} {52} (\bibinfo {year} {2023})}\BibitemShut {NoStop}%
\bibitem [{\citenamefont {M\"oller}\ \emph {et~al.}(2012)\citenamefont {M\"oller}, \citenamefont {Amuneke}, \citenamefont {Daniel}, \citenamefont {Lorenz}, \citenamefont {de~la Cruz}, \citenamefont {Gooch},\ and\ \citenamefont {Chu}}]{Moller_2012}%
  \BibitemOpen
  \bibfield  {author} {\bibinfo {author} {\bibfnamefont {A.}~\bibnamefont {M\"oller}}, \bibinfo {author} {\bibfnamefont {N.~E.}\ \bibnamefont {Amuneke}}, \bibinfo {author} {\bibfnamefont {P.}~\bibnamefont {Daniel}}, \bibinfo {author} {\bibfnamefont {B.}~\bibnamefont {Lorenz}}, \bibinfo {author} {\bibfnamefont {C.~R.}\ \bibnamefont {de~la Cruz}}, \bibinfo {author} {\bibfnamefont {M.}~\bibnamefont {Gooch}},\ and\ \bibinfo {author} {\bibfnamefont {P.~C.~W.}\ \bibnamefont {Chu}},\ }\bibfield  {title} {\bibinfo {title} {{$A$Ag${}_{2}M$[VO${}_{4}{]}_{2}$ ($A=\mathrm{Ba},\phantom{\rule{0.16em}{0ex}}\mathrm{Sr}$; $M\phantom{\rule{0.16em}{0ex}}=\phantom{\rule{0.16em}{0ex}}\mathrm{Co},\phantom{\rule{0.16em}{0ex}}\mathrm{Ni}$): A series of ferromagnetic insulators}},\ }\href {https://doi.org/10.1103/PhysRevB.85.214422} {\bibfield  {journal} {\bibinfo  {journal} {Phys. Rev. B}\ }\textbf {\bibinfo {volume} {85}},\ \bibinfo {pages} {214422} (\bibinfo {year} {2012})}\BibitemShut {NoStop}%
\bibitem [{\citenamefont {Rawl}\ \emph {et~al.}(2017)\citenamefont {Rawl}, \citenamefont {Lee}, \citenamefont {Choi}, \citenamefont {Li}, \citenamefont {Chen}, \citenamefont {Baumbach}, \citenamefont {dela Cruz}, \citenamefont {Ma},\ and\ \citenamefont {Zhou}}]{Rawl_2017}%
  \BibitemOpen
  \bibfield  {author} {\bibinfo {author} {\bibfnamefont {R.}~\bibnamefont {Rawl}}, \bibinfo {author} {\bibfnamefont {M.}~\bibnamefont {Lee}}, \bibinfo {author} {\bibfnamefont {E.~S.}\ \bibnamefont {Choi}}, \bibinfo {author} {\bibfnamefont {G.}~\bibnamefont {Li}}, \bibinfo {author} {\bibfnamefont {K.~W.}\ \bibnamefont {Chen}}, \bibinfo {author} {\bibfnamefont {R.}~\bibnamefont {Baumbach}}, \bibinfo {author} {\bibfnamefont {C.~R.}\ \bibnamefont {dela Cruz}}, \bibinfo {author} {\bibfnamefont {J.}~\bibnamefont {Ma}},\ and\ \bibinfo {author} {\bibfnamefont {H.~D.}\ \bibnamefont {Zhou}},\ }\bibfield  {title} {\bibinfo {title} {{Magnetic properties of the triangular lattice magnets ${A}_{4}{B}^{\ensuremath{'}}{B}_{2}{\mathrm{O}}_{12}$ ($A=\mathrm{Ba}$, Sr, La; ${B}^{\ensuremath{'}}=\mathrm{Co}$, Ni, Mn; $B=\mathrm{W}$, Re)}},\ }\href {https://doi.org/10.1103/PhysRevB.95.174438} {\bibfield  {journal} {\bibinfo  {journal} {Phys. Rev. B}\ }\textbf {\bibinfo {volume} {95}},\ \bibinfo {pages} {174438} (\bibinfo {year}
  {2017})}\BibitemShut {NoStop}%
\bibitem [{\citenamefont {Becker}\ \emph {et~al.}(2015)\citenamefont {Becker}, \citenamefont {Hermanns}, \citenamefont {Bauer}, \citenamefont {Garst},\ and\ \citenamefont {Trebst}}]{Trebst_tr}%
  \BibitemOpen
  \bibfield  {author} {\bibinfo {author} {\bibfnamefont {M.}~\bibnamefont {Becker}}, \bibinfo {author} {\bibfnamefont {M.}~\bibnamefont {Hermanns}}, \bibinfo {author} {\bibfnamefont {B.}~\bibnamefont {Bauer}}, \bibinfo {author} {\bibfnamefont {M.}~\bibnamefont {Garst}},\ and\ \bibinfo {author} {\bibfnamefont {S.}~\bibnamefont {Trebst}},\ }\bibfield  {title} {\bibinfo {title} {{Spin-orbit physics of $j=\frac{1}{2}$ Mott insulators on the triangular lattice}},\ }\href {https://doi.org/10.1103/PhysRevB.91.155135} {\bibfield  {journal} {\bibinfo  {journal} {Phys. Rev. B}\ }\textbf {\bibinfo {volume} {91}},\ \bibinfo {pages} {155135} (\bibinfo {year} {2015})}\BibitemShut {NoStop}%
\bibitem [{\citenamefont {Jackeli}\ and\ \citenamefont {Avella}(2015)}]{Avella}%
  \BibitemOpen
  \bibfield  {author} {\bibinfo {author} {\bibfnamefont {G.}~\bibnamefont {Jackeli}}\ and\ \bibinfo {author} {\bibfnamefont {A.}~\bibnamefont {Avella}},\ }\bibfield  {title} {\bibinfo {title} {{Quantum order by disorder in the Kitaev model on a triangular lattice}},\ }\href {https://doi.org/10.1103/PhysRevB.92.184416} {\bibfield  {journal} {\bibinfo  {journal} {Phys. Rev. B}\ }\textbf {\bibinfo {volume} {92}},\ \bibinfo {pages} {184416} (\bibinfo {year} {2015})}\BibitemShut {NoStop}%
\bibitem [{\citenamefont {Jiang}\ \emph {et~al.}(2023)\citenamefont {Jiang}, \citenamefont {White},\ and\ \citenamefont {Chernyshev}}]{shengtao_j1j3}%
  \BibitemOpen
  \bibfield  {author} {\bibinfo {author} {\bibfnamefont {S.}~\bibnamefont {Jiang}}, \bibinfo {author} {\bibfnamefont {S.~R.}\ \bibnamefont {White}},\ and\ \bibinfo {author} {\bibfnamefont {A.~L.}\ \bibnamefont {Chernyshev}},\ }\bibfield  {title} {\bibinfo {title} {{Quantum phases in the honeycomb-lattice ${J}_{1}$--${J}_{3}$ ferro-antiferromagnetic model}},\ }\href {https://doi.org/10.1103/PhysRevB.108.L180406} {\bibfield  {journal} {\bibinfo  {journal} {Phys. Rev. B}\ }\textbf {\bibinfo {volume} {108}},\ \bibinfo {pages} {L180406} (\bibinfo {year} {2023})}\BibitemShut {NoStop}%
\bibitem [{\citenamefont {White}(1992)}]{white_density_1992}%
  \BibitemOpen
  \bibfield  {author} {\bibinfo {author} {\bibfnamefont {S.~R.}\ \bibnamefont {White}},\ }\bibfield  {title} {\bibinfo {title} {Density matrix formulation for quantum renormalization groups},\ }\href {https://doi.org/10.1103/PhysRevLett.69.2863} {\bibfield  {journal} {\bibinfo  {journal} {Phys. Rev. Lett.}\ }\textbf {\bibinfo {volume} {69}},\ \bibinfo {pages} {2863} (\bibinfo {year} {1992})}\BibitemShut {NoStop}%
\bibitem [{\citenamefont {Holstein}\ and\ \citenamefont {Primakoff}(1940)}]{hp1940}%
  \BibitemOpen
  \bibfield  {author} {\bibinfo {author} {\bibfnamefont {T.}~\bibnamefont {Holstein}}\ and\ \bibinfo {author} {\bibfnamefont {H.}~\bibnamefont {Primakoff}},\ }\bibfield  {title} {\bibinfo {title} {Field dependence of the intrinsic domain magnetization of a ferromagnet},\ }\href {https://doi.org/10.1103/PhysRev.58.1098} {\bibfield  {journal} {\bibinfo  {journal} {Phys. Rev.}\ }\textbf {\bibinfo {volume} {58}},\ \bibinfo {pages} {1098} (\bibinfo {year} {1940})}\BibitemShut {NoStop}%
\bibitem [{Note1()}]{Note1}%
  \BibitemOpen
  \bibinfo {note} {Note that there is a typo in Ref.\cite {prx_anisotropic}: wrong sign in front of $(1-\Delta )$ in $A_\protect \mathbf {k}$.}\BibitemShut {Stop}%
\bibitem [{\citenamefont {Long}(1989)}]{Long_RSPT}%
  \BibitemOpen
  \bibfield  {author} {\bibinfo {author} {\bibfnamefont {M.~W.}\ \bibnamefont {Long}},\ }\bibfield  {title} {\bibinfo {title} {Effects that can stabilise multiple spin-density waves},\ }\href {https://doi.org/10.1088/0953-8984/1/17/008} {\bibfield  {journal} {\bibinfo  {journal} {J. Phys.: Condens. Matter}\ }\textbf {\bibinfo {volume} {1}},\ \bibinfo {pages} {2857} (\bibinfo {year} {1989})}\BibitemShut {NoStop}%
\bibitem [{\citenamefont {Bergman}\ \emph {et~al.}(2007)\citenamefont {Bergman}, \citenamefont {Shindou}, \citenamefont {Fiete},\ and\ \citenamefont {Balents}}]{Bergman_RSPT}%
  \BibitemOpen
  \bibfield  {author} {\bibinfo {author} {\bibfnamefont {D.~L.}\ \bibnamefont {Bergman}}, \bibinfo {author} {\bibfnamefont {R.}~\bibnamefont {Shindou}}, \bibinfo {author} {\bibfnamefont {G.~A.}\ \bibnamefont {Fiete}},\ and\ \bibinfo {author} {\bibfnamefont {L.}~\bibnamefont {Balents}},\ }\bibfield  {title} {\bibinfo {title} {Degenerate perturbation theory of quantum fluctuations in a pyrochlore antiferromagnet},\ }\href {https://doi.org/10.1103/PhysRevB.75.094403} {\bibfield  {journal} {\bibinfo  {journal} {Phys. Rev. B}\ }\textbf {\bibinfo {volume} {75}},\ \bibinfo {pages} {094403} (\bibinfo {year} {2007})}\BibitemShut {NoStop}%
\bibitem [{\citenamefont {Zhitomirsky}(2015)}]{Zhitomirsky_RSPT}%
  \BibitemOpen
  \bibfield  {author} {\bibinfo {author} {\bibfnamefont {M.~E.}\ \bibnamefont {Zhitomirsky}},\ }\bibfield  {title} {\bibinfo {title} {Real-space perturbation theory for frustrated magnets: application to magnetization plateaus},\ }\href {https://doi.org/10.1088/1742-6596/592/1/012110} {\bibfield  {journal} {\bibinfo  {journal} {J. Phys.: Conf. Ser.}\ }\textbf {\bibinfo {volume} {592}},\ \bibinfo {pages} {012110} (\bibinfo {year} {2015})}\BibitemShut {NoStop}%
\bibitem [{\citenamefont {Wenzel}\ \emph {et~al.}(2012)\citenamefont {Wenzel}, \citenamefont {Coletta}, \citenamefont {Korshunov},\ and\ \citenamefont {Mila}}]{Coletta1}%
  \BibitemOpen
  \bibfield  {author} {\bibinfo {author} {\bibfnamefont {S.}~\bibnamefont {Wenzel}}, \bibinfo {author} {\bibfnamefont {T.}~\bibnamefont {Coletta}}, \bibinfo {author} {\bibfnamefont {S.~E.}\ \bibnamefont {Korshunov}},\ and\ \bibinfo {author} {\bibfnamefont {F.}~\bibnamefont {Mila}},\ }\bibfield  {title} {\bibinfo {title} {{Evidence for Columnar Order in the Fully Frustrated Transverse Field Ising Model on the Square Lattice}},\ }\href {https://doi.org/10.1103/PhysRevLett.109.187202} {\bibfield  {journal} {\bibinfo  {journal} {Phys. Rev. Lett.}\ }\textbf {\bibinfo {volume} {109}},\ \bibinfo {pages} {187202} (\bibinfo {year} {2012})}\BibitemShut {NoStop}%
\bibitem [{\citenamefont {Coletta}\ \emph {et~al.}(2014)\citenamefont {Coletta}, \citenamefont {Korshunov},\ and\ \citenamefont {Mila}}]{Coletta2}%
  \BibitemOpen
  \bibfield  {author} {\bibinfo {author} {\bibfnamefont {T.}~\bibnamefont {Coletta}}, \bibinfo {author} {\bibfnamefont {S.~E.}\ \bibnamefont {Korshunov}},\ and\ \bibinfo {author} {\bibfnamefont {F.}~\bibnamefont {Mila}},\ }\bibfield  {title} {\bibinfo {title} {{Semiclassical evidence of columnar order in the fully frustrated transverse-field Ising model on the square lattice}},\ }\href {https://doi.org/10.1103/PhysRevB.90.205109} {\bibfield  {journal} {\bibinfo  {journal} {Phys. Rev. B}\ }\textbf {\bibinfo {volume} {90}},\ \bibinfo {pages} {205109} (\bibinfo {year} {2014})}\BibitemShut {NoStop}%
\bibitem [{\citenamefont {Coletta}\ \emph {et~al.}(2013)\citenamefont {Coletta}, \citenamefont {Zhitomirsky},\ and\ \citenamefont {Mila}}]{Coletta3}%
  \BibitemOpen
  \bibfield  {author} {\bibinfo {author} {\bibfnamefont {T.}~\bibnamefont {Coletta}}, \bibinfo {author} {\bibfnamefont {M.~E.}\ \bibnamefont {Zhitomirsky}},\ and\ \bibinfo {author} {\bibfnamefont {F.}~\bibnamefont {Mila}},\ }\bibfield  {title} {\bibinfo {title} {{Quantum stabilization of classically unstable plateau structures}},\ }\href {https://doi.org/10.1103/PhysRevB.87.060407} {\bibfield  {journal} {\bibinfo  {journal} {Phys. Rev. B}\ }\textbf {\bibinfo {volume} {87}},\ \bibinfo {pages} {060407} (\bibinfo {year} {2013})}\BibitemShut {NoStop}%
\bibitem [{\citenamefont {Zhitomirsky}\ and\ \citenamefont {Nikuni}(1998)}]{Nikuni_1998}%
  \BibitemOpen
  \bibfield  {author} {\bibinfo {author} {\bibfnamefont {M.~E.}\ \bibnamefont {Zhitomirsky}}\ and\ \bibinfo {author} {\bibfnamefont {T.}~\bibnamefont {Nikuni}},\ }\bibfield  {title} {\bibinfo {title} {{Magnetization curve of a square-lattice Heisenberg antiferromagnet}},\ }\href {https://doi.org/10.1103/PhysRevB.57.5013} {\bibfield  {journal} {\bibinfo  {journal} {Phys. Rev. B}\ }\textbf {\bibinfo {volume} {57}},\ \bibinfo {pages} {5013} (\bibinfo {year} {1998})}\BibitemShut {NoStop}%
\bibitem [{\citenamefont {Thalmeier}\ \emph {et~al.}(2008)\citenamefont {Thalmeier}, \citenamefont {Zhitomirsky}, \citenamefont {Schmidt},\ and\ \citenamefont {Shannon}}]{Shannon_2008}%
  \BibitemOpen
  \bibfield  {author} {\bibinfo {author} {\bibfnamefont {P.}~\bibnamefont {Thalmeier}}, \bibinfo {author} {\bibfnamefont {M.~E.}\ \bibnamefont {Zhitomirsky}}, \bibinfo {author} {\bibfnamefont {B.}~\bibnamefont {Schmidt}},\ and\ \bibinfo {author} {\bibfnamefont {N.}~\bibnamefont {Shannon}},\ }\bibfield  {title} {\bibinfo {title} {{Quantum effects in magnetization of ${J}_{1}\text{\ensuremath{-}}{J}_{2}$ square lattice antiferromagnet}},\ }\href {https://doi.org/10.1103/PhysRevB.77.104441} {\bibfield  {journal} {\bibinfo  {journal} {Phys. Rev. B}\ }\textbf {\bibinfo {volume} {77}},\ \bibinfo {pages} {104441} (\bibinfo {year} {2008})}\BibitemShut {NoStop}%
\bibitem [{\citenamefont {Mourigal}\ \emph {et~al.}(2010)\citenamefont {Mourigal}, \citenamefont {Zhitomirsky},\ and\ \citenamefont {Chernyshev}}]{Mourigal10}%
  \BibitemOpen
  \bibfield  {author} {\bibinfo {author} {\bibfnamefont {M.}~\bibnamefont {Mourigal}}, \bibinfo {author} {\bibfnamefont {M.~E.}\ \bibnamefont {Zhitomirsky}},\ and\ \bibinfo {author} {\bibfnamefont {A.~L.}\ \bibnamefont {Chernyshev}},\ }\bibfield  {title} {\bibinfo {title} {{Field-induced decay dynamics in square-lattice antiferromagnets}},\ }\href {https://doi.org/10.1103/PhysRevB.82.144402} {\bibfield  {journal} {\bibinfo  {journal} {Phys. Rev. B}\ }\textbf {\bibinfo {volume} {82}},\ \bibinfo {pages} {144402} (\bibinfo {year} {2010})}\BibitemShut {NoStop}%
\bibitem [{\citenamefont {Maksimov}\ \emph {et~al.}(2016)\citenamefont {Maksimov}, \citenamefont {Zhitomirsky},\ and\ \citenamefont {Chernyshev}}]{umbrella}%
  \BibitemOpen
  \bibfield  {author} {\bibinfo {author} {\bibfnamefont {P.~A.}\ \bibnamefont {Maksimov}}, \bibinfo {author} {\bibfnamefont {M.~E.}\ \bibnamefont {Zhitomirsky}},\ and\ \bibinfo {author} {\bibfnamefont {A.~L.}\ \bibnamefont {Chernyshev}},\ }\bibfield  {title} {\bibinfo {title} {{Field-induced decays in XXZ triangular-lattice antiferromagnets}},\ }\href {https://doi.org/10.1103/PhysRevB.94.140407} {\bibfield  {journal} {\bibinfo  {journal} {Phys. Rev. B}\ }\textbf {\bibinfo {volume} {94}},\ \bibinfo {pages} {140407} (\bibinfo {year} {2016})}\BibitemShut {NoStop}%
\bibitem [{\citenamefont {Fishman}\ \emph {et~al.}(2020)\citenamefont {Fishman}, \citenamefont {White},\ and\ \citenamefont {Stoudenmire}}]{itensor}%
  \BibitemOpen
  \bibfield  {author} {\bibinfo {author} {\bibfnamefont {M.}~\bibnamefont {Fishman}}, \bibinfo {author} {\bibfnamefont {S.~R.}\ \bibnamefont {White}},\ and\ \bibinfo {author} {\bibfnamefont {E.~M.}\ \bibnamefont {Stoudenmire}},\ }\href@noop {} {\bibinfo {title} {The \mbox{ITensor} software library for tensor network calculations}} (\bibinfo {year} {2020}),\ \Eprint {https://arxiv.org/abs/2007.14822} {arXiv:2007.14822} \BibitemShut {NoStop}%
\bibitem [{\citenamefont {Chernyshev}\ and\ \citenamefont {Maksimov}(2016)}]{us_kagome}%
  \BibitemOpen
  \bibfield  {author} {\bibinfo {author} {\bibfnamefont {A.~L.}\ \bibnamefont {Chernyshev}}\ and\ \bibinfo {author} {\bibfnamefont {P.~A.}\ \bibnamefont {Maksimov}},\ }\bibfield  {title} {\bibinfo {title} {{Damped Topological Magnons in the Kagome-Lattice Ferromagnets}},\ }\href {https://doi.org/10.1103/PhysRevLett.117.187203} {\bibfield  {journal} {\bibinfo  {journal} {Phys. Rev. Lett.}\ }\textbf {\bibinfo {volume} {117}},\ \bibinfo {pages} {187203} (\bibinfo {year} {2016})}\BibitemShut {NoStop}%
\bibitem [{\citenamefont {Rau}\ \emph {et~al.}(2019)\citenamefont {Rau}, \citenamefont {Moessner},\ and\ \citenamefont {McClarty}}]{rau_pyrochlore}%
  \BibitemOpen
  \bibfield  {author} {\bibinfo {author} {\bibfnamefont {J.~G.}\ \bibnamefont {Rau}}, \bibinfo {author} {\bibfnamefont {R.}~\bibnamefont {Moessner}},\ and\ \bibinfo {author} {\bibfnamefont {P.~A.}\ \bibnamefont {McClarty}},\ }\bibfield  {title} {\bibinfo {title} {{Magnon interactions in the frustrated pyrochlore ferromagnet ${\mathrm{Yb}}_{2}{\mathrm{Ti}}_{2}{\mathrm{O}}_{7}$}},\ }\href {https://doi.org/10.1103/PhysRevB.100.104423} {\bibfield  {journal} {\bibinfo  {journal} {Phys. Rev. B}\ }\textbf {\bibinfo {volume} {100}},\ \bibinfo {pages} {104423} (\bibinfo {year} {2019})}\BibitemShut {NoStop}%
\bibitem [{\citenamefont {Mook}\ \emph {et~al.}(2021)\citenamefont {Mook}, \citenamefont {Plekhanov}, \citenamefont {Klinovaja},\ and\ \citenamefont {Loss}}]{Mook_honeycomb_2021}%
  \BibitemOpen
  \bibfield  {author} {\bibinfo {author} {\bibfnamefont {A.}~\bibnamefont {Mook}}, \bibinfo {author} {\bibfnamefont {K.}~\bibnamefont {Plekhanov}}, \bibinfo {author} {\bibfnamefont {J.}~\bibnamefont {Klinovaja}},\ and\ \bibinfo {author} {\bibfnamefont {D.}~\bibnamefont {Loss}},\ }\bibfield  {title} {\bibinfo {title} {{Interaction-Stabilized Topological Magnon Insulator in Ferromagnets}},\ }\href {https://doi.org/10.1103/PhysRevX.11.021061} {\bibfield  {journal} {\bibinfo  {journal} {Phys. Rev. X}\ }\textbf {\bibinfo {volume} {11}},\ \bibinfo {pages} {021061} (\bibinfo {year} {2021})}\BibitemShut {NoStop}%
\bibitem [{\citenamefont {Hickey}\ \emph {et~al.}(2025)\citenamefont {Hickey}, \citenamefont {Lozano-Gómez},\ and\ \citenamefont {Gingras}}]{Hickey_pyrochlore_2025}%
  \BibitemOpen
  \bibfield  {author} {\bibinfo {author} {\bibfnamefont {A.}~\bibnamefont {Hickey}}, \bibinfo {author} {\bibfnamefont {D.}~\bibnamefont {Lozano-Gómez}},\ and\ \bibinfo {author} {\bibfnamefont {M.~J.~P.}\ \bibnamefont {Gingras}},\ }\href {https://arxiv.org/abs/2403.02391} {\bibinfo {title} {{Order-by-disorder without quantum zero-point fluctuations in the pyrochlore Heisenberg ferromagnet with Dzyaloshinskii-Moriya interactions}}} (\bibinfo {year} {2025}),\ \Eprint {https://arxiv.org/abs/2403.02391} {arXiv:2403.02391} \BibitemShut {NoStop}%
\bibitem [{\citenamefont {McClarty}\ \emph {et~al.}(2018)\citenamefont {McClarty}, \citenamefont {Dong}, \citenamefont {Gohlke}, \citenamefont {Rau}, \citenamefont {Pollmann}, \citenamefont {Moessner},\ and\ \citenamefont {Penc}}]{McClarty2018}%
  \BibitemOpen
  \bibfield  {author} {\bibinfo {author} {\bibfnamefont {P.~A.}\ \bibnamefont {McClarty}}, \bibinfo {author} {\bibfnamefont {X.-Y.}\ \bibnamefont {Dong}}, \bibinfo {author} {\bibfnamefont {M.}~\bibnamefont {Gohlke}}, \bibinfo {author} {\bibfnamefont {J.~G.}\ \bibnamefont {Rau}}, \bibinfo {author} {\bibfnamefont {F.}~\bibnamefont {Pollmann}}, \bibinfo {author} {\bibfnamefont {R.}~\bibnamefont {Moessner}},\ and\ \bibinfo {author} {\bibfnamefont {K.}~\bibnamefont {Penc}},\ }\bibfield  {title} {\bibinfo {title} {{Topological magnons in Kitaev magnets at high fields}},\ }\href {https://doi.org/10.1103/PhysRevB.98.060404} {\bibfield  {journal} {\bibinfo  {journal} {Phys. Rev. B}\ }\textbf {\bibinfo {volume} {98}},\ \bibinfo {pages} {060404} (\bibinfo {year} {2018})}\BibitemShut {NoStop}%
\bibitem [{\citenamefont {C\^onsoli}\ \emph {et~al.}(2020)\citenamefont {C\^onsoli}, \citenamefont {Janssen}, \citenamefont {Vojta},\ and\ \citenamefont {Andrade}}]{Vojta2020_NLSWT}%
  \BibitemOpen
  \bibfield  {author} {\bibinfo {author} {\bibfnamefont {P.~M.}\ \bibnamefont {C\^onsoli}}, \bibinfo {author} {\bibfnamefont {L.}~\bibnamefont {Janssen}}, \bibinfo {author} {\bibfnamefont {M.}~\bibnamefont {Vojta}},\ and\ \bibinfo {author} {\bibfnamefont {E.~C.}\ \bibnamefont {Andrade}},\ }\bibfield  {title} {\bibinfo {title} {{Heisenberg-Kitaev model in a magnetic field: $1/S$ expansion}},\ }\href {https://doi.org/10.1103/PhysRevB.102.155134} {\bibfield  {journal} {\bibinfo  {journal} {Phys. Rev. B}\ }\textbf {\bibinfo {volume} {102}},\ \bibinfo {pages} {155134} (\bibinfo {year} {2020})}\BibitemShut {NoStop}%
\bibitem [{\citenamefont {Koyama}\ and\ \citenamefont {Nasu}(2023)}]{Koyama_2023}%
  \BibitemOpen
  \bibfield  {author} {\bibinfo {author} {\bibfnamefont {S.}~\bibnamefont {Koyama}}\ and\ \bibinfo {author} {\bibfnamefont {J.}~\bibnamefont {Nasu}},\ }\bibfield  {title} {\bibinfo {title} {{Flavor-wave theory with quasiparticle damping at finite temperatures: Application to chiral edge modes in the Kitaev model}},\ }\href {https://doi.org/10.1103/PhysRevB.108.235162} {\bibfield  {journal} {\bibinfo  {journal} {Phys. Rev. B}\ }\textbf {\bibinfo {volume} {108}},\ \bibinfo {pages} {235162} (\bibinfo {year} {2023})}\BibitemShut {NoStop}%
\bibitem [{\citenamefont {Gallegos}\ and\ \citenamefont {Chernyshev}(2024)}]{Gallegos_CoNb2O6}%
  \BibitemOpen
  \bibfield  {author} {\bibinfo {author} {\bibfnamefont {C.~A.}\ \bibnamefont {Gallegos}}\ and\ \bibinfo {author} {\bibfnamefont {A.~L.}\ \bibnamefont {Chernyshev}},\ }\bibfield  {title} {\bibinfo {title} {{Magnon interactions in the quantum paramagnetic phase of ${\mathrm{CoNb}}_{2}{\mathrm{O}}_{6}$}},\ }\href {https://doi.org/10.1103/PhysRevB.109.014424} {\bibfield  {journal} {\bibinfo  {journal} {Phys. Rev. B}\ }\textbf {\bibinfo {volume} {109}},\ \bibinfo {pages} {014424} (\bibinfo {year} {2024})}\BibitemShut {NoStop}%
\bibitem [{\citenamefont {Maksimov}\ and\ \citenamefont {Chernyshev}(2022)}]{us_jpp}%
  \BibitemOpen
  \bibfield  {author} {\bibinfo {author} {\bibfnamefont {P.~A.}\ \bibnamefont {Maksimov}}\ and\ \bibinfo {author} {\bibfnamefont {A.~L.}\ \bibnamefont {Chernyshev}},\ }\bibfield  {title} {\bibinfo {title} {{Easy-plane anisotropic-exchange magnets on a honeycomb lattice: Quantum effects and dealing with them}},\ }\href {https://doi.org/10.1103/PhysRevB.106.214411} {\bibfield  {journal} {\bibinfo  {journal} {Phys. Rev. B}\ }\textbf {\bibinfo {volume} {106}},\ \bibinfo {pages} {214411} (\bibinfo {year} {2022})}\BibitemShut {NoStop}%
\bibitem [{\citenamefont {Chernyshev}\ and\ \citenamefont {Zhitomirsky}(2009)}]{triangle2}%
  \BibitemOpen
  \bibfield  {author} {\bibinfo {author} {\bibfnamefont {A.~L.}\ \bibnamefont {Chernyshev}}\ and\ \bibinfo {author} {\bibfnamefont {M.~E.}\ \bibnamefont {Zhitomirsky}},\ }\bibfield  {title} {\bibinfo {title} {{Spin waves in a triangular lattice antiferromagnet: Decays, spectrum renormalization, and singularities}},\ }\href {https://doi.org/10.1103/PhysRevB.79.144416} {\bibfield  {journal} {\bibinfo  {journal} {Phys. Rev. B}\ }\textbf {\bibinfo {volume} {79}},\ \bibinfo {pages} {144416} (\bibinfo {year} {2009})}\BibitemShut {NoStop}%
\bibitem [{\citenamefont {Zhitomirsky}\ and\ \citenamefont {Chernyshev}(2013)}]{decay_review}%
  \BibitemOpen
  \bibfield  {author} {\bibinfo {author} {\bibfnamefont {M.~E.}\ \bibnamefont {Zhitomirsky}}\ and\ \bibinfo {author} {\bibfnamefont {A.~L.}\ \bibnamefont {Chernyshev}},\ }\bibfield  {title} {\bibinfo {title} {{Colloquium: Spontaneous magnon decays}},\ }\href {https://doi.org/10.1103/RevModPhys.85.219} {\bibfield  {journal} {\bibinfo  {journal} {Rev. Mod. Phys.}\ }\textbf {\bibinfo {volume} {85}},\ \bibinfo {pages} {219} (\bibinfo {year} {2013})}\BibitemShut {NoStop}%
\end{thebibliography}%

\end{document}